\documentclass[a4paper,12pt,colorlinks=true,urlcolor=blue,anchorcolor=blue,citecolor=blue,filecolor=blue,
               linkcolor=blue,menucolor=blue,linktocpage=true,pdfproduc er=medialab,pagebackref=true]{article}
\pdfoutput=1
\usepackage{setspace}
\usepackage{jheppub_X}
\usepackage[T1]{fontenc}
\usepackage{graphicx}
\usepackage{amsmath,mathrsfs,booktabs}
\usepackage{amssymb}
\usepackage[dvipsnames,table]{xcolor}
\usepackage{bbm}
\usepackage{autobreak}
\usepackage{tensor}
\usepackage{bbold}
\usepackage{multicol}
\usepackage{multirow}
\usepackage{enumitem}
\usepackage{cleveref}
\crefname{table}{Table}{Tables}
\crefname{equation}{Eq.}{Eqs.}
\crefname{appendix}{App.}{Apps.}
\crefname{section}{Sec.}{Secs.}
\crefname{figure}{Fig.}{Figs.}
\usepackage[normalem]{ulem}
\usepackage{physics}
\usepackage{slashed}

\usepackage{bm}
\usepackage{dsfont}
\usepackage{pifont}
\usepackage[most]{tcolorbox}
\definecolor{light-gray}{gray}{0.9}
\usepackage{orcidlink}
\usepackage[centertableaux]{ytableau}
\ytableausetup{smalltableaux}
\usepackage{longtable}
\usepackage{lieart}
\usepackage{cancel}
\usepackage{verbatim}

\usepackage{mmacells}

\newcommand{\pd}{{\partial}}
\newcommand{\Lag}{{\mathcal{L}}}
\newcommand{\HS}{{\mathcal{H}}}
\newcommand{\I}{\text{Inv}}

\newcommand{\ring}{\mathbb{r}}
\newcommand{\rI}{{\ring_\I}}

\newcommand{\Mod}{\mathcal{M}}
\newcommand{\trless}[1]{{\overline{#1}}}

\newcommand{\Op}{Q}
\newcommand{\hc}{\text{h.c.}}

\newcommand{\Irrep}{\text{Irrep}}
\newcommand{\tdim}{\text{dim}}
\definecolor{brightpink}{rgb}{1.0, 0.0, 0.5}

\def\eg{\textit{e.g.}}
\def\ie{\textit{i.e.}}

\usepackage{fontawesome}
\def\Mathematica{\texttt{Mathematica}}
\def\HilbCalc{\texttt{HilbCalc}}
\def\link{https://github.com/HilbertSeries/Group_Invariants_and_Covariants}

\definecolor{mygreen}{rgb}{0.0, 0.5, 0.0}

\preprint{}

\title{
Most general EFTs from spurion analysis \\
{\large Hilbert series and Minimal Lepton Flavor Violation}
}

\author[a]{Benjamín Grinstein\orcidlink{0000-0003-2447-4756},}
\author[a]{Xiaochuan Lu\orcidlink{0000-0001-8821-2574},}
\author[b]{Carlos Miró\orcidlink{0000-0003-0336-9025},}
\author[a]{and Pablo Quílez\orcidlink{0000-0002-4327-2706}}

\affiliation[a]{Department of Physics, University of California, San Diego, La Jolla, CA 92093, USA}
\affiliation[b]{Departament de Física Teòrica and Instituto de Física Corpuscular (IFIC),\\ Universitat de València -- CSIC, E-46100 Valencia, Spain}

\emailAdd{bgrinstein@ucsd.edu}
\emailAdd{xil224@ucsd.edu}
\emailAdd{carlos.miro@uv.es}
\emailAdd{pquilezlasanta@ucsd.edu}

\abstract{We derive a saturation theorem for general Effective Field Theories (EFTs) constructed using spurion analysis. Let $S$ be a set of spurion fields introduced to organize the breaking of a global symmetry $G_f$, and $H_S$ be the subgroup of $G_f$ that remains unbroken under a generic vacuum expectation value $\langle S\rangle$; we show that the EFT Lagrangian constructed from the spurion analysis $saturates$ the EFT Lagrangian without the spurions but restricted to $H_S$ invariance, provided that arbitrary powers of the spurion fields are allowed. As examples, we study several implementations of the Minimal Lepton Flavor Violation (MLFV) principle, corresponding to various origins of the neutrino masses. In each scenario, we compute the Hilbert series to obtain the numbers of independent lepton flavor covariants that appear in the corresponding EFT at mass dimension 6. These numbers agree with the number of $H_S$ invariants in the EFT without the spurions, demonstrating the saturation theorem. Motivated by phenomenological connections, we provide linearly independent spurion polynomials for selected lepton flavor covariants. An ancillary file supplies \href{\link}{HilbCalc} \href{\link}{\faGithub}, a \Mathematica\ notebook that provides functions for computing general  Hilbert series of  invariants and covariants of compact classical groups. It presents examples demonstrating the use of the code, including the Hilbert series for our MLFV scenarios.}

\begin{document}
\maketitle
\flushbottom
\setcounter{page}{2}
\newpage
\begin{spacing}{1.1}
\parskip=0ex

\section{Introduction}
\label{sec:Introduction}

The Standard Model (SM) of electroweak interactions very successfully incorporates the three generations of quarks and leptons. The flavor dependence of quark interactions is parametrized in its entirety by two $3\times 3$ Yukawa matrices $Y_u$ and $Y_d$, or equivalently by the six quark masses and the Cabibbo-Kobayashi-Maskawa (CKM) matrix \cite{Cabibbo:1963yz,Kobayashi:1973fv}. The model exhibits the Glashow-Iliopoulos-Maiani (GIM) mechanism \cite{Glashow:1970gm}: an approximate flavor symmetry is responsible for highly suppressed Flavor Changing Neutral Currents (FCNCs). To date, no experimental test of quark flavor interactions conclusively deviates from the predictions of this model, although some ``flavor anomalies'' persist \cite{Capdevila:2023yhq}.

Neither have deviations from the SM been detected in its spontaneous symmetry breaking sector, \ie, the sector of the Higgs and gauge bosons. To test the accurate validity of the SM, a convenient parametrization of possible deviations is the SM Effective Field Theory (SMEFT), where one extends the Lagrangian to include non-renormalizable interactions, \ie, effective operators with mass dimension 5 and above, among the SM fields and subject to the SM symmetries. These non-renormalizable interactions (\eg\ four-quark operators at mass dimension 6) generally give rise to FCNCs, which would contradict experimental evidences~\cite{Cerri:2018ypt}.

The Minimal Flavor Violation (MFV)~\cite{DAmbrosio:2002vsn, Cirigliano:2005ck} hypothesis is designed to describe the class of theories that automatically suppress FCNCs induced by higher dimension operators, through the incorporation of the GIM mechanism in the SMEFT. In this approach, the Yukawas $Y_u$ and $Y_d$ are treated as spurions, that is, as if they were space-time independent fields. In their absence, the SM would display an enlarged global ``quark flavor symmetry'': $G_f = U(3)_q \times U(3)_u \times U(3)_d$, where each unitary group factor acts on the three generations of quarks with common quantum numbers:
\begin{equation}
q_L \;\;\longrightarrow\;\; U_q\, q_L \,,\qquad
u_R \;\;\longrightarrow\;\; U_u\, u_R \,,\qquad
d_R \;\;\longrightarrow\;\; U_d\, d_R \,,
\end{equation}
with $U_q \in U(3)_q$, $U_u \in U(3)_u$, $U_d \in U(3)_d$. When the Yukawa matrices are restored, this quark flavor symmetry can be formally preserved in the SM by requiring the spurions to transform accordingly:
\begin{equation}
Y^{\phantom{\dagger}}_u \;\;\longrightarrow\;\; U^{\phantom{\dagger}}_q\, Y^{\phantom{\dagger}}_u\, U_u^\dagger \,,\qquad
Y^{\phantom{\dagger}}_d \;\;\longrightarrow\;\; U^{\phantom{\dagger}}_q\, Y^{\phantom{\dagger}}_d\, U_d^\dagger \,.
\end{equation}
Now, when it comes to higher dimension operators in SMEFT, the MFV strategy is to require the operator coefficients be built as polynomials in the spurions $Y_u$ and $Y_d$, subject to the condition of formally preserving the quark flavor symmetry. Given these restrictions in the MFV approach, one may ask, how many truly independent coefficients one can construct for an effective operator of a given irreducible representation (irrep) of the flavor group? Said differently, does the MFV principle restrict the coefficients to be in any specific direction in  flavor space?

To study the above questions, three of us with collaborators undertook a systematic investigation on spurion covariants in Ref.~\cite{Grinstein:2023njq}. In particular, we employed and extended the Hilbert series technology \cite{Benvenuti:2006qr, Feng:2007ur, Gray:2008yu, Jenkins:2009dy, Hanany:2010vu, Lehman:2015via, Henning:2015daa, Lehman:2015coa, Henning:2015alf, Henning:2017fpj, Graf:2020yxt, Graf:2022rco} to analyze group covariants in general, and established that the number of linearly independent operators under a given irrep of a flavor group $G_f$ (denoted ``$\Irrep_{G_f}$'') is the rank of the module $_{\rI}\Mod_{\Irrep_{G_f}}$, which can be computed by taking a ratio of two Hilbert series:
\begin{align}
\rank \left(_{\rI}\Mod_{\Irrep_{G_f}} \right) &= \frac{\HS_{\Irrep_{G_f}}(q)}{\HS_{\I_{G_f}}(q)}\Bigg|_{q\to1} \,,
\label{eqn:RankModHSIntro}
\end{align}
where ``$\I_{G_f}$'' denote the invariant irrep of the flavor group. Applying this general analysis to the case of MFV SMEFT, it is found in Ref.~\cite{Grinstein:2023njq} that every direction in the quark flavor space can in fact be fulfilled by the spurion polynomials, for any irrep of the quark flavor group, \ie, ``(quark) MFV SMEFT = SMEFT''. This phenomenon is called ``saturation'' in Ref.~\cite{Grinstein:2023njq}. It is fairly nontrivial, as this is not guaranteed to happen for any combinations of flavor groups and spurions. For instance, a single Yukawa $Y_d$ under the flavor group $U(3)_q \times U(3)_d$ would not display such a saturation. Note that this imagined example of a single bi-fundamental Yukawa spurion under  $U(3)^2$ becomes reality in one of the implementations of the MFV principle to the lepton sector --- in the strict SMEFT scenario (\ie, massless neutrinos), there is indeed only one Yukawa matrix $Y_e$ available.

In this paper, we generalize the saturation observed in Ref.~\cite{Grinstein:2023njq} into a \emph{saturation theorem} (summarized in \cref{eqn:Saturation} below) that holds for general EFTs constructed with spurion analysis. More concretely, we will elaborate in \cref{sec:Saturation} that although the spurion EFT may not always saturate the original EFT, it always saturates that EFT restricted by requiring invariance under  $H_S \subset G_f$,  the  subgroup that leaves invariant a generic vacuum expectation value (vev) of the spurion $S$:\footnote{The subgroup $H_S$ is sometimes called the stabilizer subgroup (of a generic vev), little group, or isotropy group.}
\begin{equation}
G_f \;\;\xrightarrow[]{\;\;\text{generic } \langle S\rangle\;\;}\;\; H_S \subset G_f \,.
\label{eqn:HSdefIntro}
\end{equation}

In \cref{sec:MLFV}, we study the implementations of the MFV principle to the lepton sector. Minimal Lepton Flavor Violation (MLFV) is interesting in its own right \cite{Cirigliano:2005ck, Cirigliano:2006su, Grinstein:2006cg, Raidal:2008jk}. Moreover, it is not a straightforward extension of the quark sector analysis. This is because the origin of neutrino masses can be ascribed to different mechanisms that translate into distinct assumed flavor structures, and correspondingly, different flavor groups and spurions. In this paper, we study the four cases of MLFV summarized in \cref{tab:Cases} below, which cover a variety of origins of the neutrino masses. For each of these cases, we make use of \cref{eqn:RankModHSIntro} to compute the number of independent operators for lepton flavor covariants that appear at mass dimension 6 in the spurion EFT, and verify our new saturation theorem in \cref{eqn:Saturation}. For a selected set of lepton flavor covariants, we also provide the explicit linearly independent spurion polynomials, motivated by phenomenological connections.

We conclude and discuss some future directions in \cref{sec:Outlook}. Some of our MLFV Hilbert series are lengthy and relegated to \cref{appsec:HSMLFV}. They are also collected in a \Mathematica\ notebook \href{\link}{HilbCalc} \href{\link}{\faGithub}, provided as an ancillary file. The notebook has packaged general functions for computing the Haar measures, Weyl characters, and Hilbert series for group invariants and covariants, which can be readily used by the reader for other calculations. A detailed description of how to use these functions is provided in \cref{appsec:Ancillary}.

\section{Spurion analysis for EFTs and the saturation theorem}
\label{sec:Saturation}

The main result of this section is the saturation theorem. It is presented in \cref{subsec:Saturation}. We begin by reviewing the standard spurion analysis in \cref{subsec:Spurion}, which also serves the purpose of introducing notations.

\subsection{EFTs from spurion analysis}
\label{subsec:Spurion}

Consider an EFT specified by a set of dynamical fields $\phi(x)$ and a Lagrangian $\Lag_\text{\,EFT}\big([\phi]\big)$.\footnote{The Lagrangian is a polynomial in $\phi$ and their derivatives; here we are using the shorthand $[\phi] = \big\{ \phi,\, \pd_\mu\phi,\, \pd_\mu\pd_\nu\phi,\, \cdots \big\}$.}
As usual, its definition comes with a power counting scheme and a set of imposed global and/or gauge symmetries. On top of these, let us consider a global symmetry $G_f$ that is not imposed as part of the definition of $\Lag_\text{\,EFT}\big([\phi]\big)$ (\eg\ the lepton flavor group in SMEFT). Sometimes, however, we are interested in gaining some control over the breaking of $G_f$ in $\Lag_\text{\,EFT}\big([\phi]\big)$, and a typical approach is the spurion analysis.

In this analysis, we begin by examining the $G_f$ violations in $\Lag_\text{\,EFT}\big([\phi]\big)$ at its low orders, typically the leading order, and sometimes also the next-to-leading order. We identify the interactions (operators) at these orders that break the symmetry $G_f$, and promote their couplings $\lambda_S$ into a set of non-dynamical \emph{spurion fields} $S$ (or simply \emph{spurions}):
\begin{equation}
\lambda_S \quad\longrightarrow\quad S \,.
\label{eqn:lambdaStoS}
\end{equation}
We require $S$ to transform under $G_f$ in such a way that this symmetry is restored at the low orders of $\Lag_\text{\,EFT}\big([\phi]\big)$. The $G_f$-violating couplings $\lambda_S$ can then be viewed as vevs of the spurion fields: $\lambda_S = \langle S\rangle$.

Next, we consider higher-order interactions in the EFT, which contain the dynamical fields $\phi(x)$ and their derivatives, together with the spurion fields $S$, satisfying $G_f$ invariance. These interactions are codified by a Lagrangian:
\begin{equation}
\Lag_\text{\,EFT} \big( [\phi] , S \big)^{G_f} \,,
\end{equation}
where the superscript indicates symmetry under $G_f$. Finally, the \emph{EFT from spurion analysis} is obtained by replacing the spurion fields $S$ by their vevs:
\begin{equation}
\Lag_\text{\,EFT,\,Spurion} \equiv \Lag_\text{\,EFT} \big( [\phi] , S \big)^{G_f}
\Bigr|_{S = \langle S\rangle = \lambda_S} \,.
\label{eqn:SpurionEFT}
\end{equation}
In this EFT, the $G_f$ invariance is formally respected, and all its breaking effects are attributed to the vevs of $S$, giving us control over the breaking of $G_f$.

Before moving on, let us mention that promoting certain $G_f$-violating coupling $\lambda_S$ in the EFT is a typical way of introducing a spurion field, but it is not the only way. For example, in the QCD Chiral Lagrangian, the mass spurion fields (and their transformations) are motivated from the UV theory (\ie, the QCD Lagrangian), not from promoting a coupling in the Chiral Lagrangian itself.

\subsection{Saturation theorem}
\label{subsec:Saturation}

How much is the EFT from spurion analysis in \cref{eqn:SpurionEFT} more restricted relative to the original EFT? To answer this question, let us denote by $H_S$ the subgroup of $G_f$ that remains unbroken under a \emph{generic} vev $\langle S\rangle$. Since all the $G_f$ breaking effects in $\Lag_\text{\,EFT,\,Spurion}$ are due to a vev of $S$, either generic or special, $\Lag_\text{\,EFT,\,Spurion}$ is guaranteed to respect $H_S$ --- it is \emph{contained} in $\Lag_\text{\,EFT}\big( [\phi] \big)^{H_S}$:
\begin{equation}
\Lag_\text{\,EFT,\,Spurion} \;\subset\; \Lag_\text{\,EFT}\big( [\phi] \big)^{H_S} \,.
\label{eqn:Bound}
\end{equation}
This containing bound ``$\,\subset\,$'' means that the effective operators allowed in $\Lag_\text{\,EFT,\,Spurion}$ is a subset of the effective operators allowed in $\Lag_\text{\,EFT}\big( [\phi] \big)^{H_S}$. In other words, the linear space of Wilson coefficients in $\Lag_\text{\,EFT,\,Spurion}$ is a subspace of that in $\Lag_\text{\,EFT}\big( [\phi] \big)^{H_S}$.

A key result in this paper is a \emph{saturation theorem} --- the bound in \cref{eqn:Bound} is saturated when arbitrary powers of $S$ are included in $\Lag_\text{\,EFT,\,Spurion}$ and when $\lambda_S=\langle S\rangle$ fulfills a generic vev.
\begin{tcolorbox}[colback=light-gray]
\begin{center}
\begin{minipage}{5.5in}\vspace{3pt}
\textbf{Saturation theorem:} Let $H_S$ be the subgroup of $G_f$ that remains unbroken under a \emph{generic} vev of the spurion fields $S$:
\begin{equation}
G_f \;\;\xrightarrow[]{\;\;\text{generic } \langle S\rangle\;\;}\;\; H_S \subset G_f \,,
\label{eqn:HSdef}
\end{equation}
then the EFT from spurion analysis \emph{saturates} the original EFT with $H_S$ imposed:
\begin{equation}
\Lag_\text{\,EFT,\,Spurion} \equiv
\Lag_\text{\,EFT} \big( [\phi] , S \big)^{G_f} \Bigr|_{S = \langle S\rangle = \lambda_S}
\cong \Lag_\text{\,EFT} \big( [\phi] \big)^{H_S} \,,
\label{eqn:Saturation}
\end{equation}
when all powers of $S$ are included in $\Lag_\text{\,EFT}^\text{\,Spurion}$ and $\lambda_S=\langle S\rangle$ is a generic vev.
\end{minipage}
\end{center}
\end{tcolorbox}
\noindent
The equivalence symbol ``$\,\cong\,$'' in \cref{eqn:Saturation} means that the two EFTs have the same set of allowed effective operators, \ie, $\Lag_\text{\,EFT,\,Spurion}$ can reproduce the most general $\Lag_\text{\,EFT} \big( [\phi] \big)^{H_S}$ --- their Wilson coefficients form the same linear space.

Before getting into the technical proof of this saturation theorem, let us first check two extreme scenarios to better see its implications:
\begin{itemize}
\item \textbf{Extreme scenario 1:} All the spurion fields $S$ are invariants under $G_f$. In this case, there is no $G_f$ breaking at all and we expect the EFT from spurion analysis to be equivalent with the original EFT restricted to $G_f$ invariance: $\Lag_\text{\,EFT,\,Spurion} \cong \Lag_\text{\,EFT} \big( [\phi] \big)^{G_f}$. This scenario is consistent with the saturation theorem in \cref{eqn:Saturation}, because $H_S=G_f$ in this case.
\item \textbf{Extreme scenario 2:} We have sufficiently many spurion fields $S$ transforming non-trivially under $G_f$, such that $H_S = \{\mathbf{e}\}$ (\ie, the identity element only). In this case, the saturation theorem in \cref{eqn:Saturation} tells us that the EFT from spurion analysis is equivalent with the original EFT: $\Lag_\text{\,EFT,\,Spurion} \cong \Lag_\text{\,EFT} \big( [\phi] \big)$, namely that with so many spurions, one can reconstruct the most general $\Lag_\text{\,EFT} \big( [\phi] \big)$. This is consistent with the saturation discussed in Ref.~\cite{Grinstein:2023njq} --- it is a special case of our general saturation theorem in \cref{eqn:Saturation}.
\end{itemize}
In the generic case, the EFT from spurion analysis, $\Lag_\text{\,EFT,\,Spurion}$, will be more restricted compared to the original EFT, $\Lag_\text{\,EFT} \big( [\phi] \big)$, but not as restricted as imposing the full $G_f$ on it:
\begin{equation}
\Lag_\text{\,EFT} \big( [\phi] \big)^{G_f}
\quad\subset\quad
\Lag_\text{\,EFT,\,Spurion} = \Lag_\text{\,EFT} \big( [\phi] \big)^{H_S}
\quad\subset\quad
\Lag_\text{\,EFT} \big( [\phi] \big) \,.
\end{equation}
In other words, it is in between the two extreme scenarios discussed above, because the unbroken subgroup $H_S\subset G_f$ is generically bigger than the trivial group, and smaller than the full group $G_f$.

Now let us prove the saturation theorem in \cref{eqn:Saturation}. The EFT from spurion analysis, \ie, $\Lag_\text{\,EFT,\,Spurion}$ constructed in \cref{eqn:SpurionEFT}, consists of a list of effective operators $\Op_i[\phi]$ made out of the dynamical fields $\phi(x)$ and their derivatives (subject to the usual gauge and space-time symmetry requirements and redundancies). Their corresponding Wilson coefficients $C_i(S)$ are polynomials in the spurion fields $S$, evaluated at their vevs $\langle S\rangle = \lambda_S$. One can group the Wilson coefficients into irreps of $G_f$. For each $G_f$-irrep, denoted ``$\Irrep_{G_f}$'' below, the number of independent Wilson coefficients, $C_{\Irrep_{G_f}}(S)$, is given by the rank of the module $_{\rI} \Mod_{\Irrep_{G_f}}^{G_f,\, S}$, where $\rI$ is the ring of the $G_f$-invariant polynomials in $S$; see Ref.~\cite{Grinstein:2023njq} for details. According to a theorem by Brion \cite{brion1993modules, broer1994hilbert}, this rank satisfies
\begin{equation}
\rank\left( _{\rI} \Mod_{\Irrep_{G_f}}^{G_f,\, S} \right) = \tdim \Big( \Irrep_{G_f}^{H_S} \Big) \,,
\label{eqn:RankMod}
\end{equation}
where $\Irrep_{G_f}^{H_S}$ denotes the subspace of $\Irrep_{G_f}$ that is invariant under $H_S$, and hence $\tdim \big( \Irrep_{G_f}^{H_S} \big)$ gives the number of $H_S$-invariants (\ie, $\I_{H_S}$) in the branching rule of $\Irrep_{G_f}$:
\begin{equation}
\Irrep_{G_f} = \tdim \Big( \Irrep_{G_f}^{H_S} \Big)\, \I_{H_S}
\;\oplus\; \text{non-invariant $H_S$ irreps} \,.
\end{equation}
On the other hand, let us check the number of independent operators $\Op_i[\phi]$ in the EFT without spurion but restricted to $H_S$ invariance, \ie, $\Lag_\text{\,EFT} \big( [\phi] \big)^{H_S}$. Following a similar approach, we first organize the Wilson coefficients into $G_f$-irreps. Then for each $\Irrep_{G_f}$, only the $H_S$-invariant component are retained in $\Lag_\text{\,EFT} \big( [\phi] \big)^{H_S}$, so the number precisely agrees with the right-hand side of \cref{eqn:RankMod}. 
Therefore, the two EFTs, $\Lag_\text{\,EFT,\,Spurion}$ and $\Lag_\text{\,EFT} \big( [\phi] \big)^{H_S}$, have the same set of allowed effective operators. This proves the saturation theorem in \cref{eqn:Saturation}.

\section{Minimal Lepton Flavor Violation}
\label{sec:MLFV}

In this section, we discuss four cases of implementations of the Minimal Lepton Flavor Violation (MLFV) principle, which are summarized in \cref{tab:Cases}. 
For  these  we perform spurion analyses, that account for how certain couplings  break the lepton flavor groups $G_{Lf}$. 
We  compute Hilbert series to study the group covariants, and use these to   demonstrate, in each case, the saturation theorem in \cref{eqn:Saturation}. We begin with the SMEFT Cases I and II in \cref{subsec:SMEFT}, and then move on to the $\nu$SMEFT Cases III and IV in \cref{subsec:nuSMEFT}. Calculation details of all the Hilbert series presented in this section (and those gathered in \cref{appsec:HSMLFV}) are available in a \Mathematica\ notebook \href{\link}{HilbCalc} \href{\link}{\faGithub} as an  ancillary file.

In the remainder of this section we often use the same symbols for spurion fields, their generic expectation values and their specific values from Lagrangian parameters. This makes for a more compact and readable presentation and we hope the distinction can be drawn from the context.

\begin{table}[t]
\begin{center}
\renewcommand{\arraystretch}{1.3}
\setlength{\tabcolsep}{0.6em}
\begin{tabular}{ccccc}
\toprule
Case & $\phi$ & $G_{Lf}$ & $S$ & $H_S$ \\
\midrule
I & \multirow{2}{*}{SMEFT} & \multirow{2}{*}{$U(3)_\ell \times U(3)_e$} & $Y_e$ 
                       & $U(1)_e\times U(1)_\mu \times U(1)_\tau$ \\
II &                   & & $Y_e,\, C_5 $ 
                       & $\mathbb{Z}_2=\{1,\,\mathcal{P}_\text{LN}\}$ \\
\midrule
III & \multirow{2}{*}{$\nu$SMEFT} & \multirow{2}{*}{$U(3)_\ell \times U(3)_e \times U(3)_\nu$}
                            & $Y_e,\, Y_\nu $ & $U(1)_\text{LN}$ \\
IV &                        & & $Y_e,\, Y_\nu,\, m_R^{} $ & $\mathbb{Z}_2=\{1,\, \mathcal{P}_\text{LN}\}$ \\
\bottomrule
\end{tabular}
\end{center}
\caption{Four cases of MLFV implementations that we study throughout this paper. They differ in the content of the dynamical fields $\phi$ and the spurion fields $S$. The SMEFT Cases I and II are discussed in \cref{subsubsec:Ye,subsubsec:YeC5}. $\nu$SMEFT supplements the SMEFT by adding three flavors of right-handed neutrinos $\nu$, which are SM gauge singlets. This introduces an additional $U(3)_\nu$ factor in the lepton flavor group $G_{Lf}$. These Cases, III and IV, are discussed in \cref{subsubsec:YeYnu,subsubsec:YeYnumR}. The spurion fields $S$ implicitly include also the hermitian conjugates of the listed matrices in the table. The subgroup of $G_{Lf}$ that is left unbroken by a generic vev of $S$ is indicated under $H_S$, where the lepton parity is $\mathcal{P}_\text{LN} = (-1)^\text{LN}$, with ``LN'' denoting the lepton number.}
\label{tab:Cases}
\end{table}

\subsection{SMEFT with lepton flavor group $G_{Lf,\,\mathrm{SM}} = U(3)_\ell \times U(3)_e$}
\label{subsec:SMEFT}

In this subsection, we consider the case of SMEFT, in which the dynamical fields $\phi(x)$ are the SM fields
\begin{align}
\phi_\text{SM} &\in \Big\{ q,\, u,\, d,\, \ell,\, e,\, H,\, G_{\mu\nu}^A,\, W_{\mu\nu}^a,\, B_{\mu\nu} \Big\}
\notag\\[5pt]
&\quad
+ \Big\{ q^c,\, u^c,\, d^c,\, \ell^c,\, e^c,\, H^\dagger,\, \widetilde{G}_{\mu\nu}^A,\, \widetilde{W}_{\mu\nu}^a,\, \widetilde{B}_{\mu\nu} \Big\} \,.
\label{eqn:phiSMEFT}
\end{align}
Here we follow the \emph{Warsaw basis} notation \cite{Grzadkowski:2010es} for the fermionic fields and drop the chirality indices ``$L$'' and ``$R$'':
\begin{equation}
q = q_L \,,\quad
u = u_R \,,\quad
d = d_R \,,\quad
\ell = \ell_L \,,\quad
e = e_R \,.
\end{equation}
The flavor group in the leptonic sector is
\begin{equation}
G_{Lf,\,\text{SM}}= U(3)_\ell \times U(3)_e \,,
\label{eqn:GfSMEFT}
\end{equation}
under which only $\ell$ and $e$ among the dynamical fields transform:
\begin{equation}
\ell \;\;\longrightarrow\;\; U_\ell\, \ell \,,\qquad
e    \;\;\longrightarrow\;\; U_e\,    e \,.
\label{eqn:TransSMEFT}
\end{equation}
At the leading order of $\Lag_\text{\,SMEFT}$, \ie, $\Lag_\text{\,SM}$, this flavor group is broken by the lepton Yukawa matrices $Y_e^{\phantom{\dagger}}$ and $Y_e^\dagger$:
\begin{equation}
\Lag_\text{\,SM} \supset - \bar\ell\; Y_e\, e\, H + \hc
\label{eqn:Yedef}
\end{equation}
These Yukawas will be considered as the spurion fields in \cref{subsubsec:Ye}. This is the Case I listed in \cref{tab:Cases}.

In \cref{subsubsec:YeC5}, we will move on to the Case II in \cref{tab:Cases}, where we extend the spurion fields $S$ to also include the coupling matrix $C_5$ (and its hermitian conjugate) of Weinberg's dim-5 SMEFT effective operator \cite{Weinberg:1979sa}:
\begin{equation}
\Lag_\text{\,SMEFT}^\text{\,dim-5} = -\frac12\, \big( \overline{\ell^c}\, \widetilde{H}^* \big)\, C_5\, \big( \widetilde{H}^\dagger \ell \big) + \hc
= \frac12\, \epsilon^{ik} \epsilon^{jl} H_i H_j \big( \ell_k^T\, i\gamma^2\gamma^0\, C_5\, \ell_l \big) + \hc \,,
\label{Eq:dim-5Neutrino}
\end{equation}
where $\widetilde{H} \equiv i\sigma^2 H^*$, and the charge conjugation on Dirac spinors is given by $\psi^c = -i\gamma^2 \psi^*$. Here $i,j,k,l$ are $SU(2)_L$ indices, and flavor indices are suppressed, just as in \cref{eqn:Yedef}. From \cref{eqn:TransSMEFT}, we obtain the transformation law of the spurions $Y_e$ and $C_5$ under the lepton flavor group $G_{Lf,\,\text{SM}}= U(3)_\ell \times U(3)_e$:
\begin{subequations}
\begin{alignat}{2}
Y_e   &\;\;\longrightarrow\;\; 
U_\ell^{\phantom{\dagger}} \, Y_e\,   U_e^\dagger \,,\quad
& Y_e   &\;\sim\; (\bm{3}, \bm{\bar{3}}) \;\;\text{with $U(1)_{\ell,\, e}$ charges}\;\; (+1,-1)\,, \\[5pt]
C_5   &\;\;\longrightarrow\;\; U_\ell^*\, C_5\,   U_\ell^\dagger \,,\quad
& C_5   &\;\sim\; (\bm{\bar{6}}, \bm{1}) \;\;\text{with $U(1)_{\ell,\, e}$ charges}\;\; (-2,0) \,.
\end{alignat}
\end{subequations}

\subsubsection{SMEFT spurion analysis with $S_\mathrm{I}=\{Y_e^{\phantom{\dagger}},Y_e^\dagger\}$}
\label{subsubsec:Ye}

Due to the smallness of neutrino masses, one may be interested in studying the class of theories that are captured by the MLFV hypothesis in the massless neutrino limit. In these cases, and within the SM matter content, the only spurions responsible for breaking $G_{Lf,\,\text{SM}}$ are the charged lepton Yukawas $S_\text{I}=\{Y_e^{\phantom{\dagger}},Y_e^\dagger\}$. This is the Case I listed in \cref{tab:Cases}. In order to verify the saturation theorem in \cref{sec:Saturation}, we are going to independently study the operators arising within the EFT Lagrangian from spurion analysis (the LHS of \cref{eqn:Saturation}) and those arising in the $H_{S_\text{I}}$-restricted EFT Lagrangian (the RHS of \cref{eqn:Saturation}). For concreteness, we will focus on the dim-6 operators in SMEFT, and adopt the Warsaw basis \cite{Grzadkowski:2010es}, in which the lepton number preserving operators are those summarized in \cref{tab:MLFVdim6SMEFTLNP}, and the lepton number violating operators \cite{Alonso:2014zka} are those  in \cref{tab:MLFVdim6SMEFTLNV}.

\subsubsection*{EFT from spurion analysis, $\Lag_{\mathrm{\,SMEFT},\,S_\mathrm{I}}$}

In \cref{tab:MLFVdim6SMEFTLNP,tab:MLFVdim6SMEFTLNV}, we show the irrep decompositions of the Wilson coefficients under the lepton flavor group, $U(3)_\ell \times U(3)_e \times U(3)_\nu$, in order for the operator to satisfy the MLFV criterion.\footnote{The transformation properties under $U(3)_\nu$ can be ignored for Cases I and II, with flavor group  $U(3)_\ell \times U(3)_e$, but not for Cases III and IV.}
Our goal is to count and obtain the different flavor structures that can contribute to MLFV SMEFT operators. As explained in \cite{Grinstein:2023njq}, a systematic approach to achieve this involves calculating the Hilbert series for each of the lepton flavor covariants.

Taking the unified grading scheme
\begin{equation}
Y_e^{\phantom{\dagger}} \sim q \,,\quad
Y_e^\dagger \sim q \,,
\end{equation}
we obtain the following Hilbert series for the Lepton Number (LN) preserving covariants in \cref{tab:MLFVdim6SMEFTLNP} (computed with the \Mathematica\ notebook \href{\link}{HilbCalc} \href{\link}{\faGithub})
\allowdisplaybreaks{
\begin{subequations}\label{Eq:HSSMYeLNP1}
\begin{align}
\HS_{\mathbf{(1,1)}} &= \frac{1}{D(q)}
= \frac{1}{\left(1-q^2\right) \left(1-q^4\right) \left(1-q^6\right)} \,, \\[8pt]
\mathcal{H}_{\mathbf{(3,\bar{3})}} &= \frac{1}{D(q)}\, q\, 
\mqty( 1 + q^2 + q^4 ) \,, \\[8pt]
\mathcal{H}_{\mathbf{(8,1)}} &= \frac{1}{D(q)}\, q^2\, 
\mqty( 1 + q^2 ) \,, \\[8pt]
\mathcal{H}_{\mathbf{(8,8)}} &= \frac{1}{D(q)}\, q^2\, 
\mqty( 1 + 3q^2 + 4q^4 + 2q^6 ) \,, \\[8pt]
\mathcal{H}_{\mathbf{(27,1)}} &= \frac{1}{D(q)}\, q^4\, 
\mqty( 1 + q^2 + q^4 ) \,,
\end{align}
\end{subequations}}
as well as
\begin{equation}
\HS_{\mathbf{(1,8)}} = \HS_{\mathbf{(8,1)}} \,,\qquad
\HS_{\mathbf{(1,27)}} = \HS_{\mathbf{(27,1)}} \,.
\label{Eq:HSSMYeLNP2}
\end{equation}
For the lepton number violating covariants appearing in \cref{tab:MLFVdim6SMEFTLNV}, the Hilbert series vanish:
\begin{equation}
\HS_\mathbf{(\bar 3, 1)} = \HS_\mathbf{(1, \bar 3)} = 0 \,.
\label{eqn:HSSMYeLNV}
\end{equation}

As elaborated in Ref.~\cite{Grinstein:2023njq}, in general, the number of linearly independent operators in a $G$-irrep is given by the rank of the module of the $G$-irrep covariants, $_{\rI} \Mod_{\Irrep_G}$, which can be computed by taking the ratio of the two corresponding Hilbert series:
\begin{align}
\rank \left(_{\rI}\Mod_{\Irrep_G} \right) &= \frac{\HS_{\Irrep_G}(q)}{\HS_{\I_G}(q)}\Bigg|_{q\to1} \,.
\label{Eq:RankModHS}
\end{align}
Applying to our case at hand, we evaluate the numerators of each Hilbert series in \cref{Eq:HSSMYeLNP1,Eq:HSSMYeLNP2,eqn:HSSMYeLNV} at $q\to1$ to obtain the ranks:
\begin{subequations}\label{eqn:RankYe}
\begin{align}
\rank \left(_{\rI}\Mod_{\mathbf{(3,\bar{3})}} \right) &= 3 \,, \\[5pt]
\rank \left(_{\rI}\Mod_{\mathbf{(8,1)}} \right) = \rank \left(_{\rI}\Mod_{\mathbf{(1,8)}} \right) &= 2 \,, \\[5pt]
\rank \left(_{\rI}\Mod_{\mathbf{(8,8)}} \right) &= 10 \,, \\[5pt]
\rank \left(_{\rI}\Mod_{\mathbf{(27,1)}} \right) = \rank \left(_{\rI}\Mod_{\mathbf{(1,27)}} \right) &= 3 \,, \\[5pt]
\rank \left(_{\rI}\Mod_{\mathbf{(\bar{3},1)}} \right) = \rank \left(_{\rI}\Mod_{\mathbf{(1,\bar{3})}} \right) &= 0 \,. 
\end{align}
\end{subequations}
Finally, we provide (a choice of) linearly independent covariants for a few modules, namely,  ${}_\rI\Mod_\mathbf{(3,\bar{3})}$, ${}_\rI\Mod_\mathbf{(8,1)}$, and ${}_\rI\Mod_\mathbf{(1,8)}$, in \cref{tab:YeBasis33bar,tab:YeBasis81,tab:YeBasis18}.\footnote{It can be shown that these modules are free modules, and the linearly independent covariants presented in each table actually form a basis of the module --- a complete set of linearly independent covariants that generate the module. We refer the reader to Ref.~\cite{Grinstein:2023njq} for further details on free modules and their bases.}
For convenience, we have introduced the hermitian matrix $h_e^{\phantom{\dagger}} \equiv Y_e^{\phantom{\dagger}}\, Y_e^\dagger$, and used an overline to denote the traceless component of a matrix:
\begin{equation}
\trless{A} \equiv A - \tfrac13 \tr(A)\, I_3 \,,
\label{eqn:Traceless}
\end{equation}
where $I_3$ denotes the $3\times 3$ identity matrix.

\begin{table}[t]
\renewcommand{\arraystretch}{1.2}
\setlength{\arrayrulewidth}{.2mm}
\setlength{\tabcolsep}{1em}
\centering
\begin{tabular}{cc}
\toprule
\(q\) & \( Y_e \) \\
\midrule
\(q^3\) & \( h_e Y_e \) \\
\midrule
\(q^5\) & \( h_e^2 Y_e \) \\
\bottomrule
\end{tabular}
\caption{Basis of the module ${}_\rI\Mod_\mathbf{(3,\bar{3})}$ in Case I MLFV listed in \cref{tab:Cases}. This module has rank 3, as shown in \cref{eqn:RankYe}.}
\label{tab:YeBasis33bar}
\end{table}

\begin{table}[t]
\renewcommand{\arraystretch}{1.2}
\setlength{\arrayrulewidth}{.2mm}
\setlength{\tabcolsep}{1em}
\centering
\begin{tabular}{cc}
\toprule
\(q^2\) & \( \trless{h_e} \) \\
\midrule
\(q^4\) & \( \trless{h_e^2} \) \\
\bottomrule
\end{tabular}
\caption{Basis of the module ${}_\rI\Mod_\mathbf{(8,1)}$ in Case I MLFV listed in \cref{tab:Cases}. This module has rank 2, as shown in \cref{eqn:RankYe}. An overline denotes the traceless component of a matrix; see \cref{eqn:Traceless} for details.}
\label{tab:YeBasis81}
\end{table}

\begin{table}[t]
\renewcommand{\arraystretch}{1.2}
\setlength{\arrayrulewidth}{.2mm}
\setlength{\tabcolsep}{1em}
\centering
\begin{tabular}{cc}
\toprule
\(q^2\) & \( \trless{Y_e^\dag Y_e^{\phantom{\dagger}}} \) \\
\midrule
\(q^4\) & \( \trless{(Y_e^\dag Y_e^{\phantom{\dagger}})^2} \) \\
\bottomrule
\end{tabular}
\caption{Basis of the module ${}_\rI\Mod_\mathbf{(1,8)}$ in Case I MLFV listed in \cref{tab:Cases}. This module has rank 2, as shown in \cref{eqn:RankYe}. An overline denotes the traceless component of a matrix; see \cref{eqn:Traceless} for details.}
\label{tab:YeBasis18}
\end{table}

\subsubsection*{Lepton Flavor Universality Violating EFT, $\Lag_\mathrm{\,SMEFT}^{U(1)_e \times U(1)_\mu \times U(1)_\tau}$}

To see the manifestation of the saturation theorem in \cref{eqn:Saturation} in this example, we work out the unbroken subgroup $H_{S_\text{I}}\subset G_{Lf,\,\text{SMEFT}}$ for a generic vev of $S_\text{I} = \{Y_e^{\phantom{\dagger}},\,Y_e^\dagger\}$:
\begin{align}
U(3)_\ell \times U(3)_{e}
\quad\xrightarrow[]{\quad\langle Y_e^{\phantom{\dagger}},\, Y_e^\dagger\rangle\quad}\quad
U(1)_e \times U(1)_\mu \times U(1)_\tau \equiv H_{S_\text{I}}\,,
\label{Eq:PatternSMEFTYe}
\end{align}
where $U(1)_e\times U(1)_\mu \times U(1)_\tau$ correspond to the three flavor lepton numbers, with $U(1)_\tau$ being the vectorial component of $U(1)_{\tau_L}\times U(1)_{\tau_R} $ and analogously for $e$ and $\mu$. Imposing this symmetry results in a theory with Lepton Flavor Universality Violation (LFUV), in which there is no flavor changing processes nor any lepton number violation. The lepton number violating operators in \cref{tab:MLFVdim6SMEFTLNV} are hence immediately ruled out. The lepton number preserving operators in \cref{tab:MLFVdim6SMEFTLNP} receive further restrictions on their lepton flavor structures, as summarized separately in \cref{tab:MLFVdim6SMEFTLNPH}, where we also provide the resulting numbers of allowed operators. Now it is straightforward to verify that these numbers match the sum of the ranks of the modules listed in \cref{eqn:RankYe}. This confirms the saturation theorem for Case I MLFV in \cref{tab:Cases} at mass dimension 6:
\begin{equation}
\Lag_{\text{\,SMEFT},\,S_\text{I}} \Big|_{\text{dim-}6} \;\cong\;
\Lag_\text{\,SMEFT}^{U(1)_e \times U(1)_\mu \times U(1)_\tau} \Big|_{\text{dim-}6} \,.
\end{equation}
Importantly, the saturation theorem is not limited to this explicit verification; it applies to arbitrary mass dimensions in the SMEFT expansion.

\subsubsection{SMEFT spurion analysis with $S_\mathrm{II} = \{Y_e^{\phantom{\dagger}},Y_e^\dagger, C_5^{\phantom{\dagger}}, C_5^{\dagger}\}$}
\label{subsubsec:YeC5}

Restricting to the SM matter content, neutrino masses can be generated through the dimension-5 operator  in \cref{Eq:dim-5Neutrino}. Hence, it interesting to study the MLFV SMEFT scenario in which the coefficients of the dimension-5 operator are also included into the set of spurions fields: $S_\text{II} = \{Y_e^{\phantom{\dagger}},Y_e^\dagger, C_5^{\phantom{\dagger}}, C_5^{\dagger}\}$.\footnote{Note that $C_5^\dagger=C_5^*$, as the matrix of the dimension-5 operator in \cref{Eq:dim-5Neutrino} is symmetric.}
This is the Case II listed in \cref{tab:Cases}. We will verify the saturation theorem for this case, following an approach similar to that in \cref{subsubsec:Ye}.

\subsubsection*{EFT from spurion analysis, $\Lag_{\text{\,SMEFT},\, S_\text{II}}$}

With the new set of spurions $S_\text{II} = \{Y_e^{\phantom{\dagger}},Y_e^\dagger, C_5^{\phantom{\dagger}}, C_5^{\dagger}\}$, we again compute the Hilbert series for each lepton flavor covariant listed in \cref{tab:MLFVdim6SMEFTLNP,tab:MLFVdim6SMEFTLNV}. Taking the unified grading scheme\footnote{Multi-graded Hilbert series are gathered in \cref{appsubsec:YeC5}.}
\begin{equation}
Y_e^{\phantom{\dagger}}         \sim q \,,\qquad
Y_e^\dagger \sim q \,,\qquad
C_5^{\phantom{\dagger}}           \sim q \,,\qquad
C_5^\dagger \sim q \,,
\end{equation}
we obtain the Hilbert series of lepton number preserving covariants in \cref{tab:MLFVdim6SMEFTLNP} as
\begin{allowdisplaybreaks}
\begin{subequations}\label{Eq:HSSMYeC5LNP}
\begin{align}
\HS_{\mathbf{(1,1)}} &= \frac{1}{D(q)}\, 
\begin{footnotesize}
\mqty( 1 + q^6 + 2 q^8 + 4 q^{10} + 8 q^{12} + 7 q^{14} + 9 q^{16} + 10 q^{18} \\[4pt]
+ 9 q^{20} + 7 q^{22} + 8 q^{24} + 4 q^{26} + 2 q^{28} + q^{30} + q^{36} )
\end{footnotesize} \,, \\[8pt]
\HS_{\mathbf{(3,\bar{3})}} &= \frac{1}{D(q)}\, q\,
\begin{footnotesize}
\mqty( 1 + 2 q^2 + 5 q^4 + 10 q^6 + 21 q^8 + 36 q^{10} + 55 q^{12} \\[4pt]
+ 72 q^{14} + 86 q^{16} + 90 q^{18} + 86 q^{20} + 72 q^{22} + 55 q^{24} \\[4pt]
+ 36 q^{26} + 21 q^{28} + 10 q^{30} + 5 q^{32} + 2 q^{34} + q^{36} )
\end{footnotesize} \,, \\[8pt]
\HS_{\mathbf{(8,1)}} &= \frac{1}{D(q)}\, q^2\,
\begin{footnotesize}
\mqty( 2 + 5 q^2 + 9 q^4 + 19 q^6 + 32 q^8 + 47 q^{10} + 65 q^{12} + 77 q^{14} + 80 q^{16} \\[4pt]
+ 77 q^{18} + 65 q^{20} + 47 q^{22} + 32 q^{24} + 19 q^{26} + 9 q^{28} + 5 q^{30} + 2 q^{32} )
\end{footnotesize} \,, \\[8pt]
\HS_{\mathbf{(1,8)}} &= \frac{1}{D(q)}\, q^2\,
\begin{footnotesize}
\mqty( 1 + 2 q^2 + 4 q^4 + 10 q^6 + 21 q^8 + 35 q^{10} + 53 q^{12} \\[4pt]
+ 68 q^{14} + 78 q^{16} + 83 q^{18} + 77 q^{20} + 62 q^{22} + 46 q^{24} \\[4pt]
+ 29 q^{26} + 13 q^{28} + 6 q^{30} + 3 q^{32} + q^{34} + q^{36} - q^{40} )
\end{footnotesize} \,, \\[8pt]
\HS_{\mathbf{(8,8)}} &= \frac{1}{D(q)}\, q^2\,
\begin{footnotesize}
\mqty( 1 + 7 q^2 + 26 q^4 + 69 q^6 + 150 q^8 + 273 q^{10} + 420 q^{12} \\[4pt]
+ 563 q^{14} + 664 q^{16} + 689 q^{18} + 634 q^{20} + 515 q^{22} + 362 q^{24} \\[4pt]
+ 217 q^{26} + 108 q^{28} + 39 q^{30} + 7 q^{32} - 2 q^{34} - 4 q^{36} - 2 q^{38} )
\end{footnotesize} \,, \\[8pt]
\HS_{\mathbf{(27,1)}} &= \frac{1}{D(q)}\, q^2\,
\begin{footnotesize}
\mqty( 1 + 7 q^2 + 21 q^4 + 52 q^6 + 98 q^8 + 164 q^{10} \\[4pt]
+ 227 q^{12} + 281 q^{14} + 296 q^{16} + 281 q^{18} + 227 q^{20} \\[4pt]
+ 164 q^{22} + 98 q^{24} + 52 q^{26} + 21 q^{28} + 7 q^{30} + q^{32} )
\end{footnotesize} \,, \\[8pt]
\HS_{\mathbf{(1,27)}} &= \frac{1}{D(q)}\, q^4\, 
\begin{footnotesize}
\mqty( 1 + 3 q^2 + 11 q^4 + 29 q^6 + 68 q^8 + 124 q^{10} + 198 q^{12} \\[4pt]
+ 263 q^{14} + 312 q^{16} + 314 q^{18} + 281 q^{20}
+ 209 q^{22} + 133 q^{24} \\[4pt]
+ 62 q^{26} + 18 q^{28} - 5 q^{30} - 9 q^{32} - 7 q^{34} - 4 q^{36} - 2 q^{38} - q^{40} )
\end{footnotesize} \,,
\end{align}
\end{subequations}
\end{allowdisplaybreaks}%
where the denominator factor is
\begin{equation}
\frac{1}{D(q)} = \frac{1}{\left(1 - q^2\right)^2 \left(1 - q^4\right)^3 \left(1 - q^6\right)^4 \left(1 - q^8\right)^2 \left(1 - q^{10}\right)} \,.
\end{equation}
For the lepton number violating covariants in \cref{tab:MLFVdim6SMEFTLNV}, the Hilbert series vanish:
\begin{equation}
\HS_\mathbf{(\bar 3, 1)} = \HS_\mathbf{(1, \bar 3)} = 0 \,.
\label{Eq:HSSMYeC5LNV}
\end{equation}

Again, the number of independent operators is given by the rank of the corresponding module of flavor covariants \cite{Grinstein:2023njq}, which we can compute with the above Hilbert series using \cref{Eq:RankModHS}. Note that the Hilbert series for invariants $\HS_{\mathbf{(1,1)}}$ now has a nontrivial numerator, which evaluates to $74$ under the $q=1$ limit. Therefore, \cref{Eq:RankModHS} is not simply given by evaluating the numerator of covariant Hilbert series at $q=1$. Carrying out the ratio, we find that the rank of a flavor irrep in \cref{Eq:HSSMYeC5LNP} (\ie, lepton number preserving irrep) equals the dimension of that irrep, while the rank of each irrep in \cref{Eq:HSSMYeC5LNV} (\ie, lepton number violating irrep) vanishes:
\begin{subequations}\label{eqn:RankC5}
\begin{alignat}{2}
\rank \left(_{\rI}\Mod_{\Irrep_G} \right) &= \dim (\Irrep_G) \qquad
&&\text{for } \Irrep_G \text{ in \cref{Eq:HSSMYeC5LNP}} \,, \\[5pt]
\rank \left(_{\rI}\Mod_{\Irrep_G} \right) &= 0 \qquad 
&&\text{for } \Irrep_G \text{ in \cref{Eq:HSSMYeC5LNV}} \,.
\end{alignat}
\end{subequations}

\subsubsection*{Lepton parity conserving EFT, $\Lag_\mathrm{\,SMEFT}^{\mathcal{P}_\mathrm{LN}}$}

Under a generic vev of the spurions $S_\text{II} = \{Y_e^{\phantom{\dagger}}, Y_e^\dagger, C_5^{\phantom{\dagger}}, C_5^{\dagger}\}$, the lepton flavor group is broken to the lepton parity:
\begin{equation}
U(3)_\ell \times U(3)_e
\quad\xrightarrow[]{\quad\langle Y_e^{\phantom{\dagger}},\, Y_e^\dagger,\, C_5^{\phantom{\dagger}},\, C_5^{\dagger}\rangle\quad}\quad
\mathbb{Z}_2 = \{ 1, \mathcal{P}_\text{LN} \} \equiv H_{S_\text{II}} \,,
\label{Eq:PatternSMEFTYeC5}
\end{equation}
with $\mathcal{P}_\text{LN} = (-1)^\text{LN}$. This is because the dimension-5 operator breaks the lepton number by two units. Considering the most general Lagrangian that respects this $\mathbb{Z}_2$ symmetry leads to a theory, $\Lag_\text{\,SMEFT}^{\mathcal{P}_\text{LN}}$, with Charged Lepton Flavor Violation (CLFV), yielding non-zero cross sections for processes such as $\mu \to e\gamma$. Additionally, the theory permits lepton number violating processes, provided that the violation is by an even number of units.

In terms of the dim-6 operators in SMEFT, the lepton parity does not place further restrictions on the lepton number preserving operators in \cref{tab:MLFVdim6SMEFTLNP}, because it is a subgroup of the lepton number $U(1)_\text{LN}$. Consequently, all flavor combinations are allowed and the number of operators matches the dimension of the representation indicated in \cref{tab:MLFVdim6SMEFTLNP}. On the other hand, all the lepton number violating dim-6 operators in \cref{tab:MLFVdim6SMEFTLNV} are forbidden by the lepton parity, because they all violate the lepton number by one unit~\cite{Kobach:2016ami,Helset:2019eyc}. In summary, we see that the resulting numbers of dim-6 operators from imposing the lepton parity precisely agree with those from the ranks computed in \cref{eqn:RankC5}:
\begin{equation}
\Lag_{\text{\,SMEFT},\,S_\text{II}} \Big|_{\text{dim-}6} \;\cong\;
\Lag_\text{\,SMEFT}^{\mathcal{P}_\text{LN}} \Big|_{\text{dim-}6} \,.
\end{equation}
This verifies the saturation theorem in \cref{eqn:Saturation} at dim-6 for Case II MLFV in \cref{tab:Cases}.

\subsection{$\nu$SMEFT with lepton flavor group $G_{Lf,\,\nu{\rm SM}} = U(3)_\ell \times U(3)_e \times U(3)_\nu$}
\label{subsec:nuSMEFT}

In this subsection, we move on to consider the case of $\nu$SMEFT, in which the SMEFT field content is supplemented by three flavors of right-handed neutrino fields $\nu$:
\begin{equation}
\phi_{\nu\text{SMEFT}} \in \{\phi_\text{SM} \;,\; \nu\} \,.
\label{eqn:phinuSMEFT}
\end{equation}
These new fields $\nu$ are SM gauge singlets. The flavor group in the leptonic sector is now extended to
\begin{equation}
G_{Lf,\,\nu\text{SM}} = U(3)_\ell \times U(3)_e \times U(3)_\nu \,,
\end{equation}
with the following transformation law of the dynamical fields
\begin{equation}
\ell \;\;\longrightarrow\;\; U_\ell\, \ell \,,\qquad
e    \;\;\longrightarrow\;\; U_e\,    e \,,\qquad
\nu  \;\;\longrightarrow\;\; U_\nu\,  \nu \,.
\label{eqn:TransnuSMEFT}
\end{equation}
$G_{Lf,\,\nu\text{SM}}$ violating couplings in the renormalizable level of the Lagrangian now include the Yukawa couplings $Y_e, Y_\nu$, as well as the complex symmetric Majorana mass matrix $m_R^{}$ for the right-handed neutrinos:
\begin{equation}
\Lag_{\,\nu\text{SM}} \supset -\frac12\, \bar\nu\, m_R\, \nu^c
- \bar\ell\; Y_e\, e\, H - \bar\ell\; Y_\nu\, \nu\, \widetilde{H} + \hc
\label{Eq:NeutrinoMassLagrangian}
\end{equation}
In \cref{subsubsec:YeYnu}, we will consider the spurion fields $S$ only as the Yukawa matrices $Y_e, Y_\nu$ (and their hermitian conjugates), which corresponds to Case III in \cref{tab:Cases}. In this case, lepton number is preserved and the generation of lepton masses is similar to that of the quarks. In \cref{subsubsec:YeYnumR}, we will include also the Majorana mass $m_R^{}$ (and its hermitian conjugate) into $S$, which is the Case IV in \cref{tab:Cases}. The transformation laws of these spurions under $G_{Lf,\,\nu\text{SM}} = U(3)_\ell \times U(3)_e \times U(3)_\nu$ are
\begin{subequations}
\begin{alignat}{2}
Y_e   &\;\longrightarrow\; U_\ell\, Y_e\,   U_e^\dagger \,,
& Y_e   &\sim (\bm{3}, \bm{\bar{3}}, \bm{1}) \;\;\text{with $U(1)_{\ell,\, e,\, \nu}$ charges}\;\; (+1,-1,0) \,, \\[5pt]
Y_\nu &\;\longrightarrow\; U_\ell\, Y_\nu\, U_\nu^\dagger \,,
& Y_\nu &\sim (\bm{3}, \bm{1}, \bm{\bar{3}}) \;\;\text{with $U(1)_{\ell,\, e,\, \nu}$ charges}\;\; (+1,0,-1) \,, \\[5pt]
m_R^{} &\;\longrightarrow\; U_\nu\,  m_R^{}\, U_\nu^T \,,\quad
& m_R^{} &\sim (\bm{1}, \bm{1}, \bm{6}) \;\;\text{with $U(1)_{\ell,\, e,\, \nu}$ charges}\;\; (0,0,+2) \,.
\end{alignat}
\end{subequations}

\subsubsection{$\nu$SMEFT spurion analysis with $S_\mathrm{III} = \{Y_e^{\phantom{\dagger}}, Y_e^\dagger, Y_\nu^{\phantom{\dagger}}, Y_\nu^\dagger\}$}
\label{subsubsec:YeYnu}

This is the Case III in \cref{tab:Cases}, where we take the spurion fields $S_\text{III} = \{Y_e^{\phantom{\dagger}}, Y_e^\dagger, Y_\nu^{\phantom{\dagger}}, Y_\nu^\dagger\}$ as responsible for the breaking of $G_{Lf,\,\nu\text{SMEFT}}$. With the lepton number preserved, one gets Dirac neutrino masses from the Yukawa coupling $Y_\nu$ in \cref{Eq:NeutrinoMassLagrangian}. This case is in close parallelism with the quark sector of SMEFT, to which all the results presented below apply upon the replacements $Y_e \to Y_d$ and $Y_\nu \to Y_u$.

Since $\nu$SMEFT extends the SMEFT field content with 3 right-handed neutrino fields $\nu$, it has more dim-6 operators in addition to the SMEFT ones listed in \cref{tab:MLFVdim6SMEFTLNP,tab:MLFVdim6SMEFTLNV}. They are summarized in \cref{tab:MLFVdim6nuSMEFTLNP} for lepton number preserving ones and in \cref{tab:MLFVdim6nuSMEFTLNV} for lepton number violating ones. In below, we verify the saturation theorem by comparing the numbers of dim-6 $\nu$SMEFT operators in the EFT from spurion analysis and the $H_S$-restricted EFT.

\subsubsection*{EFT from spurion analysis, $\Lag_{\text{\,$\nu$SMEFT},\,S_\text{III}}$}

Taking the unified grading scheme\footnote{Multi-graded Hilbert series are gathered in \cref{appsubsec:YeYnu}.}
\begin{equation}
Y_e^{\phantom{\dagger}} \sim q \,,\qquad
Y_e^\dagger \sim q \,,\qquad
Y_\nu^{\phantom{\dagger}} \sim q \,,\qquad
Y_\nu^\dagger \sim q \,.
\end{equation}
we obtain the following Hilbert series for the lepton number preserving covariants that appear in \cref{tab:MLFVdim6SMEFTLNP,tab:MLFVdim6nuSMEFTLNP}:
\begin{subequations}\label{Eq:HSnuSMYeYnuLNP1}
\begingroup
\allowdisplaybreaks
\begin{align}
\HS_{\mathbf{(1,1,1)}} &= \frac{1}{D(q)}\, 
\mqty( 1 + q^{12} ) \,, \\[8pt]
\HS_{\mathbf{(3,\bar{3},1)}} &= \frac{1}{D(q)}\, q\, 
\mqty( 1 + 2 q^2 + 4 q^4 + 4 q^6 + 4 q^8 + 2 q^{10} + q^{12} ) \,, \\[8pt]
\HS_{\mathbf{(1,\bar{3},3)}} &= \frac{1}{D(q)}\, q^2\, 
\mqty( 1 + 2q^2 + 4q^4 + 4q^6 + 4q^8 + 2q^{10} + q^{12} ) \,, \\[8pt]
\HS_{\mathbf{(\bar{3},\bar{3},\bar{3})}} &= \frac{1}{D(q)}\, q^2\, 
\mqty( 1 + 4q^2 + 9q^4 + 14q^6 + 15q^8 + 12q^{10} \\[4pt]
+ 5q^{12} - 3q^{16} - 2q^{18} - q^{20} ) \,, \\[8pt]
\HS_{\mathbf{(6,\bar{3},\bar{3})}} &= \frac{1}{D(q)}\, q^2\, 
\mqty( 1 + 4q^2 + 12q^4 + 22q^6 + 32q^8 + 32q^{10} + 24q^{12} \\[4pt]
+ 8q^{14} - 4q^{16} - 10q^{18} - 8q^{20} - 4q^{22} - q^{24} ) \,, \\[8pt]
\HS_{\mathbf{(8,1,1)}} &= \frac{1}{D(q)}\, 2q^2\, 
\mqty( 1 + 2q^2 + 2q^4 + 2q^6 + q^8 ) \,, \\[8pt]
\HS_{\mathbf{(1,8,1)}} &= \frac{1}{D(q)}\, q^2\, 
\mqty( 1 + 2q^2 + 3q^4 + 4q^6 + 4q^8 + 2q^{10} + q^{12} - q^{16} ) \,, \\[8pt]
\HS_{\mathbf{(8,8,1)}} &= \frac{1}{D(q)}\, q^2\, 
\mqty( 1 + 6q^2 + 17q^4 + 30q^6 + 39q^8 + 38q^{10} + 24q^{12}  \\[4pt]
+ 6q^{14} - 7q^{16} - 12q^{18} - 9q^{20} - 4q^{22} - q^{24} ) \,, \\[8pt]
\HS_{\mathbf{(1,8,8)}} &= \frac{1}{D(q)}\, q^4\, 
\mqty( 2 + 8q^2 + 19q^4 + 32q^6 + 40q^8 + 36q^{10} + 21q^{12} \\[4pt]
+ 4q^{14} - 9q^{16} - 12q^{18} - 8q^{20} - 4q^{22} - q^{24} ) \,, \\[8pt]
\HS_{\mathbf{(27,1,1)}} &= \frac{1}{D(q)}\, q^4\, 
\mqty( 3 + 8q^2 + 17q^4 + 20q^6 + 19q^8 + 8q^{10} \\[4pt]
- q^{12} - 8q^{14} - 7q^{16} - 4q^{18} - q^{20} ) \,, \\[8pt]
\HS_{\mathbf{(1,27,1)}} &= \frac{1}{D(q)}\, q^4\, 
\mqty( 1 + 2q^2 + 6q^4 + 10q^6 + 17q^8 + 18q^{10} + 16q^{12} \\[4pt]
+ 6q^{14} - 2q^{16} - 8q^{18} - 7q^{20} - 4q^{22} - q^{24} ) \,,
\end{align}
\endgroup
\end{subequations}
together with
\begin{subequations}\label{Eq:HSnuSMYeYnuLNP2}
\begin{align}
\HS_{\mathbf{(3,1,\bar{3})}} = \HS_{\mathbf{(\bar{3},1,3)}} &= \HS_{\mathbf{(3,\bar{3},1)}} \,, \\[5pt]
\HS_{\mathbf{(1,1,8)}} &= \HS_{\mathbf{(1,8,1)}} \,, \\[5pt]
\HS_{\mathbf{(8,1,8)}} &= \HS_{\mathbf{(8,8,1)}} \,, \\[5pt]
\HS_{\mathbf{(1,1,27)}} &= \HS_{\mathbf{(1,27,1)}} \,,
\end{align}
\end{subequations}
where the denominator factor is given by
\begin{equation}
\frac{1}{D(q)} = \frac{1}{\left(1 - q^2\right)^2 \left(1 - q^4\right)^3 \left(1 - q^6\right)^4 \left(1 - q^8\right)} \,.
\end{equation}
For the lepton number violating covariants appearing in \cref{tab:MLFVdim6SMEFTLNV,tab:MLFVdim6nuSMEFTLNV}, the Hilbert series vanish:
\begin{equation} \label{Eq:HSnuSMYeYnuLNV}
\HS_\mathbf{(\bar 3, 1, 1)} = \HS_\mathbf{(1, \bar 3, 1)} = \HS_\mathbf{(1, 1, \bar 3)}
= \HS_\mathbf{(1, 1 , 6)} = 0 \,.
\end{equation}

As before, to get the number of independent operators, we compute the rank of the corresponding modules with \cref{Eq:RankModHS} using the above Hilbert series. We find that the rank of a flavor irrep in \cref{Eq:HSnuSMYeYnuLNP1,Eq:HSnuSMYeYnuLNP2} (\ie, lepton number preserving irrep) equals the dimension of that irrep, while the rank of an irrep in \cref{Eq:HSnuSMYeYnuLNV} (\ie, lepton number violating irrep) vanish:
\begin{subequations}\label{eqn:RankYeYnu}
\begin{alignat}{2}
\rank \left(_{\rI}\Mod_{\Irrep_G} \right) &= \dim (\Irrep_G) \qquad
&&\text{for } \Irrep_G \text{ in \cref{Eq:HSnuSMYeYnuLNP1,Eq:HSnuSMYeYnuLNP2}} \,, \\[5pt]
\rank \left(_{\rI}\Mod_{\Irrep_G} \right) &= 0 \qquad 
&&\text{for } \Irrep_G \text{ in \cref{Eq:HSnuSMYeYnuLNV}} \,.
\end{alignat}
\end{subequations}
Finally, we provide (a choice of) linearly independent covariants for a few modules ${}_\rI\Mod_\mathbf{(3,\bar{3},1)}$, ${}_\rI\Mod_\mathbf{(8,1,1)}$, ${}_\rI\Mod_\mathbf{(3,1,\bar{3})}$, ${}_\rI\Mod_\mathbf{(1,3,\bar{3})}$, and ${}_\rI\Mod_\mathbf{(1,8,1)}$ in \cref{tab:YnuYeBasis33bar1,tab:YnuYeBasis811,tab:YnuYeBasis313bar,tab:YnuYeBasis133bar,tab:YnuYeBasis181},\footnote{Different from those in \cref{tab:YeBasis33bar,tab:YeBasis81,tab:YeBasis18}, the linearly independent covariants here do not form a basis of the module, because the modules are not free and they do not possess a basis.}
where we have introduced a hermitian matrix $h_\nu \equiv Y_\nu^{\phantom{\dagger}}\, Y_\nu^\dagger$ for convenience.

\begin{table}[t]
\renewcommand{\arraystretch}{1.5}
\setlength{\arrayrulewidth}{.2mm}
\setlength{\tabcolsep}{1em}
\centering
\begin{tabular}{cc}
\toprule
\(q^2\) & \( \trless{h_\nu}\quad,\quad \trless{h_e} \) \\
\midrule
\(q^4\) & \( \trless{h_\nu^2}\quad,\quad \trless{h_e^2}\quad,\quad \trless{h_\nu h_e + \hc } \) \\
\midrule
\(q^6\) & \( \trless{h_\nu^2 h_e + \hc}\quad,\quad \trless{h_\nu h_e^2 + \hc} \) \\
\midrule
\(q^8\) & \( \trless{h_\nu^2 h_e^2 + \hc} \) \\
\bottomrule
\end{tabular}
\caption{A linearly independent set of covariants in the module ${}_\rI\Mod_\mathbf{(8,1,1)}$ in Case III MLFV listed in \cref{tab:Cases}. An overline denotes the traceless component of a matrix; see \cref{eqn:Traceless} for details.}
\label{tab:YnuYeBasis811}
\end{table}

\begin{table}[t]
\renewcommand{\arraystretch}{1.2}
\setlength{\arrayrulewidth}{.2mm}
\setlength{\tabcolsep}{1em}
\centering
\begin{tabular}{cc}
\toprule
\(q\) & \(  Y_\nu \) \\
\midrule
\(q^3\) & \( h_\nu\, Y_\nu\quad,\quad h_e\, Y_\nu \) \\
\midrule
\(q^5\) & \( h_\nu^2\, Y_\nu\quad,\quad   h_e^2\, Y_\nu\quad,\quad   (h_\nu h_e + \hc)\, Y_\nu \) \\
\midrule
\(q^7\) & \( (h_\nu^2 h_e + \hc)\, Y_\nu\quad,\quad   (h_\nu h_e^2 + \hc)\, Y_\nu \) \\
\midrule
\(q^9\) & \( (h_\nu^2 h_e^2 + \hc)\, Y_\nu \) \\
\bottomrule
\end{tabular}
\caption{A linearly independent set of covariants in the module ${}_\rI\Mod_\mathbf{(3,\bar 3,1)}$ in Case III MLFV listed in \cref{tab:Cases}.}
\label{tab:YnuYeBasis33bar1}
\end{table}

\begin{table}[t]
\renewcommand{\arraystretch}{1.2}
\setlength{\arrayrulewidth}{.2mm}
\setlength{\tabcolsep}{1em}
\centering
\begin{tabular}{cc}
\toprule
\(q\) & \(  Y_e \) \\
\midrule
\(q^3\) & \( h_\nu\, Y_e \quad,\quad h_e\, Y_e \) \\
\midrule
\(q^5\) & \( h_\nu^2\, Y_e \quad,\quad   h_e^2\, Y_e \quad,\quad   (h_\nu h_e + \hc)\, Y_e \) \\
\midrule
\(q^7\) & \( (h_\nu^2 h_e + \hc)\, Y_e \quad,\quad   (h_\nu h_e^2 + \hc)\, Y_e \) \\
\midrule
\(q^9\) & \( (h_\nu^2 h_e^2 + \hc)\, Y_e \) \\
\bottomrule
\end{tabular}
\caption{A linearly independent set of covariants in the module ${}_\rI\Mod_\mathbf{(3,1,\bar 3)}$ in Case III MLFV listed in \cref{tab:Cases}.}
\label{tab:YnuYeBasis313bar}
\end{table}

\begin{table}[t]
\renewcommand{\arraystretch}{1.2}
\setlength{\arrayrulewidth}{.2mm}
\setlength{\tabcolsep}{1em}
\centering
\begin{tabular}{cc}
\toprule
\(q^2\) & \( Y_\nu^\dag Y_e \) \\
\midrule
\(q^4\) & \( Y_\nu^\dag h_\nu^{\phantom{\dagger}}\, Y_e^{\phantom{\dagger}} \quad,\quad Y_\nu^\dag h_e\, Y_e^{\phantom{\dagger}} \) \\
\midrule
\(q^6\) & \( Y_\nu^\dag h_\nu^2\, Y_e^{\phantom{\dagger}} \quad,\quad  Y_\nu^\dag h_e^2\, Y_e^{\phantom{\dagger}} \quad,\quad  Y_\nu^\dag (h_\nu h_e + \hc)\, Y_e \) \\
\midrule
\(q^8\) & \( Y_\nu^\dag (h_\nu^2 h_e^{\phantom{\dagger}} + \hc)\, Y_e^{\phantom{\dagger}} \quad,\quad  Y_\nu^\dag (h_\nu^{\phantom{\dagger}} h_e^2 + \hc)\, Y_e^{\phantom{\dagger}} \) \\
\midrule
\(q^{12}\) & \( Y_\nu^\dag (h_\nu^2 h_e^2 + \hc)\, Y_e^{\phantom{\dagger}} \) \\
\bottomrule
\end{tabular}
\caption{A linearly independent set of covariants in the module ${}_\rI\Mod_\mathbf{(1,3,\bar 3)}$ in Case III MLFV listed in \cref{tab:Cases}.}
\label{tab:YnuYeBasis133bar}
\end{table}

\begin{table}[t]
\renewcommand{\arraystretch}{1.2}
\setlength{\arrayrulewidth}{.2mm}
\setlength{\tabcolsep}{1em}
\centering
\begin{tabular}{cc}
\toprule
\(q^2\) & \( \trless{Y_\nu^\dag Y_\nu^{\phantom{\dagger}}} \) \\
\midrule
\(q^4\) & \( \trless{Y_\nu^\dag h_\nu\, Y_\nu^{\phantom{\dagger}}} \quad,\quad \trless{Y_\nu^\dag h_e^{\phantom{\dagger}}\, Y_\nu^{\phantom{\dagger}}} \) \\
\midrule
\(q^6\) & \( \trless{Y_\nu^\dag h_e^2\, Y_\nu} \quad,\quad  \trless{Y_\nu^\dag (h_\nu h_e + \hc)\, Y_\nu} \) \\
\midrule
\(q^8\) & \( \trless{Y_\nu^\dag (h_\nu^2 h_e + \hc)\, Y_\nu} \quad,\quad  \trless{Y_\nu^\dag (h_\nu h_e^2 + \hc)\, Y_\nu} \) \\
\midrule
\(q^{12}\) & \( \trless{Y_\nu^\dag (h_\nu^2 h_e^2 + \hc)\, Y_\nu} \) \\
\bottomrule
\end{tabular}
\caption{A linearly independent set of covariants in the module ${}_\rI\Mod_\mathbf{(1,8,1)}$ in Case III MLFV listed in \cref{tab:Cases}. An overline denotes the traceless component of a matrix; see \cref{eqn:Traceless} for details.}
\label{tab:YnuYeBasis181}
\end{table}

\subsubsection*{Lepton number conserving EFT, $\Lag_{\,\nu\mathrm{SMEFT}}^{U(1)_\mathrm{LN}}$}

Similarly to the case of quark MFV, the lepton flavor group breaks to lepton number once the spurion Yukawas $S_{\rm III} = \{Y_e^{\phantom{\dagger}}, Y_e^\dagger, Y_\nu^{\phantom{\dagger}}, Y_\nu^\dagger\}$ take a generic vev:
\begin{align}
U(3)_{\ell} \times U(3)_{e} \times U(3)_{\nu}
\quad\xrightarrow[]{\quad\langle Y_e^{\phantom{\dagger}},\, Y_e^\dagger,\, Y_\nu^{\phantom{\dagger}},\, Y_\nu^\dagger \rangle\quad}\quad
U(1)_\text{LN} \equiv H_{S_{\rm III}}\,.
\label{Eq:PatternSMEFTYeYnu}
\end{align}
The most general Lagrangian at mass dimension 6 respecting lepton number includes all the flavor combinations in \cref{tab:MLFVdim6SMEFTLNP,tab:MLFVdim6nuSMEFTLNP}, and none of the operators in \cref{tab:MLFVdim6SMEFTLNV} or \cref{tab:MLFVdim6nuSMEFTLNV}. As a consequence, the number of $U(1)_\text{LN}$-invariant dim-6 operators exactly match the ranks computed from the spurion analysis in \cref{eqn:RankYeYnu}:
\begin{equation}
\Lag_{\text{\,$\nu$SMEFT},\,S_\text{III}} \Big|_{\text{dim-}6} \;\cong\;
\Lag_\text{\,$\nu$SMEFT}^{U(1)_\text{LN}} \Big|_{\text{dim-}6} \,.
\end{equation}
This verifies the saturation theorem in \cref{eqn:Saturation} at mass dimension 6 for Case III MLFV in \cref{tab:Cases}.

\subsubsection{$\nu$SMEFT spurion analysis with $S_\mathrm{IV} = \{Y_e^{\phantom{\dagger}}, Y_e^\dagger, Y_\nu^{\phantom{\dagger}}, Y_\nu^\dagger, m_R^{}, m_R^\dagger\}$}
\label{subsubsec:YeYnumR}

Since right-handed neutrinos are singlets under the SM gauge group, a Majorana mass term for them is not forbidden. Consequently, the most general renormalizable Lagrangian for the lepton sector with three right-handed neutrinos includes all the terms in \cref{Eq:NeutrinoMassLagrangian}. This motivates the study of MFLV with spurions $S_{\rm IV} = \{Y_e^{\phantom{\dagger}}, Y_e^\dagger, Y_\nu^{\phantom{\dagger}}, Y_\nu^\dagger, m_R^{}, m_R^\dagger\}$. Let us verify the saturation theorem in \cref{eqn:Saturation} by studying the dim-6 operators in \cref{tab:MLFVdim6SMEFTLNP,tab:MLFVdim6SMEFTLNV,tab:MLFVdim6nuSMEFTLNP,tab:MLFVdim6nuSMEFTLNV}.

\subsubsection*{EFT from spurion analysis, $\Lag_{\,\nu\mathrm{SMEFT},\,S_\mathrm{IV}}$}

Taking the unified grading scheme
\begin{equation}
Y_e^{\phantom{\dagger}} \sim q \,,\qquad
Y_e^\dagger \sim q \,,\qquad
Y_\nu^{\phantom{\dagger}} \sim q \,,\qquad
Y_\nu^\dagger \sim q \,\qquad
m_R^{\phantom{\dagger}} \sim q \,,\qquad
m_R^\dagger \sim q \,,
\end{equation}
we have computed the Hilbert series for all the lepton flavor covariants listed in \cref{tab:MLFVdim6SMEFTLNP,tab:MLFVdim6SMEFTLNV,tab:MLFVdim6nuSMEFTLNP,tab:MLFVdim6nuSMEFTLNV}. For example, the Hilbert series for the invariants and the $\mathbf{(1,1,6)}$-covariants read
\begin{subequations}
\begin{align}
\HS_{\mathbf{(1,1,1)}} &= \frac{1}{D(q)}\,
\begin{footnotesize}
\mqty( 1 + q^4 + 5q^6 + 9q^8 + 22q^{10} + 61q^{12} + 126q^{14} \\[4pt]
+ 273q^{16} + 552q^{18} + 1038q^{20} + 1880q^{22} + 3293q^{24} + 5441q^{26} \\[4pt]
+ 8712q^{28} + 13417q^{30} + 19867q^{32} + 28414q^{34} + 39351q^{36} \\[4pt]
+ 52604q^{38} + 68220q^{40} + 85783q^{42} + 104588q^{44} + 123852q^{46} \\[4pt]
+ 142559q^{48} + 159328q^{50} + 173201q^{52} + 183138q^{54} + 188232q^{56} \\[4pt]
+ 188232q^{58} + 183138q^{60} + 173201q^{62} + 159328q^{64} + 142559q^{66} \\[4pt]
+ 123852q^{68} + 104588q^{70} + 85783q^{72} + 68220q^{74} + 52604q^{76} \\[4pt]
+ 39351q^{78} + 28414q^{80} + 19867q^{82} + 13417q^{84} + 8712q^{86} \\[4pt]
+ 5441q^{88} + 3293q^{90} + 1880q^{92} + 1038q^{94} + 552q^{96} + 273q^{98} \\[4pt]
+ 126q^{100} + 61q^{102} + 22q^{104} + 9q^{106} + 5q^{108} + q^{110} + q^{114} )
\end{footnotesize} \,, \\[10pt]
\HS_{\mathbf{(1,1,6)}} &= \frac{1}{D(q)}\, q^2\,
\begin{footnotesize}
\mqty( 1 + 2q^2 + 8q^4 + 24q^6 + 75q^8 + 198q^{10} + 498q^{12} \\[4pt]
+ 1135q^{14} + 2434q^{16} + 4870q^{18} + 9247q^{20} + 16622q^{22} \\[4pt]
+ 28528q^{24} + 46754q^{26} + 73518q^{28} + 111013q^{30} \\[4pt]
+ 161467q^{32} + 226415q^{34} + 306716q^{36} + 401724q^{38} \\[4pt]
+ 509439q^{40} + 625911q^{42} + 745783q^{44} + 862182q^{46} \\[4pt]
+ 967716q^{48} + 1054864q^{50} + 1117120q^{52} + 1149544q^{54} \\[4pt]
+ 1149544q^{56} + 1117120q^{58} + 1054864q^{60} + 967716q^{62} \\[4pt]
+ 862182q^{64} + 745783q^{66} + 625911q^{68} + 509439q^{70} \\[4pt]
+ 401724q^{72} + 306716q^{74} + 226415q^{76} + 161467q^{78} \\[4pt]
+ 111013q^{80} + 73518q^{82} + 46754q^{84} + 28528q^{86} \\[4pt]
+ 16622q^{88} + 9247q^{90} + 4870q^{92} + 2434q^{94} + 1135q^{96} \\[4pt]
+ 498q^{98} + 198q^{100} + 75q^{102} + 24q^{104} + 8q^{106} + 2q^{108} + q^{110} )
\end{footnotesize} \,,
\end{align}
\end{subequations}%
where the denominator factor is given by
\begin{footnotesize}
\begin{equation}
\frac{1}{D(q)} = \frac{1}{\left(1-q^2\right)^3 \left(1-q^4\right)^4 \left(1-q^6\right)^4 \left(1-q^8\right)^2 \left(1-q^{10}\right)^2 \left(1-q^{12}\right)^3 \left(1-q^{14}\right)^2 \left(1-q^{16}\right)} \,.
\end{equation}
\end{footnotesize}%
Since the Hilbert series in this MLFV scenario are lengthy, we relegate the other single-graded Hilbert series to \cref{appsubsec:YeYnumR}.\footnote{Hilbert series with a two-variable grading scheme $\{Y_e^{\phantom{\dagger}},\, Y_e^\dagger,\, Y_\nu^{\phantom{\dagger}},\, Y_\nu^\dagger\} \sim q$ and $\{m_R,\, m_R^\dagger\} \sim p$ are also available for selected covariants in our ancillary \Mathematica\ notebook \href{\link}{HilbCalc} \href{\link}{\faGithub}.}

With the Hilbert series given above and in \cref{appsubsec:YeYnumR}, we can make use of \cref{Eq:RankModHS} to compute the ranks of the modules, which give the numbers of the linearly independent operators in the EFT from spurion analysis. For lepton number preserving operators in \cref{tab:MLFVdim6SMEFTLNP,tab:MLFVdim6nuSMEFTLNP}, the rank of each flavor irrep equals the dimension of the irrep:
\begin{equation}
\rank \left(_{\rI}\Mod_{\Irrep_G} \right) = \dim (\Irrep_G)
\qquad\text{for } \Irrep_G \text{ in \cref{tab:MLFVdim6SMEFTLNP,tab:MLFVdim6nuSMEFTLNP}} \,.
\label{Eq:RankYeYnumRP}
\end{equation}
For lepton number violating operators in \cref{tab:MLFVdim6SMEFTLNV,tab:MLFVdim6nuSMEFTLNV}, the ranks vanish except for the irrep $\mathbf{(1,1,6)}$:
\begin{subequations}\label{Eq:RankYeYnumRV}
\begin{align}
\rank \left(_{\rI}\Mod_{\mathbf{(1,1,6)}} \right) &= 6 \,, \\[5pt]
\rank \left(_{\rI}\Mod_{\Irrep_G} \right) &= 0
\qquad\text{for other } \Irrep_G \text{ in \cref{tab:MLFVdim6SMEFTLNV,tab:MLFVdim6nuSMEFTLNV}} \,.
\end{align}
\end{subequations}

\subsubsection*{Lepton parity conserving EFT, $\Lag_{\,\nu\mathrm{SMEFT}}^{\mathcal{P}_\mathrm{LN}}$}

Under a generic vev of the spurions $S_\text{IV}= \{Y_e^{\phantom{\dagger}}, Y_e^\dagger, Y_\nu^{\phantom{\dagger}}, Y_\nu^\dagger, m_R^{}, m_R^\dagger\}$ the lepton flavor group breaks into the lepton parity:
\begin{align}
U(3)_{\ell} \times U(3)_{e} \times U(3)_{\nu}
\;\;\xrightarrow[]{\quad\langle Y_e^{\phantom{\dagger}},\,Y_e^\dagger,\,Y_\nu^{\phantom{\dagger}},\,Y_\nu^\dagger\,,m_R,\,m_R^\dagger\rangle\quad}\;\;
\mathbb{Z}_2 = \{ 1, \mathcal{P}_\text{LN} \} \equiv H_{S_{\rm IV}} \,,
\label{Eq:PatternSMEFTYeYnumR}
\end{align}
with $\mathcal{P}_\text{LN} = (-1)^\text{LN}$ as before. Much as the case in \cref{Eq:PatternSMEFTYeC5}, the Majorana mass term for right-handed neutrinos breaks lepton number by two units and hence lepton parity is conserved.

The most general lepton parity conserving $\nu$SMEFT Lagrangian includes all the lepton number preserving operators in \cref{tab:MLFVdim6SMEFTLNP,tab:MLFVdim6nuSMEFTLNP}. As for the lepton number violating operators in \cref{tab:MLFVdim6SMEFTLNV,tab:MLFVdim6nuSMEFTLNV}, most of them are forbidden by the lepton parity $\mathcal{P}_\text{LN} = (-1)^\text{LN}$, except for the operator $\Op_{\nu\nu\nu\nu}$ in \cref{tab:MLFVdim6nuSMEFTLNV}. Comparing with the ranks of modules obtained in \cref{Eq:RankYeYnumRP,Eq:RankYeYnumRV}, we see that 
\begin{equation}
\Lag_{\text{\,$\nu$SMEFT},\,S_\text{IV}} \Big|_{\text{dim-}6} \;\cong\;
\Lag_\text{\,$\nu$SMEFT}^{\mathcal{P}_\text{LN}} \Big|_{\text{dim-}6} \,.
\end{equation}
This verifies the saturation theorem in \cref{eqn:Saturation} at mass dimension 6 for Case IV MLFV in \cref{tab:Cases}.

\section{Conclusions and Outlook}
\label{sec:Outlook}

In this paper, we derived a saturation theorem (\cref{eqn:Saturation}) for general EFTs constructed from a spurion analysis. When a set of spurion fields $S$ are introduced to organize the breaking of a global symmetry $G_f$ in an EFT $\Lag_\text{\,EFT}\big([\phi]\big)$, the resulting EFT $\Lag_\text{\,EFT,\,Spurion}$ \emph{saturates} the original one restricted by requiring the symmetry under $H_S\subset G_f$ (where $H_S$ is the unbroken subgroup under a generic spurion vev, $\langle S\rangle$):
\begin{equation}
\Lag_\text{\,EFT,\,Spurion} \equiv \Lag_\text{\,EFT} \big( [\phi] , S \big)^{G_f}
\Bigr|_{S = \langle S\rangle = \lambda_S}
\quad\cong\quad
\Lag_\text{\,EFT}\big([\phi]\big)^{H_S} \,,
\label{eqn:SaturationReproduced}
\end{equation}
when all powers of $S$ are included in $\Lag_\text{\,EFT,\,Spurion}$ and $\lambda_S=\langle S\rangle$ is a generic vev. Here the equivalence ``$\,\cong\,$'' means that the two EFTs have the same set of allowed independent effective operators.

We have studied four implementations of the MLFV principle, as summarized in \cref{tab:Cases}, corresponding to a variety of origins of the neutrino masses in SMEFT and $\nu$SMEFT. We used these to  demonstrate explicitly the saturation theorem at lowest orders in the EFT expansion. Specifically, for each MLFV scenario, we studied the lepton flavor covariants that appear at mass dimension 6 in the EFT. We obtained the number of linearly independent operators for each of them by computing the Hilbert series and making use of \cref{eqn:RankModHSIntro}. For certain selected flavor covariants, we also provided the linearly independent polynomials in spurion fields, motivated by their connections with phenomenology.

The saturation theorem in \cref{eqn:SaturationReproduced} holds when arbitrary powers of the spurion fields $S$ are allowed in the construction of $\Lag_\text{\,EFT,\,Spurion}$. However, quite often in practical spurion analysis, $S$ are only involved up to a certain power. In such cases, $\Lag_\text{\,EFT,\,Spurion}$ may not saturate the EFT $\Lag_\text{\,EFT}\big([\phi]\big)^{H_S}$ --- the theory $\Lag_\text{\,EFT,\,Spurion}$ with limited powers of $S$ can be more restricted and hence phenomenologically predictive. This kind of truncations are motivated, for example, in the case of MLFV where the Yukawa coupling constants $Y_e,Y_\nu$ are extremely small. In addition, there might even be accidental symmetries that arise due to a power truncation in $S$, and one can exploit the method recently developed in \cite{Grinstein:2024jqt} to check if any systematic patterns exist for this.

Likewise, the saturation theorem in \cref{eqn:SaturationReproduced} holds only when the actual couplings $\lambda_S = \langle S \rangle$ does fulfill a generic vev. If this is not the case (for an extreme example, say, experimental measurements suggest that $\lambda_S=0$), then there could be a larger symmetry, $H_\text{Special}$, even with arbitrary powers of $S$ involved:
\begin{equation}
H_S \subset H_\text{Special} \subset G_f \,.
\end{equation}
It would be interesting to check if the saturation still holds in this case, that is, whether \cref{eqn:SaturationReproduced} is satisfied when substituting $H_S\to H_\text{Special}$. We leave such an investigation to future work.

In this paper, we implemented the MLFV principle on the maximal lepton flavor groups, \ie, $G_{Lf,\,\text{SMEFT}} = U(3)_\ell \times U(3)_e$ for SMEFT and $G_{Lf,\,\nu\text{SMEFT}} = U(3)_\ell \times U(3)_e \times U(3)_\nu$ for $\nu$SMEFT respectively. An interesting direction is to consider implementing the MLFV principle on a subgroup of these maximal lepton flavor groups. For example, in $\nu$SMEFT case, the presence of a heavy Majorana neutrino mass proportional to the identity matrix will break the lepton number but not flavor, leading us to consider $G_{Lf} = SU(3)_\ell \times SU(3)_e \times SO(3)_\nu$ instead. This was studied in Ref.~\cite{Cirigliano:2005ck}, and its extension to a local symmetry  was studied in \cite{Alonso:2016onw}. One could also consider the minimal massive neutrino scenario, where one extends the SM field content with only two generations of right-handed neutrinos, resulting in $G_{Lf} = SU(3)_\ell \times SU(3)_e \times SU(2)_\nu$ or $SU(3)_\ell \times SU(3)_e \times SO(2)_\nu$.

\acknowledgments
We thank Enrique Fern\'andez-Mart\'inez and Aneesh Manohar useful discussions.
The work of B.G., X.L., and P.Q. is supported by the U.S. Department of Energy under grant number DE-SC0009919.
X.L.\ is supported in part by Simons Foundation Award 568420.
C.M.\ is funded by \textit{Conselleria de Innovación, Universidades, Ciencia y Sociedad Digital} from \textit{Generalitat Valenciana} and by \textit{Fondo Social Europeo} under grants ACIF/2021/284, and CIBEFP/2023/96. C.M.\ specially thanks UC San Diego Theoretical Physics Department for the hospitality during his visit.

\section*{Appendices}
\addcontentsline{toc}{section}{\protect\numberline{}Appendices}%
\appendix

\section{Using HilbCalc to compute general Hilbert series}
\label{appsec:Ancillary}

In this appendix, we provide an introductory guide on how to use our ancillary \Mathematica\ notebook, \href{\link}{HilbCalc} \href{\link}{\faGithub}, to calculate a general Hilbert series (either single-graded or multi-graded) for group invariants and covariants, restricted to classical Lie groups $SU(n)$, $SO(n)$, $Sp(n=2r)$, and $U(1)$ abelian factors.

Suppose that we have a set of spurion fields $\Phi$, which form a representation of a symmetry group $G$, and we are interested in its polynomials. In particular, given each irrep of $G$, denoted ``$\Irrep_G$'' below, we would like to know how many of them can be made by $\Phi$ polynomials. This is encoded in the Hilbert series $\HS_{\Irrep_G}^{G,\, \Phi}$. It has several inputs to take:
\begin{itemize}
\item The group $G$, which is a product group in general.
\item The $G$-irrep of interest, \ie, the goal $\Irrep_G$.
\item The spurions $\Phi$, which we will refer to as ``building block fields''. They are assigned grading variables to keep track of their powers in the polynomials. Each $G$-irrep component of $\Phi$ can be assigned a distinct grading variable.
\end{itemize}
The key formula underlying the calculation of a Hilbert series for group invariants and covariants is the Molien-Weyl formula \cite{Grinstein:2023njq}:
\begin{equation}
\HS_{\Irrep_G}^{G,\, \Phi}(q) = \int \dd\mu_G^{}(x)\, \chi_{\Irrep_G}^*(x)\, \chi_{\ring_\Phi}(q, x)
\qquad \mathrm{with} \qquad |q|<1 \,. 
\label{eqn:MolienWeylFormula}
\end{equation}
The integrand consists of three factors:
\begin{enumerate}
\item \textbf{The Haar measure $\dd\mu_G^{}(x)$}.
\item \textbf{The Weyl character $\chi_{\Irrep_G}^*(x)$ of the goal $\Irrep_G$}.
\item \textbf{The graded (Weyl) character $\chi_{\ring_\Phi}(q, x)$ of the polynomial ring $\ring_\Phi$ in the building block fields}.
\end{enumerate}
As indicated above, the logic of the Molien-Weyl formula is to make use of character orthogonality under the Haar measure $\dd\mu_G^{}(x)$ to extract the component of our interest $\Irrep_G$ in the graded character $\chi_{\ring_\Phi}(q, x)$ of the polynomial ring $\ring_\Phi$ in the building block fields.

Our ancillary \Mathematica\ notebook \href{\link}{HilbCalc} \href{\link}{\faGithub} provides general functions for calculating each of the above three factors. More importantly, it packages them together to calculate the Hilbert series for a general set of group invariants or covariants that is asked by a user. In what follows, we navigate through it to explain how to use it for computing each of these three factors: Haar measures in \cref{appsubsec:HaarMeasureFunctions}, Weyl characters in \cref{appsubsec:WeylCharacterFunctions}, graded characters of polynomial rings in \cref{appsubsec:GradedPolyRingFunctions}, and finally their combination --- the Hilbert series in \cref{appsubsec:HilbertSeriesFunctions}.

We highlight that a more comprehensive list of demonstration examples are provided in the notebook. It also collects a more complete list of the Hilbert series for the four cases of MLFV implementations discussed in \cref{sec:MLFV}.

\subsection{Haar measure calculations}\label{appsubsec:HaarMeasureFunctions}

A Haar measure $\dd \mu_G^{}(x)$ is parametrized by $r$ phase variables, $x = \{x_1, \cdots, x_r\}$, where $r$ is the rank of the group $G$. In the Molien-Weyl formula \cref{eqn:MolienWeylFormula}, each phase variable $x_i$ is being integrated over the contour of the unit circle. To compute a Haar measure with \HilbCalc, one simply uses the function \mmaInlineCell{Code}{HaarGroup} as
\begin{mmaCell}[index=1]{Input}
\mmaDef{HaarGroup}[gp,chvars]
\end{mmaCell}
As indicated above, two input arguments are required: 
\begin{itemize} 
\item \mmaInlineCell{Code}{gp} is a \texttt{Symbol} that specifies the group $G$ among \mmaInlineCell{Code}{U1}, \mmaInlineCell{Code}{SUn}, \mmaInlineCell{Code}{SOn}, or \mmaInlineCell{Code}{Spn}, where \mmaInlineCell{Code}{n} is an integer. For instance, one can type \mmaInlineCell{Code}{SU3}, \mmaInlineCell{Code}{SO7}, \mmaInlineCell{Code}{Sp10}, and so on.
\item \mmaInlineCell{Code}{chvars} is a \texttt{List} that contains the $r$ variables parametrizing the Haar measure, \eg, \mmaInlineCell{Input}{\{\mmaSub{x}{\mmaUnd{1}},\mmaSub{x}{\mmaUnd{2}}\}} for a group with rank $r=2$.
\end{itemize}
Let us compute a simple example --- the Haar measure of $SU(3)$:
\begin{mmaCell}[index=2]{Input}
\mmaDef{HaarGroup}[SU3,\{\mmaSub{x}{\mmaUnd{1}},\mmaSub{x}{\mmaUnd{2}}\}]
\end{mmaCell}
\begin{mmaCell}[index=3]{Output}
\mmaFrac{\Big(1-\mmaFrac{1}{\mmaSub{x}{1}\mmaSubSup{x}{2}{2}}\Big)\Big(1-\mmaFrac{1}{\mmaSubSup{x}{1}{2}\mmaSub{x}{2}}\Big)\Big(1-\mmaFrac{\mmaSub{x}{1}}{\mmaSub{x}{2}}\Big)\Big(1-\mmaFrac{\mmaSub{x}{2}}{\mmaSub{x}{1}}\Big)\big(1-\mmaSubSup{x}{1}{2}\mmaSub{x}{2}\big)\big(1-\mmaSub{x}{1}\mmaSubSup{x}{2}{2}\big)}{6\mmaSub{x}{1}\mmaSub{x}{2}} 
\end{mmaCell}
In case the integrand in the Molien-Weyl formula is symmetric under the Weyl group, the Haar measures can be further simplified or \emph{reduced}. To get the reduced version of the Haar measures, one can use the function \mmaInlineCell{Code}{HaarReducedGroup} instead, \eg\
\begin{mmaCell}[index=3]{Input}
\mmaDef{HaarReducedGroup}[SU3,\{\mmaSub{x}{\mmaUnd{1}},\mmaSub{x}{\mmaUnd{2}}\}]
\end{mmaCell}
\begin{mmaCell}[index=4]{Output}
\mmaFrac{\Big(1-\mmaFrac{\mmaSub{x}{1}}{\mmaSub{x}{2}}\Big)\big(1-\mmaSubSup{x}{1}{2}\mmaSub{x}{2}\big)\big(1-\mmaSub{x}{1}\mmaSubSup{x}{2}{2}\big)}{\mmaSub{x}{1}\mmaSub{x}{2}} 
\end{mmaCell}

To deal with a product group $G = G_1\times G_2\times \cdots\times G_k$, one uses the following two extended functions
\begin{mmaCell}[index=4]{Input}
\mmaDef{HaarProductGroup}[gplist,chvarsGps]
\end{mmaCell}
\begin{mmaCell}[index=5]{Input}
\mmaDef{HaarReducedProductGroup}[gplist,chvarsGps]
\end{mmaCell}
with input arguments given by
\begin{itemize}
\item \mmaInlineCell{Code}{gplist} is a \texttt{List} of \mmaInlineCell{Code}{gp}, such as \mmaInlineCell{Input}{\{SU3,U1\}}.
\item \mmaInlineCell{Code}{chvarsGps} is a \texttt{List} of \mmaInlineCell{Code}{chvars}, such as \mmaInlineCell{Input}{\{\{\mmaSub{x}{\mmaUnd{1}},\mmaSub{x}{\mmaUnd{2}}\},\{x\}\}}.
\end{itemize}
Taking $SU(3) \times U(1)$ as an example, one can readily check that
\begin{mmaCell}[index=6]{Input}
\mmaDef{HaarReducedProductGroup}[\{SU3,U1\},\{\{\mmaSub{x}{\mmaUnd{1}},\mmaSub{x}{\mmaUnd{2}}\},\{x\}\}]
\end{mmaCell}
\begin{mmaCell}[index=7]{Output}
\mmaFrac{\Big(1-\mmaFrac{\mmaSub{x}{1}}{\mmaSub{x}{2}}\Big)\big(1-\mmaSubSup{x}{1}{2}\mmaSub{x}{2}\big)\big(1-\mmaSub{x}{1}\mmaSubSup{x}{2}{2}\big)}{x \mmaSub{x}{1}\mmaSub{x}{2}} 
\end{mmaCell}

\subsection{Weyl character calculations}
\label{appsubsec:WeylCharacterFunctions}

The Weyl character $\chi_{\text{Irrep}_G}^{} (x)$ depends on the irrep specified. In \HilbCalc, we use the highest weight vectors to label irreps of a Lie algebra. We provide auxiliary functions to automatically translate between a highest weight vector and the Dynkin labels of an irrep, as the latter is also frequently used for referring to an irrep. Furthermore, we also provide a function for computing the dimension of irreps. For user convenience, we include a function that generates a table summarizing the Dynkin labels, dimensions, and highest weight vectors for the irreps of a given group, up to a specified maximum Dynkin label. These auxiliary functions are
\begin{mmaCell}[index=7]{Input}
\mmaDef{DynkinLabelsFromIrrephwt}[gp, irrephwt]
\end{mmaCell}
\begin{mmaCell}[index=8]{Input}
\mmaDef{IrrephwtFromDynkinLabels}[gp, dynkinlabels]
\end{mmaCell}
\begin{mmaCell}[index=9]{Input}
\mmaDef{DimIrrephwt}[gp, irrephwt]
\end{mmaCell}
\begin{mmaCell}[index=10]{Input}
\mmaDef{IrrepTableGroup}[gp, maxdynkinlabel]
\end{mmaCell}
In all cases, the first input argument \mmaInlineCell{Code}{gp} specifies the group of interest, as explained in \cref{appsubsec:HaarMeasureFunctions}. In addition, a second argument is required:
\begin{itemize}
\item \mmaInlineCell{Code}{irrephwt} is a \texttt{List} of $r$ components, being $r$ the rank of the group. Taking $SU(3)$ for example, we have \mmaInlineCell{Input}{\{1,0\}} for the fundamental irrep, \mmaInlineCell{Input}{\{1,1\}} for the anti-fundamental, and \mmaInlineCell{Input}{\{2,1\}} for the adjoint.\footnote{For $SU(n=r+1)$ groups, we embed the weight lattice in the $(r+1)$-dimensional space, such that the components of the weight vectors are all integers. The $(r+1)$-th component of a highest weight vector is zero, and we can view it as a vector with $r$ components, as in the $SU(3)$ examples given above. With this convention, for a given $SU(n=r+1)$ irrep, there is a convenient map between its highest weight vector and its Young diagram --- the $k$-th component of \texttt{irrephwt} gives the number of boxes in the $k$-th row of the Young diagram.}
\item \mmaInlineCell{Code}{dynkinlabels} is a \texttt{List} of $r$ non-negative \texttt{Integers}.
\item \mmaInlineCell{Code}{maxdynkinlabel} is a non-negative \texttt{Integer} that that sets the maximum value for each component of the Dynkin labels of the irreps included in the table.
\end{itemize}
As an illustrative example, one can check the conversion from highest weight vector to Dynkin labels, and viceversa, for the adjoint irrep of $SU(3)$:
\begin{mmaCell}[index=11]{Input}
\mmaDef{DynkinLabelsFromIrrephwt}[SU3,\{2,1\}]
\end{mmaCell}
\begin{mmaCell}[index=12]{Output}
\{1,1\}
\end{mmaCell}
\begin{mmaCell}[index=12]{Input}
\mmaDef{IrrephwtFromDynkinLabels}[SU3,\{1,1\}]
\end{mmaCell}
\begin{mmaCell}[index=13]{Output}
\{2,1\}
\end{mmaCell}
One can also generate a table of the $SU(3)$ irreps with Dynkin labels smaller than 2:
\begin{mmaCell}[index=13]{Input}
\mmaDef{IrrepTableGroup}[SU3,2]
\end{mmaCell}
\vspace{-15pt}
\begin{minipage}[t]{0.01\textwidth}
\begin{mmaCell}[index=14]{Output}
\,
\end{mmaCell}
\end{minipage}
\begin{minipage}{0.87\textwidth}
\flushleft\begin{align*}
&\textbf{\sffamily Summary of SU3 irreps }\\
&\begin{array}{|l|l|l|}
\hline 
\text{\sffamily Dynkin Label\quad } & \text{\sffamily Dimension\quad } & \text{\sffamily Highest weight vector\quad } \\
\hline
\{0,0\} & 1 & \{0,0\} \\
\hline
\{0,1\} & 3 & \{1,1\} \\
\hline
\{0,2\} & 6 & \{2,2\} \\
\hline
\{1,0\} & 3 & \{1,0\} \\
\hline
\{1,1\} & 8 & \{2,1\} \\
\hline
\{1,2\} & 15 & \{3,2\} \\
\hline
\{2,0\} & 6 & \{2,0\} \\
\hline
\{2,1\} & 15 & \{3,1\} \\
\hline
\{2,2\} & 27 & \{4,2\} \\
\hline
\end{array}
\end{align*}\hspace{20cm}
\end{minipage}

With the irrep convention established, one can compute a Weyl character with the function
\begin{mmaCell}[index=14]{Input}
\mmaDef{WeylChGroup}[gp, chvars, irrephwt]
\end{mmaCell}
for a single group, and 
\begin{mmaCell}[index=15]{Input}
\mmaDef{WeylChProductGroup}[gplist, chvarsGps, irrephwtGps]
\end{mmaCell}
for product groups, where \mmaInlineCell{Code}{irrephwtGps} is a \texttt{List} of \mmaInlineCell{Code}{irrephwt}, such as \mmaInlineCell{Input}{\{\{1,0\},\{+1\}\}}. Coming back to our previous $SU(3) \times U(1)$ example, the Weyl character for the fundamental irrep with charge $+1$ reads
\begin{mmaCell}[index=16]{Input}
\mmaDef{WeylChProductGroup}[\{SU3,U1\},\{\{\mmaSub{x}{\mmaUnd{1}},\mmaSub{x}{\mmaUnd{2}}\},\{x\}\}, \{\{1,0\},\{+1\}\}]
\end{mmaCell}
\begin{mmaCell}[index=17]{Output}
x \mmaSub{x}{1} + \mmaFrac{x}{\mmaSub{x}{1}\mmaSub{x}{2}} + x \mmaSub{x}{2}
\end{mmaCell}

\subsection{Graded characters of the polynomial ring in building block fields}
\label{appsubsec:GradedPolyRingFunctions}
The remaining piece of the integrand in the Molien-Weyl formula, namely $\chi_{\ring_\Phi}(q, x)$, is the graded character of the polynomial ring in a set of building block fields \mmaInlineCell{Code}{BBFlist}. It is a \texttt{List} of \mmaInlineCell{Code}{BBF}, and each \mmaInlineCell{Code}{BBF} is a \texttt{List} with three components:
\begin{itemize}
\item \mmaInlineCell{Code}{BBF[[1]]} is a \texttt{Symbol} referring to the grading variable of the building block field, \eg, \mmaInlineCell{Input}{\mmaUnd{\(\phi\)}} or \mmaInlineCell{Code}{q}.
\item \mmaInlineCell{Code}{BBF[[2]]} is an \texttt{Integer} that takes values \mmaInlineCell{Code}{0} or \mmaInlineCell{Code}{1} for Grassmann even and Grassmann odd fields, respectively.
\item \mmaInlineCell{Code}{BBF[[3]]} has the format of the \mmaInlineCell{Code}{irrephwtGps} presented before, which is a \texttt{List} of highest weight vector \mmaInlineCell{Code}{irrephwt}.
\end{itemize}
Then, the graded character of a polynomial ring in a given set of building block fields can be readily computed for a single group or a product group by using
\begin{mmaCell}[index=17]{Input}
\mmaDef{GradedChProductGroup}[gp, chvars, BBFlist]
\end{mmaCell}
\begin{mmaCell}[index=18]{Input}
\mmaDef{GradedRingChProductGroup}[gplist, chvarsGps, BBFlist]
\end{mmaCell}
Let us consider our recurrent example $SU(3)\times U(1)$ from bosonic fundamental \mmaInlineCell{Input}{\mmaUnd{\(\phi\)}} with charge \mmaInlineCell{Code}{+1} and anti-fundamental \mmaInlineCell{Input}{\mmaSub{\mmaUnd{\(\phi\)}}{d}} with charge \mmaInlineCell{Code}{-1}. The graded character of the polynomial ring is given by
\begin{mmaCell}[index=19]{Input}
\mmaUnd{\(\phi\)}BBF = \{\mmaUnd{\(\phi\)}, 0, \{\{1,0\}, \{+1\}\}\};
\mmaUnd{\(\phi\)}dBBF = \{\mmaSub{\mmaUnd{\(\phi\)}}{d}, 0, \{\{1,1\}, \{-1\}\}\};
\mmaDef{GradedRingChProductGroup}[\{SU3,U1\},\{\{\mmaSub{x}{\mmaUnd{1}},\mmaSub{x}{\mmaUnd{2}}\},\{x\}\},\{\mmaUnd{\(\phi\)}BBF,\mmaUnd{\(\phi\)}dBBF\}]
\end{mmaCell}
\begin{mmaCell}[index=20]{Output}
\mmaFrac{1}{\big(1-x\mmaSub{x}{1}\(\phi\)\big)\Big(1-\mmaFrac{x\(\phi\)}{\mmaSub{x}{1}\mmaSub{x}{2}}\Big)\big(1-x\mmaSub{x}{2}\(\phi\)\big)\Big(1-\mmaFrac{\mmaSub{\(\phi\)}{d}}{x\mmaSub{x}{1}}\Big)\Big(1-\mmaFrac{\mmaSub{\(\phi\)}{d}}{x\mmaSub{x}{2}}\Big)\Big(1-\mmaFrac{\mmaSub{x}{1}\mmaSub{x}{2}\mmaSub{\(\phi\)}{d}}{x}\Big)} 
\end{mmaCell}

\subsection{Hilbert series evaluations}\label{appsubsec:HilbertSeriesFunctions}

Finally, we present the main functions of \HilbCalc. These are
\begin{mmaCell}[index=20]{Input}
\mmaDef{HilbertSeriesBBFGRCh}[gplist, chvarsGps, irrephwtGps, BBFgrch]
\end{mmaCell}
\begin{mmaCell}[index=21]{Input}
\mmaDef{HilbertSeriesBBFlist}[gplist, chvarsGps, irrephwtGps, BBFlist]
\end{mmaCell}
The sole difference between these two functions is in their last input argument. The former requires to take the graded character of the polynomial ring \mmaInlineCell{Code}{BBFgrch} computed separately as an input; the latter directly takes the list of building block fields \mmaInlineCell{Code}{BBFlist} as an input. One should notice that the input argument \mmaInlineCell{Code}{irrephwtGps} here refers to the goal irrep for which we are computing the Hilbert --- it is not about the irreps of the building block fields.

It is instructive to present a simple example computation to clearly illustrate the different calls of our main functions --- the multi-graded Hilbert series for $SU(3)$ invariants made out of bosonic fundamental \mmaInlineCell{Input}{\mmaUnd{\(\phi\)}} and anti-fundamental \mmaInlineCell{Input}{\mmaSub{\mmaUnd{\(\phi\)}}{d}}. We have
\begin{mmaCell}[index=22]{Input}
\mmaUnd{gplist} = \{SU3\}
\mmaUnd{chvarsGps} = \{\{\mmaSub{x}{1},\mmaSub{x}{2}\}\}
\mmaUnd{irrephwtGps} = \{\{0,0\}\}
\mmaUnd{\(\phi\)}BBF = \{\mmaUnd{\(\phi\)}, 0, \{\{1,0\}\}\}
\mmaUnd{\(\phi\)}dBBF = \{\mmaSub{\mmaUnd{\(\phi\)}}{d}, 0, \{\{1,1\}\}
\mmaUnd{BBFlist} = \{\mmaUnd{\(\phi\)}BBF, \mmaUnd{\(\phi\)}dBBF\}
\mmaUnd{BBFGRCh} = \mmaDef{GradedRingChProductGroup}[gplist,chvarsGps,BBFlist]
\end{mmaCell}
where the last line uses the function explained in \cref{appsubsec:GradedPolyRingFunctions} to compute the graded character \mmaInlineCell{Code}{BBFgrch} of the polynomial ring in \mmaInlineCell{Input}{\mmaUnd{\(\phi\)}} and \mmaInlineCell{Input}{\mmaSub{\mmaUnd{\(\phi\)}}{d}}. After evaluating the above, we can call the two main functions:
\begin{mmaCell}[index=23]{Input}
\mmaDef{HilbertSeriesBBFGRCh}[gplist,chvarsGps,irrephwtGps,BBFGRCh]
\mmaDef{HilbertSeriesBBFlist}[gplist,chvarsGps,irrephwtGps,BBFlist]
\end{mmaCell}
They both yield
\begin{mmaCell}[index=24]{Output}
-\mmaFrac{1}{-1+\(\phi\)\mmaSub{\(\phi\)}{d}} 
\end{mmaCell}
Although the second main function is clearly more direct, the first one can be especially useful when we are given a fixed set of building block fields and computing the Hilbert series for different goal irreps, as is encountered in the case of calculating the MLFV Hilbert series.

\vfill
\newpage
\section{More Hilbert Series in MLFV}
\label{appsec:HSMLFV}

In this appendix, we collect more Hilbert series results for the MLFV cases discussed in \cref{sec:MLFV}.

\subsection{Multi-graded Hilbert series for spurions $S_\text{II} = \{Y_e^{\phantom{\dagger}}, Y_e^\dagger, C_5^{\phantom{\dagger}}, C_5^{\dagger}\}$}
\label{appsubsec:YeC5}

Using the grading scheme
\begin{equation}
Y_e \sim q_e \,,\qquad
Y_e^\dagger \sim q_e \,,\qquad
C_5 \sim q_5 \,,\qquad
C_5^\dagger \sim q_5 \,,
\end{equation}
the multi-graded Hilbert series for the MLFV case II in \cref{tab:Cases} read:
\begin{allowdisplaybreaks}
\begin{footnotesize}
\begin{subequations}
\begin{align*}
\mathcal{H}_{\mathbf{(1,1)}} &= \frac{1}{D(q_5,q_e)}\,\hspace{1em} 
\mqty( 1 + 2 q_e^4 q_5^4 + 2 q_e^6 q_5^4 + 3 q_e^8 q_5^4 + q_e^{10} q_5^4 + q_e^{12} q_5^4 + 2 q_e^4 q_5^6 + 3 q_e^6 q_5^6 \\[4pt]
+ 3 q_e^8 q_5^6 + q_e^{10} q_5^6 - q_e^{14} q_5^6 - q_e^{16} q_5^6 + q_e^4 q_5^8 + q_e^6 q_5^8 + 3 q_e^8 q_5^8 \\[4pt]
+ q_e^{10} q_5^8 - 2 q_e^{14} q_5^8 - q_e^{16} q_5^8 + q_e^8 q_5^{10} + 2 q_e^{10} q_5^{10} - q_e^{14} q_5^{10} - 3 q_e^{16} q_5^{10} \\[4pt]
- q_e^{18} q_5^{10} - q_e^{20} q_5^{10} + q_e^8 q_5^{12} + q_e^{10} q_5^{12} - q_e^{14} q_5^{12} - 3 q_e^{16} q_5^{12} - 3 q_e^{18} q_5^{12} \\[4pt]
- 2 q_e^{20} q_5^{12} - q_e^{12} q_5^{14} - q_e^{14} q_5^{14} - 3 q_e^{16} q_5^{14} - 2 q_e^{18} q_5^{14} - 2 q_e^{20} q_5^{14} - q_e^{24} q_5^{18} )\,, \\[12pt]
\mathcal{H}_{\mathbf{(3,\bar{3})}} &= \frac{1}{D(q_5,q_e)}\, q_e\, 
\mqty( 1 + q_e^2 + q_e^4 + q_5^2 + 3 q_e^2 q_5^2 + 5 q_e^4 q_5^2 + 5 q_e^6 q_5^2 \\[4pt]
+ 3 q_e^8 q_5^2 + q_e^{10} q_5^2 + q_5^4 + 4 q_e^2 q_5^4 + 11 q_e^4 q_5^4 + 15 q_e^6 q_5^4 + 15 q_e^8 q_5^4 \\[4pt]
+ 8 q_e^{10} q_5^4 + 3 q_e^{12} q_5^4 + 3 q_e^2 q_5^6 + 12 q_e^4 q_5^6 + 21 q_e^6 q_5^6 + 24 q_e^8 q_5^6 + 16 q_e^{10} q_5^6 \\[4pt]
+ 5 q_e^{12} q_5^6 - 2 q_e^{14} q_5^6 - 3 q_e^{16} q_5^6 - q_e^{18} q_5^6 + q_e^2 q_5^8 + 8 q_e^4 q_5^8 \\[4pt]
+ 17 q_e^6 q_5^8 + 24 q_e^8 q_5^8 + 17 q_e^{10} q_5^8 + 3 q_e^{12} q_5^8 - 10 q_e^{14} q_5^8 - 12 q_e^{16} q_5^8 \\[4pt]
- 7 q_e^{18} q_5^8 - 2 q_e^{20} q_5^8 + 2 q_e^4 q_5^{10} + 7 q_e^6 q_5^{10} + 12 q_e^8 q_5^{10} + 10 q_e^{10} q_5^{10} \\[4pt]
- 3 q_e^{12} q_5^{10} - 17 q_e^{14} q_5^{10} - 24 q_e^{16} q_5^{10} - 17 q_e^{18} q_5^{10} - 8 q_e^{20} q_5^{10} - q_e^{22} q_5^{10} \\[4pt]
+ q_e^6 q_5^{12} + 3 q_e^8 q_5^{12} + 2 q_e^{10} q_5^{12} - 5 q_e^{12} q_5^{12} - 16 q_e^{14} q_5^{12} - 24 q_e^{16} q_5^{12} \\[4pt]
- 21 q_e^{18} q_5^{12} - 12 q_e^{20} q_5^{12} - 3 q_e^{22} q_5^{12} - 3 q_e^{12} q_5^{14} - 8 q_e^{14} q_5^{14} - 15 q_e^{16} q_5^{14} \\[4pt]
- 15 q_e^{18} q_5^{14} - 11 q_e^{20} q_5^{14} - 4 q_e^{22} q_5^{14} - q_e^{24} q_5^{14} - q_e^{14} q_5^{16} - 3 q_e^{16} q_5^{16} \\[4pt]
- 5 q_e^{18} q_5^{16} - 5 q_e^{20} q_5^{16} - 3 q_e^{22} q_5^{16} - q_e^{24} q_5^{16} - q_e^{20} q_5^{18} - q_e^{22} q_5^{18} - q_e^{24} q_5^{18} )\,, \\[12pt]
\mathcal{H}_{\mathbf{(1,8)}} &= \frac{1}{D(q_5,q_e)}\, q_e^2\, 
\mqty( 1 + q_e^2 + q_5^2 + 3 q_e^2 q_5^2 + 5 q_e^4 q_5^2 + 5 q_e^6 q_5^2 + 3 q_e^8 q_5^2 \\[4pt]
+ q_e^{10} q_5^2 + q_5^4 + 4 q_e^2 q_5^4 + 11 q_e^4 q_5^4 + 15 q_e^6 q_5^4 + 13 q_e^8 q_5^4 + 6 q_e^{10} q_5^4 \\[4pt]
- q_e^{14} q_5^4 - q_e^{16} q_5^4 + 3 q_e^2 q_5^6 + 12 q_e^4 q_5^6 + 21 q_e^6 q_5^6 + 22 q_e^8 q_5^6 + 13 q_e^{10} q_5^6 \\[4pt]
+ 2 q_e^{12} q_5^6 - 3 q_e^{14} q_5^6 - 3 q_e^{16} q_5^6 + q_e^{20} q_5^6 + q_e^2 q_5^8 + 8 q_e^4 q_5^8 \\[4pt]
+ 17 q_e^6 q_5^8 + 23 q_e^8 q_5^8 + 16 q_e^{10} q_5^8 - 11 q_e^{14} q_5^8 - 12 q_e^{16} q_5^8 - 5 q_e^{18} q_5^8 \\[4pt]
- q_e^{20} q_5^8 + 2 q_e^4 q_5^{10} + 7 q_e^6 q_5^{10} + 12 q_e^8 q_5^{10} + 10 q_e^{10} q_5^{10} - 4 q_e^{12} q_5^{10} \\[4pt]
- 19 q_e^{14} q_5^{10} - 24 q_e^{16} q_5^{10} - 16 q_e^{18} q_5^{10} - 5 q_e^{20} q_5^{10} + q_e^{24} q_5^{10} + q_e^6 q_5^{12} \\[4pt]
+ 3 q_e^8 q_5^{12} + 2 q_e^{10} q_5^{12} - 6 q_e^{12} q_5^{12} - 17 q_e^{14} q_5^{12} - 24 q_e^{16} q_5^{12} - 20 q_e^{18} q_5^{12} \\[4pt]
- 9 q_e^{20} q_5^{12} + 2 q_e^{24} q_5^{12} - 3 q_e^{12} q_5^{14} - 8 q_e^{14} q_5^{14} - 14 q_e^{16} q_5^{14} - 14 q_e^{18} q_5^{14} \\[4pt]
- 8 q_e^{20} q_5^{14} - 2 q_e^{22} q_5^{14} + q_e^{24} q_5^{14} - q_e^{14} q_5^{16} - 3 q_e^{16} q_5^{16} - 5 q_e^{18} q_5^{16} \\[4pt]
- 5 q_e^{20} q_5^{16} - 3 q_e^{22} q_5^{16} - q_e^{24} q_5^{16} - q_e^{20} q_5^{18} - q_e^{22} q_5^{18} - q_e^{24} q_5^{18} + q_e^{28} q_5^{18} )\,, \\[20pt]
\end{align*}
\end{subequations}
\end{footnotesize}
\end{allowdisplaybreaks}%
\vfill
\vspace{45pt}
\vfill
\begin{allowdisplaybreaks}
\begin{footnotesize}
\begin{subequations}
\begin{align*}
\mathcal{H}_{\mathbf{(8,1)}} &= \frac{1}{D(q_5,q_e)}\, \hspace{1em} 
\mqty( q_e^2 + q_e^4 + q_5^2 + 3 q_e^2 q_5^2 + 5 q_e^4 q_5^2 + 5 q_e^6 q_5^2 + 3 q_e^8 q_5^2 \\[4pt]
+ q_e^{10} q_5^2 + q_5^4 + 4 q_e^2 q_5^4 + 9 q_e^4 q_5^4 + 13 q_e^6 q_5^4 + 12 q_e^8 q_5^4 + 7 q_e^{10} q_5^4 \\[4pt]
+ 2 q_e^{12} q_5^4 + 3 q_e^2 q_5^6 + 10 q_e^4 q_5^6 + 18 q_e^6 q_5^6 + 21 q_e^8 q_5^6 + 15 q_e^{10} q_5^6 + 5 q_e^{12} q_5^6 \\[4pt]
- q_e^{14} q_5^6 - 2 q_e^{16} q_5^6 - q_e^{18} q_5^6 + q_e^2 q_5^8 + 7 q_e^4 q_5^8 + 16 q_e^6 q_5^8 \\[4pt]
+ 21 q_e^8 q_5^8 + 16 q_e^{10} q_5^8 + 3 q_e^{12} q_5^8 - 8 q_e^{14} q_5^8 - 11 q_e^{16} q_5^8 - 7 q_e^{18} q_5^8 \\[4pt]
- 2 q_e^{20} q_5^8 + 2 q_e^4 q_5^{10} + 7 q_e^6 q_5^{10} + 11 q_e^8 q_5^{10} + 8 q_e^{10} q_5^{10} - 3 q_e^{12} q_5^{10} \\[4pt]
- 16 q_e^{14} q_5^{10} - 21 q_e^{16} q_5^{10} - 16 q_e^{18} q_5^{10} - 7 q_e^{20} q_5^{10} - q_e^{22} q_5^{10} + q_e^6 q_5^{12} \\[4pt]
+ 2 q_e^8 q_5^{12} + q_e^{10} q_5^{12} - 5 q_e^{12} q_5^{12} - 15 q_e^{14} q_5^{12} - 21 q_e^{16} q_5^{12} - 18 q_e^{18} q_5^{12} \\[4pt]
- 10 q_e^{20} q_5^{12} - 3 q_e^{22} q_5^{12} - 2 q_e^{12} q_5^{14} - 7 q_e^{14} q_5^{14} - 12 q_e^{16} q_5^{14} - 13 q_e^{18} q_5^{14} \\[4pt]
- 9 q_e^{20} q_5^{14} - 4 q_e^{22} q_5^{14} - q_e^{24} q_5^{14} - q_e^{14} q_5^{16} - 3 q_e^{16} q_5^{16} - 5 q_e^{18} q_5^{16} \\[4pt]
- 5 q_e^{20} q_5^{16} - 3 q_e^{22} q_5^{16} - q_e^{24} q_5^{16} - q_e^{20} q_5^{18} - q_e^{22} q_5^{18} )\,, \\[45pt]
\mathcal{H}_{\mathbf{(8,8)}} &= \frac{1}{D(q_5,q_e)}\, q_e^2\, 
\mqty( 1 + 3 q_e^2 + 4 q_e^4 + 2 q_e^6 + 4 q_5^2 + 16 q_e^2 q_5^2 + 30 q_e^4 q_5^2 \\[4pt]
+ 35 q_e^6 q_5^2 + 26 q_e^8 q_5^2 + 12 q_e^{10} q_5^2 + 3 q_e^{12} q_5^2 + 6 q_5^4 + 31 q_e^2 q_5^4 + 73 q_e^4 q_5^4 \\[4pt]
+ 107 q_e^6 q_5^4 + 104 q_e^8 q_5^4 + 66 q_e^{10} q_5^4 + 24 q_e^{12} q_5^4 + q_e^{14} q_5^4 - 3 q_e^{16} q_5^4 - q_e^{18} q_5^4 \\[4pt]
+ 5 q_5^6 + 33 q_e^2 q_5^6 + 94 q_e^4 q_5^6 + 159 q_e^6 q_5^6 + 176 q_e^8 q_5^6 + 123 q_e^{10} q_5^6 \\[4pt]
+ 39 q_e^{12} q_5^6 - 18 q_e^{14} q_5^6 - 28 q_e^{16} q_5^6 - 14 q_e^{18} q_5^6 - 2 q_e^{20} q_5^6 + q_e^{22} q_5^6 \\[4pt]
+ 2 q_5^8 + 20 q_e^2 q_5^8 + 71 q_e^4 q_5^8 + 140 q_e^6 q_5^8 + 171 q_e^8 q_5^8 + 119 q_e^{10} q_5^8 \\[4pt]
+ 9 q_e^{12} q_5^8 - 82 q_e^{14} q_5^8 - 100 q_e^{16} q_5^8 - 61 q_e^{18} q_5^8 - 17 q_e^{20} q_5^8 + 2 q_e^{22} q_5^8 \\[4pt]
+ 2 q_e^{24} q_5^8 + 5 q_e^2 q_5^{10} + 28 q_e^4 q_5^{10} + 69 q_e^6 q_5^{10} + 95 q_e^8 q_5^{10} + 59 q_e^{10} q_5^{10} \\[4pt]
- 41 q_e^{12} q_5^{10} - 143 q_e^{14} q_5^{10} - 175 q_e^{16} q_5^{10} - 125 q_e^{18} q_5^{10} - 50 q_e^{20} q_5^{10} - 4 q_e^{22} q_5^{10} \\[4pt]
+ 5 q_e^{24} q_5^{10} + q_e^{26} q_5^{10} + 4 q_e^4 q_5^{12} + 15 q_e^6 q_5^{12} + 23 q_e^8 q_5^{12} + 2 q_e^{10} q_5^{12} \\[4pt]
- 62 q_e^{12} q_5^{12} - 142 q_e^{14} q_5^{12} - 181 q_e^{16} q_5^{12} - 147 q_e^{18} q_5^{12} - 73 q_e^{20} q_5^{12} - 15 q_e^{22} q_5^{12} \\[4pt]
+ 5 q_e^{24} q_5^{12} + 3 q_e^{26} q_5^{12} + q_e^6 q_5^{14} + q_e^8 q_5^{14} - 8 q_e^{10} q_5^{14} - 36 q_e^{12} q_5^{14} \\[4pt]
- 79 q_e^{14} q_5^{14} - 111 q_e^{16} q_5^{14} - 104 q_e^{18} q_5^{14} - 62 q_e^{20} q_5^{14} - 18 q_e^{22} q_5^{14} + 3 q_e^{24} q_5^{14} \\[4pt]
+ 4 q_e^{26} q_5^{14} + q_e^{28} q_5^{14} - q_e^{10} q_5^{16} - 6 q_e^{12} q_5^{16} - 17 q_e^{14} q_5^{16} - 31 q_e^{16} q_5^{16} \\[4pt]
- 37 q_e^{18} q_5^{16} - 28 q_e^{20} q_5^{16} - 11 q_e^{22} q_5^{16} + q_e^{24} q_5^{16} + 3 q_e^{26} q_5^{16} + q_e^{28} q_5^{16} \\[4pt]
- q_e^{16} q_5^{18} - 3 q_e^{18} q_5^{18} - 4 q_e^{20} q_5^{18} - 3 q_e^{22} q_5^{18} + q_e^{26} q_5^{18} )\,, \\[12pt]
\mathcal{H}_{\mathbf{(1,27)}} &= \frac{1}{D(q_5,q_e)}\, q_e^4\, 
\mqty( 1 + q_e^2 + q_e^4 + 2 q_5^2 + 6 q_e^2 q_5^2 + 12 q_e^4 q_5^2 + 13 q_e^6 q_5^2 \\[4pt]
+ 10 q_e^8 q_5^2 + 4 q_e^{10} q_5^2 + q_e^{12} q_5^2 + 4 q_5^4 + 13 q_e^2 q_5^4 + 35 q_e^4 q_5^4 + 47 q_e^6 q_5^4 \\[4pt]
+ 48 q_e^8 q_5^4 + 26 q_e^{10} q_5^4 + 9 q_e^{12} q_5^4 - 2 q_e^{14} q_5^4 - 2 q_e^{16} q_5^4 - q_e^{18} q_5^4 \\[4pt]
+ 3 q_5^6 + 15 q_e^2 q_5^6 + 46 q_e^4 q_5^6 + 75 q_e^6 q_5^6 + 82 q_e^8 q_5^6 + 52 q_e^{10} q_5^6 \\[4pt]
+ 10 q_e^{12} q_5^6 - 15 q_e^{14} q_5^6 - 18 q_e^{16} q_5^6 - 7 q_e^{18} q_5^6 - q_e^{20} q_5^6 + q_e^{22} q_5^6 \\[4pt]
+ 2 q_5^8 + 10 q_e^2 q_5^8 + 39 q_e^4 q_5^8 + 70 q_e^6 q_5^8 + 86 q_e^8 q_5^8 + 52 q_e^{10} q_5^8 \\[4pt]
- 4 q_e^{12} q_5^8 - 49 q_e^{14} q_5^8 - 52 q_e^{16} q_5^8 - 27 q_e^{18} q_5^8 - 4 q_e^{20} q_5^8 + 4 q_e^{22} q_5^8 \\[4pt]
+ 2 q_e^{24} q_5^8 + 3 q_e^2 q_5^{10} + 16 q_e^4 q_5^{10} + 37 q_e^6 q_5^{10} + 47 q_e^8 q_5^{10} + 25 q_e^{10} q_5^{10} \\[4pt]
- 32 q_e^{12} q_5^{10} - 79 q_e^{14} q_5^{10} - 91 q_e^{16} q_5^{10} - 54 q_e^{18} q_5^{10} - 15 q_e^{20} q_5^{10} + 7 q_e^{22} q_5^{10} \\[4pt]
+ 6 q_e^{24} q_5^{10} + q_e^{26} q_5^{10} + 4 q_e^4 q_5^{12} + 9 q_e^6 q_5^{12} + 13 q_e^8 q_5^{12} - 2 q_e^{10} q_5^{12} \\[4pt]
- 37 q_e^{12} q_5^{12} - 75 q_e^{14} q_5^{12} - 89 q_e^{16} q_5^{12} - 62 q_e^{18} q_5^{12} - 22 q_e^{20} q_5^{12} + 6 q_e^{22} q_5^{12} \\[4pt]
+ 9 q_e^{24} q_5^{12} + 3 q_e^{26} q_5^{12} + q_e^6 q_5^{14} - q_e^8 q_5^{14} - 6 q_e^{10} q_5^{14} - 24 q_e^{12} q_5^{14} \\[4pt]
- 41 q_e^{14} q_5^{14} - 56 q_e^{16} q_5^{14} - 43 q_e^{18} q_5^{14} - 21 q_e^{20} q_5^{14} + 2 q_e^{22} q_5^{14} + 7 q_e^{24} q_5^{14} \\[4pt]
+ 4 q_e^{26} q_5^{14} + q_e^{28} q_5^{14} - q_e^{10} q_5^{16} - 4 q_e^{12} q_5^{16} - 9 q_e^{14} q_5^{16} - 15 q_e^{16} q_5^{16} \\[4pt]
- 15 q_e^{18} q_5^{16} - 10 q_e^{20} q_5^{16} - q_e^{22} q_5^{16} + 3 q_e^{24} q_5^{16} + 3 q_e^{26} q_5^{16} + q_e^{28} q_5^{16} \\[4pt]
- q_e^{16} q_5^{18} - q_e^{18} q_5^{18} - 2 q_e^{20} q_5^{18} - q_e^{22} q_5^{18} + q_e^{26} q_5^{18} + q_e^{28} q_5^{18} )\,, \\[12pt]
\mathcal{H}_{\mathbf{(27,1)}} &= \frac{1}{D(q_5,q_e)}\,\hspace{1em} 
\mqty( q_e^4 + q_e^6 + q_e^8 + q_5^2 + 3 q_e^2 q_5^2 + 8 q_e^4 q_5^2 + 11 q_e^6 q_5^2 \\[4pt]
+ 12 q_e^8 q_5^2 + 8 q_e^{10} q_5^2 + 4 q_e^{12} q_5^2 + q_e^{14} q_5^2 + 3 q_5^4 + 9 q_e^2 q_5^4 + 25 q_e^4 q_5^4 \\[4pt]
+ 36 q_e^6 q_5^4 + 44 q_e^8 q_5^4 + 33 q_e^{10} q_5^4 + 21 q_e^{12} q_5^4 + 6 q_e^{14} q_5^4 + q_e^{16} q_5^4 - q_e^{18} q_5^4 \\[4pt]
+ 3 q_5^6 + 12 q_e^2 q_5^6 + 34 q_e^4 q_5^6 + 57 q_e^6 q_5^6 + 70 q_e^8 q_5^6 + 57 q_e^{10} q_5^6 \\[4pt]
+ 29 q_e^{12} q_5^6 + 3 q_e^{14} q_5^6 - 10 q_e^{16} q_5^6 - 8 q_e^{18} q_5^6 - 4 q_e^{20} q_5^6 + 2 q_5^8 \\[4pt]
+ 9 q_e^2 q_5^8 + 31 q_e^4 q_5^8 + 54 q_e^6 q_5^8 + 70 q_e^8 q_5^8 + 52 q_e^{10} q_5^8 + 17 q_e^{12} q_5^8 \\[4pt]
- 22 q_e^{14} q_5^8 - 37 q_e^{16} q_5^8 - 30 q_e^{18} q_5^8 - 14 q_e^{20} q_5^8 - 3 q_e^{22} q_5^8 + 3 q_e^2 q_5^{10} \\[4pt]
+ 14 q_e^4 q_5^{10} + 30 q_e^6 q_5^{10} + 37 q_e^8 q_5^{10} + 22 q_e^{10} q_5^{10} - 17 q_e^{12} q_5^{10} - 52 q_e^{14} q_5^{10} \\[4pt]
- 70 q_e^{16} q_5^{10} - 54 q_e^{18} q_5^{10} - 31 q_e^{20} q_5^{10} - 9 q_e^{22} q_5^{10} - 2 q_e^{24} q_5^{10} + 4 q_e^4 q_5^{12} \\[4pt]
+ 8 q_e^6 q_5^{12} + 10 q_e^8 q_5^{12} - 3 q_e^{10} q_5^{12} - 29 q_e^{12} q_5^{12} - 57 q_e^{14} q_5^{12} - 70 q_e^{16} q_5^{12} \\[4pt]
- 57 q_e^{18} q_5^{12} - 34 q_e^{20} q_5^{12} - 12 q_e^{22} q_5^{12} - 3 q_e^{24} q_5^{12} + q_e^6 q_5^{14} - q_e^8 q_5^{14} \\[4pt]
- 6 q_e^{10} q_5^{14} - 21 q_e^{12} q_5^{14} - 33 q_e^{14} q_5^{14} - 44 q_e^{16} q_5^{14} - 36 q_e^{18} q_5^{14} - 25 q_e^{20} q_5^{14} \\[4pt]
- 9 q_e^{22} q_5^{14} - 3 q_e^{24} q_5^{14} - q_e^{10} q_5^{16} - 4 q_e^{12} q_5^{16} - 8 q_e^{14} q_5^{16} - 12 q_e^{16} q_5^{16} \\[4pt]
- 11 q_e^{18} q_5^{16} - 8 q_e^{20} q_5^{16} - 3 q_e^{22} q_5^{16} - q_e^{24} q_5^{16} - q_e^{16} q_5^{18} - q_e^{18} q_5^{18} - q_e^{20} q_5^{18} ) \,,
\end{align*}
\end{subequations}
\end{footnotesize}
\end{allowdisplaybreaks}%
where the denominator factor is given by
\begin{align}
D(q_5,q_e) &= \left(1 - q_e^2\right) \left(1 - q_e^4\right) \left(1 - q_e^6\right) \left(1 - q_5^2\right) \left(1 - q_5^4\right) \left(1 - q_5^6\right) \left(1 - q_e^2 q_5^2\right) 
\notag\\[5pt]
&\qquad
\times \left(1 - q_e^4 q_5^2\right)^2 \left(1 - q_e^2 q_5^4\right) \left(1 - q_e^6 q_5^2\right) \left(1 - q_e^4 q_5^4\right) \left(1 - q_e^8 q_5^2\right) \,.
\end{align}

\subsection{Multi-graded Hilbert series for spurions $S_\text{III} = \{Y_e^{\phantom{\dagger}}, Y_e^\dagger, Y_\nu^{\phantom{\dagger}}, Y_\nu^\dagger\}$}
\label{appsubsec:YeYnu}

Using the grading scheme
\begin{equation}
Y_e^{\phantom{\dagger}} \sim q_e \,,\qquad
Y_e^\dagger \sim q_e \,,\qquad
Y_\nu^{\phantom{\dagger}} \sim q_\nu \,,\qquad
Y_\nu^\dagger \sim q_\nu \,,
\end{equation}
the multi-graded Hilbert series for the MLFV case III in \cref{tab:Cases} read:
\begin{allowdisplaybreaks}
\begin{footnotesize}
\begin{subequations}
\begin{align*}
\mathcal{H}_{\mathbf{(1,1,1)}} &= \frac{1}{D(q_e, q_\nu)} \,
\mqty( 1 + q_e^6 q_\nu^6 )\,, \\[8pt]
\mathcal{H}_{\mathbf{(3,\bar{3},1)}} &= \frac{1}{D(q_e, q_\nu)}\,\hspace{0.5em} q_e\, \hspace{0.5em} 
\mqty( 1 + q_e^2 + q_e^4 + q_\nu^2 + 2q_e^2 q_\nu^2 + 2q_e^4 q_\nu^2 + q_e^6 q_\nu^2 \\[4pt]
+ q_\nu^4 + 2q_e^2 q_\nu^4 + 2q_e^4 q_\nu^4 + q_e^6 q_\nu^4 + q_e^2 q_\nu^6 + q_e^4 q_\nu^6 + q_e^6 q_\nu^6 )\,, \\[8pt]
\mathcal{H}_{\mathbf{(1,\bar{3},3)}} &= \frac{1}{D(q_e, q_\nu)}\, q_\nu q_e\, 
\mqty( 1 + q_e^2 + q_e^4 + q_\nu^2 + 2 q_e^2 q_\nu^2 + 2 q_e^4 q_\nu^2 + q_e^6 q_\nu^2 \\[4pt]
+ q_\nu^4 + 2 q_e^2 q_\nu^4 + 2 q_e^4 q_\nu^4 + q_e^6 q_\nu^4 + q_e^2 q_\nu^6 + q_e^4 q_\nu^6 + q_e^6 q_\nu^6 )\,, \\[8pt]
\mathcal{H}_{\mathbf{(\bar{3},\bar{3},\bar{3})}} &= \frac{1}{D(q_e, q_\nu)}\, q_\nu q_e\, 
\mqty( 1 + 2 q_e^2 + 2 q_e^4 + q_e^6 + 2 q_\nu^2 + 5 q_e^2 q_\nu^2 + 6 q_e^4 q_\nu^2 
+ 4 q_e^6 q_\nu^2 + q_e^8 q_\nu^2 + 2 q_\nu^4 \\[4pt]
+ 6 q_e^2 q_\nu^4 + 7 q_e^4 q_\nu^4 + 5 q_e^6 q_\nu^4 + q_e^8 q_\nu^4 
+ q_\nu^6 + 4 q_e^2 q_\nu^6 + 5 q_e^4 q_\nu^6 + 3 q_e^6 q_\nu^6 \\[4pt]
- q_e^{10} q_\nu^6 + q_e^2 q_\nu^8 + q_e^4 q_\nu^8 
- q_e^8 q_\nu^8 - q_e^{10} q_\nu^8 - q_e^6 q_\nu^{10} - q_e^8 q_\nu^{10} - q_e^{10} q_\nu^{10} )\,, \\[8pt]
\mathcal{H}_{\mathbf{(6,\bar{3},\bar{3})}} &= \frac{1}{D(q_e, q_\nu)}\, q_\nu q_e\, 
\mqty( 1 + 2 q_e^2 + 3 q_e^4 + 2 q_e^6 + q_e^8 + 2 q_\nu^2 + 6 q_e^2 q_\nu^2 \\[4pt]
+ 9 q_e^4 q_\nu^2 + 8 q_e^6 q_\nu^2 + 4 q_e^8 q_\nu^2 + q_e^{10} q_\nu^2 + 3 q_\nu^4 + 9 q_e^2 q_\nu^4 + 14 q_e^4 q_\nu^4 \\[4pt]
+ 12 q_e^6 q_\nu^4 + 6 q_e^8 q_\nu^4 + q_e^{10} q_\nu^4 + 2 q_\nu^6 + 8 q_e^2 q_\nu^6 + 12 q_e^4 q_\nu^6 + 10 q_e^6 q_\nu^6 \\[4pt]
+ 3 q_e^8 q_\nu^6 - q_e^{10} q_\nu^6 + q_e^2 q_\nu^8 + q_e^4 q_\nu^8 - q_e^6 q_\nu^8 - 2 q_e^8 q_\nu^8 - 4 q_e^{10} q_\nu^8 \\[4pt]
- 2 q_e^{12} q_\nu^8 + q_e^2 q_\nu^{10} + q_e^4 q_\nu^{10} - q_e^6 q_\nu^{10} - 4 q_e^8 q_\nu^{10} - 4 q_e^{10} q_\nu^{10} - 2 q_e^{12} q_\nu^{10} \\[4pt]
- q_e^6 q_\nu^{12} - 2 q_e^8 q_\nu^{12} - 2 q_e^{10} q_\nu^{12} - q_e^{12} q_\nu^{12} )\,, \\[8pt]
\mathcal{H}_{\mathbf{(8,1,1)}} &= \frac{1}{D(q_e, q_\nu)}\, \hspace{2em}
\mqty( q_e^2 + q_e^4 + q_\nu^2 + 2 q_e^2 q_\nu^2 + 2 q_e^4 q_\nu^2 + q_e^6 q_\nu^2  \\[4pt]
+ q_\nu^4 + 2 q_e^2 q_\nu^4 + 2 q_e^4 q_\nu^4 + q_e^6 q_\nu^4 + q_e^2 q_\nu^6 + q_e^4 q_\nu^6 )\,, \\[8pt]
\mathcal{H}_{\mathbf{(1,8,1)}} &= \frac{1}{D(q_e, q_\nu)}\, \hspace{0.5em}q_e^2\, \hspace{0.5em}
\mqty( 1 + q_e^2 + q_\nu^2 + 2 q_e^2 q_\nu^2 + 2 q_e^4 q_\nu^2 + q_e^6 q_\nu^2 + q_\nu^4 \\[4pt]
+ 2 q_e^2 q_\nu^4 + 2 q_e^4 q_\nu^4 + q_e^6 q_\nu^4 + q_e^2 q_\nu^6 + q_e^4 q_\nu^6 + q_e^6 q_\nu^6 - q_e^{10} q_\nu^6 )\,, \\[8pt]
\mathcal{H}_{\mathbf{(8,8,1)}} &= \frac{1}{D(q_e, q_\nu)}\, \hspace{0.5em}q_e^2\,\hspace{0.5em} 
\mqty( 1 + 3 q_e^2 + 4 q_e^4 + 2 q_e^6 + 3 q_\nu^2 + 9 q_e^2 q_\nu^2 + 12 q_e^4 q_\nu^2 
+ 9 q_e^6 q_\nu^2 + 3 q_e^8 q_\nu^2 \\[4pt]
+ 4 q_\nu^4 + 13 q_e^2 q_\nu^4 + 18 q_e^4 q_\nu^4 + 14 q_e^6 q_\nu^4 + 5 q_e^8 q_\nu^4 
+ 3 q_\nu^6 + 11 q_e^2 q_\nu^6 \\[4pt]
+ 16 q_e^4 q_\nu^6 + 11 q_e^6 q_\nu^6 + 2 q_e^8 q_\nu^6 - 2 q_e^{10} q_\nu^6 - q_e^{12} q_\nu^6 
+ q_\nu^8 + 5 q_e^2 q_\nu^8 + 7 q_e^4 q_\nu^8 \\[4pt]
+ 3 q_e^6 q_\nu^8 - 3 q_e^8 q_\nu^8 - 5 q_e^{10} q_\nu^8 - 2 q_e^{12} q_\nu^8 
+ q_e^2 q_\nu^{10} + q_e^4 q_\nu^{10} - 2 q_e^6 q_\nu^{10} \\[4pt]
- 5 q_e^8 q_\nu^{10} - 5 q_e^{10} q_\nu^{10} - 2 q_e^{12} q_\nu^{10} - q_e^6 q_\nu^{12}
- 2 q_e^8 q_\nu^{12} - 2 q_e^{10} q_\nu^{12} - q_e^{12} q_\nu^{12} )\,, \\[8pt]
\mathcal{H}_{\mathbf{(1,8,8)}} &= \frac{1}{D(q_e, q_\nu)}\, q_\nu^2 q_e^2\, 
\mqty( 2 + 4 q_e^2 + 4 q_e^4 + 2 q_e^6 + 4 q_\nu^2 + 11 q_e^2 q_\nu^2 + 14 q_e^4 q_\nu^2 \\[4pt]
+ 10 q_e^6 q_\nu^2 + 3 q_e^8 q_\nu^2 + 4 q_\nu^4 + 14 q_e^2 q_\nu^4 + 20 q_e^4 q_\nu^4 + 15 q_e^6 q_\nu^4 + 5 q_e^8 q_\nu^4 \\[4pt]
+ 2 q_\nu^6 + 10 q_e^2 q_\nu^6 + 15 q_e^4 q_\nu^6 + 11 q_e^6 q_\nu^6 + 2 q_e^8 q_\nu^6 - 3 q_e^{10} q_\nu^6 - q_e^{12} q_\nu^6 \\[4pt]
+ 3 q_e^2 q_\nu^8 + 5 q_e^4 q_\nu^8 + 2 q_e^6 q_\nu^8 - 3 q_e^8 q_\nu^8 - 5 q_e^{10} q_\nu^8 - 2 q_e^{12} q_\nu^8 - 3 q_e^6 q_\nu^{10} \\[4pt]
- 5 q_e^8 q_\nu^{10} - 4 q_e^{10} q_\nu^{10} - 2 q_e^{12} q_\nu^{10} - q_e^6 q_\nu^{12} - 2 q_e^8 q_\nu^{12} - 2 q_e^{10} q_\nu^{12} - q_e^{12} q_\nu^{12} )\,,
\end{align*}
\end{subequations}
\end{footnotesize}
\end{allowdisplaybreaks}%

\begin{allowdisplaybreaks}
\begin{footnotesize}
\begin{subequations}
\begin{align*}
\mathcal{H}_{\mathbf{(27,1,1)}} &= \frac{1}{D(q_e, q_\nu)}\, \hspace{2em}
\mqty( q_e^4 + q_e^6 + q_e^8 + q_e^2 q_\nu^2 + 3 q_e^4 q_\nu^2 + 4 q_e^6 q_\nu^2 + 3 q_e^8 q_\nu^2 + q_e^{10} q_\nu^2 \\[4pt]
+ q_\nu^4 + 3 q_e^2 q_\nu^4 + 7 q_e^4 q_\nu^4 + 7 q_e^6 q_\nu^4 + 5 q_e^8 q_\nu^4 + q_e^{10} q_\nu^4 + q_\nu^6 \\[4pt]
+ 4 q_e^2 q_\nu^6 + 7 q_e^4 q_\nu^6 + 7 q_e^6 q_\nu^6 + 3 q_e^8 q_\nu^6 - q_e^{12} q_\nu^6 + q_\nu^8 + 3 q_e^2 q_\nu^8 \\[4pt]
+ 5 q_e^4 q_\nu^8 + 3 q_e^6 q_\nu^8 - 3 q_e^8 q_\nu^8 - 5 q_e^{10} q_\nu^8 - 2 q_e^{12} q_\nu^8 + q_e^2 q_\nu^{10} + q_e^4 q_\nu^{10} \\[4pt]
- 3 q_e^8 q_\nu^{10} - 3 q_e^{10} q_\nu^{10} - 2 q_e^{12} q_\nu^{10} - q_e^6 q_\nu^{12} - 2 q_e^8 q_\nu^{12} - 2 q_e^{10} q_\nu^{12} - q_e^{12} q_\nu^{12} )\,, \\[8pt]
\mathcal{H}_{\mathbf{(1,27,1)}} &= \frac{1}{D(q_e, q_\nu)}\,\hspace{0.5em} q_e^4\, \hspace{0.5em}
\mqty( 1 + q_e^2 + q_e^4 + q_\nu^2 + 3 q_e^2 q_\nu^2 + 4 q_e^4 q_\nu^2 + 3 q_e^6 q_\nu^2 + q_e^8 q_\nu^2 \\[4pt]
+ 2 q_\nu^4 + 5 q_e^2 q_\nu^4 + 8 q_e^4 q_\nu^4 + 6 q_e^6 q_\nu^4 + 3 q_e^8 q_\nu^4 + q_\nu^6 + 5 q_e^2 q_\nu^6 \\[4pt]
+ 8 q_e^4 q_\nu^6 + 7 q_e^6 q_\nu^6 + 2 q_e^8 q_\nu^6 - q_e^{10} q_\nu^6 - q_e^{12} q_\nu^6 + q_\nu^8 + 3 q_e^2 q_\nu^8 \\[4pt]
+ 5 q_e^4 q_\nu^8 + 3 q_e^6 q_\nu^8 - q_e^8 q_\nu^8 - 3 q_e^{10} q_\nu^8 - 2 q_e^{12} q_\nu^8 + q_e^2 q_\nu^{10} + q_e^4 q_\nu^{10} \\[4pt]
- 3 q_e^8 q_\nu^{10} - 3 q_e^{10} q_\nu^{10} - 2 q_e^{12} q_\nu^{10} - q_e^6 q_\nu^{12} - 2 q_e^8 q_\nu^{12} - 2 q_e^{10} q_\nu^{12} - q_e^{12} q_\nu^{12} ) \,,
\end{align*}
\end{subequations}
\end{footnotesize}
\end{allowdisplaybreaks}%
and
\begin{subequations}
\begin{align}
\HS_{\mathbf{(\bar{3},1,3)}} = \HS_{\mathbf{(3,1,\bar{3})}} &= \HS_{\mathbf{(3,\bar{3},1)}} (q_\nu \leftrightarrow q_e) \,,\qquad
\HS_{\mathbf{(1,1,8)}} = \HS_{\mathbf{(1,8,1)}} (q_\nu \leftrightarrow q_e) \,, \\[5pt]
\HS_{\mathbf{(8,1,8)}} &= \HS_{\mathbf{(8,8,1)}} (q_\nu \leftrightarrow q_e) \,,\qquad
\HS_{\mathbf{(1,1,27)}} = \HS_{\mathbf{(1,27,1)}} (q_\nu \leftrightarrow q_e) \,,
\end{align}
\end{subequations}
where the denominator factor is given by
\begin{align}
D(q_e, q_\nu) &= \left(1 - q_e^2\right) \left(1 - q_e^4\right) \left(1 - q_e^6\right) \left(1 - q_\nu^2\right) \left(1 - q_\nu^4\right) \left(1 - q_\nu^6\right) 
\notag\\[5pt]
&\qquad
\times \left(1 - q_e^2 q_\nu^2 \right) \left(1 - q_e^2 q_\nu^4\right) \left(1 - q_e^4 q_\nu^2\right) \left(1 - q_e^4 q_\nu^4\right) \,.
\end{align}

\subsection{More Hilbert series for spurions $S_\text{IV} = \{Y_e^{\phantom{\dagger}}, Y_e^\dagger, Y_\nu^{\phantom{\dagger}}, Y_\nu^\dagger, m_R^{\phantom{\dagger}}, m_R^\dagger\}$}
\label{appsubsec:YeYnumR}

A few single-graded Hilbert series for the MLFV case IV in \cref{tab:Cases} are already presented in \cref{subsubsec:YeYnumR}. Those for the remaining relevant covariants read:
\begin{allowdisplaybreaks}
\begin{footnotesize}
\begin{align*}
\mathcal{H}_{\mathbf{(3,\bar{3},1)}} &= \frac{1}{D(q)}\, q\, 
\mqty( 1 + 2q^2 + 6q^4 + 17q^6 + 49q^8 \\[4pt]
+ 134q^{10} + 350q^{12} + 840q^{14} + 1884q^{16} + 3945q^{18} \\[4pt]
+ 7804q^{20} + 14617q^{22} + 26075q^{24} + 44374q^{26} + 72312q^{28} \\[4pt]
+ 113032q^{30} + 169916q^{32} + 246022q^{34} + 343750q^{36} + 464054q^{38} \\[4pt]
+ 606096q^{40} + 766599q^{42} + 939899q^{44} + 1117832q^{46} + 1290454q^{48} \\[4pt]
+ 1446692q^{50} + 1575652q^{52} + 1667681q^{54} + 1715623q^{56} + 1715623q^{58} \\[4pt]
+ 1667681q^{60} + 1575652q^{62} + 1446692q^{64} + 1290454q^{66} + 1117832q^{68} \\[4pt]
+ 939899q^{70} + 766599q^{72} + 606096q^{74} + 464054q^{76} + 343750q^{78} \\[4pt]
+ 246022q^{80} + 169916q^{82} + 113032q^{84} + 72312q^{86} + 44374q^{88} \\[4pt]
+ 26075q^{90} + 14617q^{92} + 7804q^{94} + 3945q^{96} + 1884q^{98} \\[4pt]
+ 840q^{100} + 350q^{102} + 134q^{104} + 49q^{106} + 17q^{108} \\[4pt]
+ 6q^{110} + 2q^{112} + q^{114} ) \,,
\end{align*}
\end{footnotesize}
\end{allowdisplaybreaks}%

\begin{allowdisplaybreaks}
\begin{footnotesize}
\begin{subequations}
\begin{align*}
\mathcal{H}_{\mathbf{(3,1,\bar{3})}} &= \frac{1}{D(q)}\, q\, 
\mqty( 1 + 3q^2 + 10q^4 + 29q^6 + 85q^8 \\[4pt]
+ 227q^{10} + 565q^{12} + 1301q^{14} + 2808q^{16} + 5694q^{18} \\[4pt]
+ 10927q^{20} + 19911q^{22} + 34597q^{24} + 57471q^{26} + 91526q^{28} \\[4pt]
+ 140038q^{30} + 206279q^{32} + 293018q^{34} + 401993q^{36} + 533306q^{38} \\[4pt]
+ 684938q^{40} + 852425q^{42} + 1028814q^{44} + 1204999q^{46} + 1370374q^{48} \\[4pt]
+ 1513873q^{50} + 1625094q^{52} + 1695561q^{54} + 1719690q^{56} + 1695561q^{58} \\[4pt]
+ 1625094q^{60} + 1513873q^{62} + 1370374q^{64} + 1204999q^{66} + 1028814q^{68} \\[4pt]
+ 852425q^{70} + 684938q^{72} + 533306q^{74} + 401993q^{76} + 293018q^{78} \\[4pt]
+ 206279q^{80} + 140038q^{82} + 91526q^{84} + 57471q^{86} + 34597q^{88} \\[4pt]
+ 19911q^{90} + 10927q^{92} + 5694q^{94} + 2808q^{96} + 1301q^{98} \\[4pt]
+ 565q^{100} + 227q^{102} + 85q^{104} + 29q^{106} + 10q^{108} \\[4pt]
+ 3q^{110} + q^{112} )\,,\\[8pt]
\mathcal{H}_{\mathbf{(1,\bar{3},3)}} &= \frac{1}{D(q)}\, q^2\, 
\mqty( 1 + 3q^2 + 10q^4 + 29q^6 + 85q^8 \\[4pt]
+ 227q^{10} + 565q^{12} + 1301q^{14} + 2808q^{16} + 5694q^{18} \\[4pt]
+ 10927q^{20} + 19911q^{22} + 34597q^{24} + 57471q^{26} + 91526q^{28} \\[4pt]
+ 140038q^{30} + 206279q^{32} + 293018q^{34} + 401993q^{36} + 533306q^{38} \\[4pt]
+ 684938q^{40} + 852425q^{42} + 1028814q^{44} + 1204999q^{46} + 1370374q^{48} \\[4pt]
+ 1513873q^{50} + 1625094q^{52} + 1695561q^{54} + 1719690q^{56} + 1695561q^{58} \\[4pt]
+ 1625094q^{60} + 1513873q^{62} + 1370374q^{64} + 1204999q^{66} + 1028814q^{68} \\[4pt]
+ 852425q^{70} + 684938q^{72} + 533306q^{74} + 401993q^{76} + 293018q^{78} \\[4pt]
+ 206279q^{80} + 140038q^{82} + 91526q^{84} + 57471q^{86} + 34597q^{88} \\[4pt]
+ 19911q^{90} + 10927q^{92} + 5694q^{94} + 2808q^{96} + 1301q^{98} \\[4pt]
+ 565q^{100} + 227q^{102} + 85q^{104} + 29q^{106} + 10q^{108} \\[4pt]
+ 3q^{110} + q^{112} )\,, \\[8pt]
\mathcal{H}_{\mathbf{(\bar{3},\bar{3},\bar{3})}} &= \frac{1}{D(q)}\, q^2\, 
\mqty( 1 + 5q^2 + 18q^4 + 60q^6 + 187q^8 \\[4pt]
+ 534q^{10} + 1391q^{12} + 3326q^{14} + 7385q^{16} + 15337q^{18} \\[4pt]
+ 30005q^{20} + 55565q^{22} + 97839q^{24} + 164360q^{26} + 264230q^{28} \\[4pt]
+ 407535q^{30} + 604396q^{32} + 863519q^{34} + 1190500q^{36} + 1585985q^{38} \\[4pt]
+ 2044114q^{40} + 2551484q^{42} + 3087009q^{44} + 3622873q^{46} + 4126606q^{48} \\[4pt]
+ 4564184q^{50} + 4903632q^{52} + 5118801q^{54} + 5192510q^{56} + 5118801q^{58} \\[4pt]
+ 4903632q^{60} + 4564184q^{62} + 4126606q^{64} + 3622873q^{66} + 3087009q^{68} \\[4pt]
+ 2551484q^{70} + 2044114q^{72} + 1585985q^{74} + 1190500q^{76} + 863519q^{78} \\[4pt]
+ 604396q^{80} + 407535q^{82} + 264230q^{84} + 164360q^{86} + 97839q^{88} \\[4pt]
+ 55565q^{90} + 30005q^{92} + 15337q^{94} + 7385q^{96} + 3326q^{98} \\[4pt]
+ 1391q^{100} + 534q^{102} + 187q^{104} + 60q^{106} + 18q^{108} \\[4pt]
+ 5q^{110} + q^{112} )\,, \\[8pt]
\mathcal{H}_{\mathbf{(6,\bar{3},\bar{3})}} &= \frac{1}{D(q)}\, q^2\, 
\mqty( 1 + 5q^2 + 22q^4 + 79q^6 + 263q^8 \\[4pt]
+ 789q^{10} + 2153q^{12} + 5362q^{14} + 12317q^{16} + 26332q^{18} \\[4pt]
+ 52792q^{20} + 99846q^{22} + 179004q^{24} + 305426q^{26} + 497663q^{28} \\[4pt]
+ 776610q^{30} + 1163530q^{32} + 1677151q^{34} + 2330050q^{36} + 3124799q^{38} \\[4pt]
+ 4050510q^{40} + 5080558q^{42} + 6172107q^{44} + 7267977q^{46} + 8300935q^{48} \\[4pt]
+ 9200124q^{50} + 9898773q^{52} + 10342078q^{54} + 10494032q^{56} + 10342078q^{58} \\[4pt]
+ 9898773q^{60} + 9200124q^{62} + 8300935q^{64} + 7267977q^{66} + 6172107q^{68} \\[4pt]
+ 5080558q^{70} + 4050510q^{72} + 3124799q^{74} + 2330050q^{76} + 1677151q^{78} \\[4pt]
+ 1163530q^{80} + 776610q^{82} + 497663q^{84} + 305426q^{86} + 179004q^{88} \\[4pt]
+ 99846q^{90} + 52792q^{92} + 26332q^{94} + 12317q^{96} + 5362q^{98} \\[4pt]
+ 2153q^{100} + 789q^{102} + 263q^{104} + 79q^{106} + 22q^{108} \\[4pt]
+ 5q^{110} + q^{112} )\,, \\[8pt]
\mathcal{H}_{\mathbf{(8,1,1)}} &= \frac{1}{D(q)}\, q^2\, 
\mqty( 2 + 5q^2 + 12q^4 + 40q^6 + 112q^8 \\[4pt]
+ 289q^{10} + 714q^{12} + 1611q^{14} + 3393q^{16} + 6766q^{18} \\[4pt]
+ 12737q^{20} + 22782q^{22} + 38933q^{24} + 63600q^{26} + 99615q^{28} \\[4pt]
+ 150049q^{30} + 217608q^{32} + 304399q^{34} + 411450q^{36} + 537876q^{38} \\[4pt]
+ 680816q^{40} + 835311q^{42} + 993980q^{44} + 1147895q^{46} + 1287364q^{48} \\[4pt]
+ 1402451q^{50} + 1484543q^{52} + 1527391q^{54} + 1527391q^{56} + 1484543q^{58} \\[4pt]
+ 1402451q^{60} + 1287364q^{62} + 1147895q^{64} + 993980q^{66} + 835311q^{68} \\[4pt]
+ 680816q^{70} + 537876q^{72} + 411450q^{74} + 304399q^{76} + 217608q^{78} \\[4pt]
+ 150049q^{80} + 99615q^{82} + 63600q^{84} + 38933q^{86} + 22782q^{88} \\[4pt]
+ 12737q^{90} + 6766q^{92} + 3393q^{94} + 1611q^{96} + 714q^{98} \\[4pt]
+ 289q^{100} + 112q^{102} + 40q^{104} + 12q^{106} + 5q^{108} \\[4pt]
+ 2q^{110} )\,, \\[8pt]
\mathcal{H}_{\mathbf{(1,8,1)}} &= \frac{1}{D(q)}\, q^2\, 
\mqty( 1 + 2q^2 + 5q^4 + 17q^6 + 48q^8 \\[4pt]
+ 129q^{10} + 341q^{12} + 818q^{14} + 1823q^{16} + 3819q^{18} \\[4pt]
+ 7531q^{20} + 14065q^{22} + 25037q^{24} + 42494q^{26} + 69019q^{28} \\[4pt]
+ 107591q^{30} + 161204q^{32} + 232605q^{34} + 323883q^{36} + 435640q^{38} \\[4pt]
+ 566745q^{40} + 713995q^{42} + 871679q^{44} + 1032049q^{46} + 1185866q^{48} \\[4pt]
+ 1322840q^{50} + 1433093q^{52} + 1508353q^{54} + 1542422q^{56} + 1532485q^{58} \\[4pt]
+ 1479449q^{60} + 1387420q^{62} + 1263554q^{64} + 1117253q^{66} + 958504q^{68} \\[4pt]
+ 797340q^{70} + 642747q^{72} + 501508q^{74} + 378271q^{76} + 275530q^{78} \\[4pt]
+ 193418q^{80} + 130565q^{82} + 84618q^{84} + 52445q^{86} + 30957q^{88} \\[4pt]
+ 17363q^{90} + 9176q^{92} + 4511q^{94} + 2065q^{96} + 846q^{98} \\[4pt]
+ 288q^{100} + 77q^{102} + 8q^{104} - 12q^{106} - 5q^{108} \\[4pt]
- 3q^{110} - 3q^{112} - q^{118} )\,, \\[8pt]
\mathcal{H}_{\mathbf{(1,1,8)}} &= \frac{1}{D(q)}\, q^2\, 
\mqty( 2 + 6q^2 + 17q^4 + 55q^6 + 147q^8 \\[4pt]
+ 362q^{10} + 855q^{12} + 1866q^{14} + 3828q^{16} + 7470q^{18} \\[4pt]
+ 13812q^{20} + 24335q^{22} + 41056q^{24} + 66350q^{26} + 102994q^{28} \\[4pt]
+ 153978q^{30} + 221909q^{32} + 308789q^{34} + 415557q^{36} + 541264q^{38} \\[4pt]
+ 683030q^{40} + 835922q^{42} + 992658q^{44} + 1144473q^{46} + 1281880q^{48} \\[4pt]
+ 1395164q^{50} + 1475916q^{52} + 1518049q^{54} + 1518049q^{56} + 1475916q^{58} \\[4pt]
+ 1395164q^{60} + 1281880q^{62} + 1144473q^{64} + 992658q^{66} + 835922q^{68} \\[4pt]
+ 683030q^{70} + 541264q^{72} + 415557q^{74} + 308789q^{76} + 221909q^{78} \\[4pt]
+ 153978q^{80} + 102994q^{82} + 66350q^{84} + 41056q^{86} + 24335q^{88} \\[4pt]
+ 13812q^{90} + 7470q^{92} + 3828q^{94} + 1866q^{96} + 855q^{98} \\[4pt]
+ 362q^{100} + 147q^{102} + 55q^{104} + 17q^{106} + 6q^{108} + 2q^{110} )\,, \\[8pt]
\mathcal{H}_{\mathbf{(8,8,1)}} &= \frac{1}{D(q)}\, q^2\, 
\mqty( 1 + 6q^2 + 21q^4 + 66q^6 + 211q^8 \\[4pt]
+ 642q^{10} + 1794q^{12} + 4605q^{14} + 10879q^{16} + 23850q^{18} \\[4pt]
+ 48978q^{20} + 94762q^{22} + 173526q^{24} + 302042q^{26} + 501409q^{28} \\[4pt]
+ 796121q^{30} + 1212242q^{32} + 1774100q^{34} + 2500075q^{36} + 3398116q^{38} \\[4pt]
+ 4461005q^{40} + 5662924q^{42} + 6958390q^{44} + 8283152q^{46} + 9558406q^{48} \\[4pt]
+ 10698365q^{50} + 11619006q^{52} + 12247840q^{54} + 12533352q^{56} + 12451260q^{58} \\[4pt]
+ 12007813q^{60} + 11239510q^{62} + 10207802q^{64} + 8991266q^{66} + 7676604q^{68} \\[4pt]
+ 6348371q^{70} + 5080394q^{72} + 3930093q^{74} + 2934908q^{76} + 2112248q^{78} \\[4pt]
+ 1462265q^{80} + 971443q^{82} + 617446q^{84} + 374142q^{86} + 215164q^{88} \\[4pt]
+ 116692q^{90} + 59222q^{92} + 27807q^{94} + 11827q^{96} + 4402q^{98} \\[4pt]
+ 1324q^{100} + 223q^{102} - 60q^{104} - 73q^{106} - 44q^{108} - 19q^{110} \\[4pt]
- 6q^{112} - 4q^{114} - 2q^{116} )\,, \\[8pt]
\mathcal{H}_{\mathbf{(8,1,8)}} &= \frac{1}{D(q)}\, q^2\, 
\mqty( 1 + 11q^2 + 53q^4 + 192q^6 + 618q^8 \\[4pt]
+ 1765q^{10} + 4531q^{12} + 10692q^{14} + 23397q^{16} + 47841q^{18} \\[4pt]
+ 92146q^{20} + 168024q^{22} + 291252q^{24} + 481825q^{26} + 763013q^{28} \\[4pt]
+ 1159515q^{30} + 1694813q^{32} + 2387110q^{34} + 3244911q^{36} + 4263103q^{38} \\[4pt]
+ 5419291q^{40} + 6672187q^{42} + 7962912q^{44} + 9218234q^{46} + 10356836q^{48} \\[4pt]
+ 11298070q^{50} + 11970601q^{52} + 12321008q^{54} + 12321008q^{56} + 11970601q^{58} \\[4pt]
+ 11298070q^{60} + 10356836q^{62} + 9218234q^{64} + 7962912q^{66} + 6672187q^{68} \\[4pt]
+ 5419291q^{70} + 4263103q^{72} + 3244911q^{74} + 2387110q^{76} + 1694813q^{78} \\[4pt]
+ 1159515q^{80} + 763013q^{82} + 481825q^{84} + 291252q^{86} + 168024q^{88} \\[4pt]
+ 92146q^{90} + 47841q^{92} + 23397q^{94} + 10692q^{96} + 4531q^{98} \\[4pt]
+ 1765q^{100} + 618q^{102} + 192q^{104} + 53q^{106} + 11q^{108} + q^{110} )\,, \\[8pt]
\mathcal{H}_{\mathbf{(1,8,8)}} &= \frac{1}{D(q)}\, q^4\, 
\mqty( 3 + 17q^2 + 68q^4 + 241q^6 + 748q^8 \\[4pt]
+ 2072q^{10} + 5239q^{12} + 12196q^{14} + 26370q^{16} + 53445q^{18} \\[4pt]
+ 102130q^{20} + 184889q^{22} + 318496q^{24} + 523840q^{26} + 824951q^{28} \\[4pt]
+ 1247143q^{30} + 1813728q^{32} + 2541921q^{34} + 3438559q^{36} + 4495578q^{38} \\[4pt]
+ 5686764q^{40} + 6966845q^{42} + 8272540q^{44} + 9526785q^{46} + 10646058q^{48} \\[4pt]
+ 11548761q^{50} + 12164637q^{52} + 12443893q^{54} + 12363141q^{56} + 11928468q^{58} \\[4pt]
+ 11175185q^{60} + 10162800q^{62} + 8967543q^{64} + 7673690q^{66} + 6363636q^{68} \\[4pt]
+ 5109663q^{70} + 3968445q^{72} + 2977438q^{74} + 2154635q^{76} + 1501165q^{78} \\[4pt]
+ 1004704q^{80} + 644098q^{82} + 394197q^{84} + 229314q^{86} + 126009q^{88} \\[4pt]
+ 64902q^{90} + 30976q^{92} + 13413q^{94} + 5088q^{96} + 1558q^{98} \\[4pt]
+ 261q^{100} - 90q^{102} - 115q^{104} - 77q^{106} - 38q^{108} \\[4pt]
- 14q^{110} - 6q^{112} - 2q^{114} )\,, \\[8pt]
\mathcal{H}_{\mathbf{(27,1,1)}} &= \frac{1}{D(q)}\, q^4\, 
\mqty( 3 + 11q^2 + 43q^4 + 144q^6 + 450q^8 \\[4pt]
+ 1247q^{10} + 3160q^{12} + 7309q^{14} + 15717q^{16} + 31534q^{18} \\[4pt]
+ 59603q^{20} + 106489q^{22} + 180873q^{24} + 292898q^{26} + 453886q^{28} \\[4pt]
+ 674573q^{30} + 963970q^{32} + 1326706q^{34} + 1761692q^{36} + 2259769q^{38} \\[4pt]
+ 2803593q^{40} + 3367158q^{42} + 3918081q^{44} + 4419761q^{46} + 4835781q^{48} \\[4pt]
+ 5133621q^{50} + 5289064q^{52} + 5289064q^{54} + 5133621q^{56} + 4835781q^{58} \\[4pt]
+ 4419761q^{60} + 3918081q^{62} + 3367158q^{64} + 2803593q^{66} + 2259769q^{68} \\[4pt]
+ 1761692q^{70} + 1326706q^{72} + 963970q^{74} + 674573q^{76} + 453886q^{78} \\[4pt]
+ 292898q^{80} + 180873q^{82} + 106489q^{84} + 59603q^{86} + 31534q^{88} \\[4pt]
+ 15717q^{90} + 7309q^{92} + 3160q^{94} + 1247q^{96} + 450q^{98} \\[4pt]
+ 144q^{100} + 43q^{102} + 11q^{104} + 3q^{106} )\,, \\[8pt]
\mathcal{H}_{\mathbf{(1,27,1)}} &= \frac{1}{D(q)}\, q^4\, 
\mqty( 1 + 2q^2 + 8q^4 + 26q^6 + 86q^8 \\[4pt]
+ 261q^{10} + 751q^{12} + 1953q^{14} + 4694q^{16} + 10414q^{18} \\[4pt]
+ 21637q^{20} + 42206q^{22} + 77874q^{24} + 136246q^{26} + 227110q^{28} \\[4pt]
+ 361556q^{30} + 551490q^{32} + 807563q^{34} + 1137804q^{36} + 1544738q^{38} \\[4pt]
+ 2024038q^{40} + 2562314q^{42} + 3137396q^{44} + 3718391q^{46} + 4268636q^{48} \\[4pt]
+ 4748621q^{50} + 5120979q^{52} + 5354610q^{54} + 5429035q^{56} + 5337006q^{58} \\[4pt]
+ 5085679q^{60} + 4695810q^{62} + 4198772q^{64} + 3632883q^{66} + 3038298q^{68} \\[4pt]
+ 2453038q^{70} + 1908536q^{72} + 1427889q^{74} + 1024161q^{76} + 701624q^{78} \\[4pt]
+ 456541q^{80} + 280052q^{82} + 159908q^{84} + 83269q^{86} + 37831q^{88} \\[4pt]
+ 13366q^{90} + 1777q^{92} - 2554q^{94} - 3363q^{96} - 2760q^{98} - 1858q^{100} \\[4pt]
- 1084q^{102} - 562q^{104} - 258q^{106} - 106q^{108} - 40q^{110} \\[4pt]
- 15q^{112} - 5q^{114} - 2q^{116} - q^{118} )\,, \\[8pt]
\mathcal{H}_{\mathbf{(1,1,27)}} &= \frac{1}{D(q)}\, q^2\, 
\mqty( 1 + 7q^2 + 26q^4 + 95q^6 + 290q^8 \\[4pt]
+ 809q^{10} + 2029q^{12} + 4728q^{14} + 10214q^{16} + 20735q^{18} \\[4pt]
+ 39649q^{20} + 71948q^{22} + 124208q^{24} + 204900q^{26} + 323684q^{28} \\[4pt]
+ 491100q^{30} + 716847q^{32} + 1008738q^{34} + 1370233q^{36} + 1799319q^{38} \\[4pt]
+ 2286404q^{40} + 2814348q^{42} + 3358130q^{44} + 3887064q^{46} + 4366822q^{48} \\[4pt]
+ 4763465q^{50} + 5046825q^{52} + 5194518q^{54} + 5194518q^{56} + 5046825q^{58} \\[4pt]
+ 4763465q^{60} + 4366822q^{62} + 3887064q^{64} + 3358130q^{66} + 2814348q^{68} \\[4pt]
+ 2286404q^{70} + 1799319q^{72} + 1370233q^{74} + 1008738q^{76} + 716847q^{78} \\[4pt]
+ 491100q^{80} + 323684q^{82} + 204900q^{84} + 124208q^{86} + 71948q^{88} \\[4pt]
+ 39649q^{90} + 20735q^{92} + 10214q^{94} + 4728q^{96} + 2029q^{98} \\[4pt]
+ 809q^{100} + 290q^{102} + 95q^{104} + 26q^{106} + 7q^{108} + q^{110} ) \,,
\end{align*}
\end{subequations}
\end{footnotesize}
\end{allowdisplaybreaks}%
where the denominator factor is given by
\begin{align}
D(q) &= \left(1-q^2\right)^3 \left(1-q^4\right)^4 \left(1-q^6\right)^4 \left(1-q^8\right)^2 \left(1-q^{10}\right)^2
\notag\\[5pt]
&\qquad
\times \left(1-q^{12}\right)^3 \left(1-q^{14}\right)^2 \left(1-q^{16}\right) \,.
\end{align}

\begin{table}[t]
\begin{center}
\fontsize{8.5pt}{11pt}\selectfont
\hspace{-3cm}
\begin{minipage}[t]{2.7cm}
\renewcommand{\arraystretch}{1.5}
\begin{tabular}[t]{c|c|c}
\multicolumn{2}{c}{$5: \psi^2H^3 + \hbox{h.c.}$}  & $SU(3)_{\ell, e, \nu}$ \\
\hline
$Q_{eH}$ & $(H^\dag H)(\bar \ell_p e_r H)$ & $(\mathbf{3},\mathbf{\bar 3},\mathbf{1})$ \\
$Q_{uH}$ & $(H^\dag H)(\bar q_p u_r \widetilde H )$ & $(\mathbf{1},\mathbf{1},\mathbf{1})$ \\
$Q_{dH}$ & $(H^\dag H)(\bar q_p d_r H)$ & $(\mathbf{1},\mathbf{1},\mathbf{1})$\\
\end{tabular}
\end{minipage}

\vspace{0.3cm}
\hspace{-2.5cm}
\begin{minipage}[t]{5.2cm}
\renewcommand{\arraystretch}{1.5}
\begin{tabular}[t]{c|c|c}
\multicolumn{2}{c}{$6:\psi^2 XH+\hbox{h.c.}$} & $SU(3)_{\ell, e, \nu}$ \\
\hline
$Q_{eW}$ & $(\bar \ell_p \sigma^{\mu\nu} e_r) \tau^I H W_{\mu\nu}^I$ & $(\mathbf{3},\mathbf{\bar 3},\mathbf{1})$ \\
$Q_{eB}$ & $(\bar \ell_p \sigma^{\mu\nu} e_r) H B_{\mu\nu}$ & $(\mathbf{3},\mathbf{\bar 3},\mathbf{1})$ \\
$Q_{uG}$ & $(\bar q_p \sigma^{\mu\nu} T^A u_r) \widetilde H \, G_{\mu\nu}^A$ & $(\mathbf{1},\mathbf{1},\mathbf{1})$ \\
$Q_{uW}$ & $(\bar q_p \sigma^{\mu\nu} u_r) \tau^I \widetilde H \, W_{\mu\nu}^I$ & $(\mathbf{1},\mathbf{1},\mathbf{1})$ \\
$Q_{uB}$ & $(\bar q_p \sigma^{\mu\nu} u_r) \widetilde H \, B_{\mu\nu}$ & $(\mathbf{1},\mathbf{1},\mathbf{1})$ \\
$Q_{dG}$ & $(\bar q_p \sigma^{\mu\nu} T^A d_r) H\, G_{\mu\nu}^A$ & $(\mathbf{1},\mathbf{1},\mathbf{1})$\\
$Q_{dW}$ & $(\bar q_p \sigma^{\mu\nu} d_r) \tau^I H\, W_{\mu\nu}^I$ & $(\mathbf{1},\mathbf{1},\mathbf{1})$\\
$Q_{dB}$ & $(\bar q_p \sigma^{\mu\nu} d_r) H\, B_{\mu\nu}$& $(\mathbf{1},\mathbf{1},\mathbf{1})$ \\
\end{tabular}
\end{minipage}
\hspace{1.5cm}
\begin{minipage}[t]{5.4cm}
\renewcommand{\arraystretch}{1.5}
\begin{tabular}[t]{c|c|c}
\multicolumn{2}{c}{$7:\psi^2H^2 D$} & $SU(3)_{\ell, e, \nu}$ \\
\hline
$Q_{H \ell}^{(1)}$ & $(H^\dag i\overleftrightarrow{D}_\mu H)(\bar \ell_p \gamma^\mu \ell_r)$ & $(\mathbf{1}\oplus\mathbf{8},\mathbf{1},\mathbf{1})$ \\
$Q_{H \ell}^{(3)}$ & $(H^\dag i\overleftrightarrow{D}^I_\mu H)(\bar \ell_p \tau^I \gamma^\mu \ell_r)$ & $(\mathbf{1}\oplus\mathbf{8},\mathbf{1},\mathbf{1})$ \\
$Q_{H e}$ & $(H^\dag i\overleftrightarrow{D}_\mu H)(\bar e_p \gamma^\mu e_r)$ & $(\mathbf{1},\mathbf{1}\oplus\mathbf{8},\mathbf{1})$ \\
$Q_{H q}^{(1)}$ & $(H^\dag i\overleftrightarrow{D}_\mu H)(\bar q_p \gamma^\mu q_r)$& $(\mathbf{1},\mathbf{1},\mathbf{1})$ \\
$Q_{H q}^{(3)}$ & $(H^\dag i\overleftrightarrow{D}^I_\mu H)(\bar q_p \tau^I \gamma^\mu q_r)$& $(\mathbf{1},\mathbf{1},\mathbf{1})$\\
$Q_{H u}$ & $(H^\dag i\overleftrightarrow{D}_\mu H)(\bar u_p \gamma^\mu u_r)$&  $(\mathbf{1},\mathbf{1},\mathbf{1})$\\
$Q_{H d}$ & $(H^\dag i\overleftrightarrow{D}_\mu H)(\bar d_p \gamma^\mu d_r)$& $(\mathbf{1},\mathbf{1},\mathbf{1})$ \\
$Q_{H u d}$ + h.c. & $i(\widetilde H ^\dag D_\mu H)(\bar u_p \gamma^\mu d_r)$& $(\mathbf{1},\mathbf{1},\mathbf{1})$\\
\end{tabular}
\end{minipage}

\vspace{0.3cm}
\hspace{-2.5cm}
\begin{minipage}[t]{4.75cm}
\renewcommand{\arraystretch}{1.5}
\begin{tabular}[t]{@{}c|c|c}
\multicolumn{2}{c}{$8:(\bar LL)(\bar LL)$} & $SU(3)_{\ell, e, \nu}$ \\
\hline
$Q_{\ell\ell}$ & $(\bar \ell_p \gamma_\mu \ell_r)(\bar \ell_s \gamma^\mu \ell_t)$ & $(\mathbf{1}\oplus\mathbf{1}\oplus\mathbf{8}\oplus\mathbf{8}\oplus\mathbf{27},\mathbf{1},\mathbf{1})$\\
$Q_{qq}^{(1)}$ & $(\bar q_p \gamma_\mu q_r)(\bar q_s \gamma^\mu q_t)$ & $(\mathbf{1},\mathbf{1},\mathbf{1})$\\
$Q_{qq}^{(3)}$ & $(\bar q_p \gamma_\mu \tau^I q_r)(\bar q_s \gamma^\mu \tau^I q_t)$ & $(\mathbf{1},\mathbf{1},\mathbf{1})$\\
$Q_{\ell q}^{(1)}$ & $(\bar \ell_p \gamma_\mu \ell_r)(\bar q_s \gamma^\mu q_t)$ & $(\mathbf{1}\oplus\mathbf{8},\mathbf{1},\mathbf{1})$\\
$Q_{\ell q}^{(3)}$ & $(\bar \ell_p \gamma_\mu \tau^I \ell_r)(\bar q_s \gamma^\mu \tau^I q_t)$ & $(\mathbf{1}\oplus\mathbf{8},\mathbf{1},\mathbf{1})$
\end{tabular}
\end{minipage}
\hspace{3.65cm}
\begin{minipage}[t]{4.75cm}
\renewcommand{\arraystretch}{1.5}
\begin{tabular}[t]{c|c|c}
\multicolumn{2}{c}{$8:(\bar LL)(\bar RR)$} & $SU(3)_{\ell, e, \nu}$ \\
\hline
$Q_{\ell e}$ & $(\bar \ell_p \gamma_\mu \ell_r)(\bar e_s \gamma^\mu e_t)$ &  $(\mathbf{1}\oplus\mathbf{8},\mathbf{1}\oplus\mathbf{8},\mathbf{1})$\\
$Q_{\ell u}$ & $(\bar \ell_p \gamma_\mu \ell_r)(\bar u_s \gamma^\mu u_t)$ & $(\mathbf{1}\oplus\mathbf{8},\mathbf{1},\mathbf{1})$ \\
$Q_{\ell d}$ & $(\bar \ell_p \gamma_\mu \ell_r)(\bar d_s \gamma^\mu d_t)$ & $(\mathbf{1}\oplus\mathbf{8},\mathbf{1},\mathbf{1})$ \\
$Q_{qe}$ & $(\bar q_p \gamma_\mu q_r)(\bar e_s \gamma^\mu e_t)$ &  $(\mathbf{1},\mathbf{1}\oplus\mathbf{8},\mathbf{1})$\\
$Q_{qu}^{(1)}$ & $(\bar q_p \gamma_\mu q_r)(\bar u_s \gamma^\mu u_t)$ &  $(\mathbf{1},\mathbf{1},\mathbf{1})$\\
$Q_{qu}^{(8)}$ & $(\bar q_p \gamma_\mu T^A q_r)(\bar u_s \gamma^\mu T^A u_t)$ &  $(\mathbf{1},\mathbf{1},\mathbf{1})$\\
$Q_{qd}^{(1)}$ & $(\bar q_p \gamma_\mu q_r)(\bar d_s \gamma^\mu d_t)$ &  $(\mathbf{1},\mathbf{1},\mathbf{1})$\\
$Q_{qd}^{(8)}$ & $(\bar q_p \gamma_\mu T^A q_r)(\bar d_s \gamma^\mu T^A d_t)$ & $(\mathbf{1},\mathbf{1},\mathbf{1})$
\end{tabular}
\end{minipage}

\vspace{0.3cm}
\hspace{-1.2cm}
\begin{minipage}[t]{5.25cm}
\renewcommand{\arraystretch}{1.5}
\begin{tabular}[t]{c|c|c}
\multicolumn{2}{c}{$8:(\bar RR)(\bar RR)$} & $SU(3)_{\ell, e, \nu}$ \\
\hline
$Q_{ee}$ & $(\bar e_p \gamma_\mu e_r)(\bar e_s \gamma^\mu e_t)$ & $(\mathbf{1},\mathbf{1}\oplus\mathbf{8}\oplus\mathbf{27},\mathbf{1})$ \\
$Q_{uu}$ & $(\bar u_p \gamma_\mu u_r)(\bar u_s \gamma^\mu u_t)$ & $(\mathbf{1},\mathbf{1},\mathbf{1})$ \\
$Q_{dd}$ & $(\bar d_p \gamma_\mu d_r)(\bar d_s \gamma^\mu d_t)$ & $(\mathbf{1},\mathbf{1},\mathbf{1})$ \\
$Q_{eu}$ & $(\bar e_p \gamma_\mu e_r)(\bar u_s \gamma^\mu u_t)$ & $(\mathbf{1},\mathbf{1}\oplus\mathbf{8},\mathbf{1})$ \\
$Q_{ed}$ & $(\bar e_p \gamma_\mu e_r)(\bar d_s\gamma^\mu d_t)$ & $(\mathbf{1},\mathbf{1}\oplus\mathbf{8},\mathbf{1})$ \\
$Q_{ud}^{(1)}$ & $(\bar u_p \gamma_\mu u_r)(\bar d_s \gamma^\mu d_t)$ & $(\mathbf{1},\mathbf{1},\mathbf{1})$ \\
$Q_{ud}^{(8)}$ & $(\bar u_p \gamma_\mu T^A u_r)(\bar d_s \gamma^\mu T^A d_t)$ & $(\mathbf{1},\mathbf{1},\mathbf{1})$ \\
\end{tabular}
\end{minipage}
\hspace{3.7cm}
\begin{minipage}[t]{5.5cm}
\renewcommand{\arraystretch}{1.5}
\begin{tabular}[t]{c|c|c}
\multicolumn{2}{c}{$8:(\bar LR)(\bar L R)+\hbox{h.c.}$} & $SU(3)_{\ell, e, \nu}$ \\
\hline
$Q_{quqd}^{(1)}$ & $(\bar q_p^j u_r) \epsilon_{jk} (\bar q_s^k d_t)$  & $(\mathbf{1},\mathbf{1},\mathbf{1})$\\
$Q_{quqd}^{(8)}$ & $(\bar q_p^j T^A u_r) \epsilon_{jk} (\bar q_s^k T^A d_t)$ &  $(\mathbf{1},\mathbf{1},\mathbf{1})$\\
$Q_{\ell equ}^{(1)}$ & $(\bar \ell_p^j e_r) \epsilon_{jk} (\bar q_s^k u_t)$  & $(\mathbf{3},\mathbf{\bar 3},\mathbf{1})$\\
$Q_{\ell equ}^{(3)}$ & $(\bar \ell_p^j \sigma_{\mu\nu} e_r) \epsilon_{jk} (\bar q_s^k \sigma^{\mu\nu} u_t)$ & $(\mathbf{3},\mathbf{\bar 3},\mathbf{1})$\\
\end{tabular}

\vspace{0.5cm}
\begin{tabular}[t]{c|c|c}
\multicolumn{2}{c}{$8:(\bar LR)(\bar RL)+\hbox{h.c.}$} & $SU(3)_{\ell, e, \nu}$ \\
\hline
$Q_{\ell edq}$ & $(\bar \ell_p^j e_r)(\bar d_s q_{tj})$ & $(\mathbf{3},\mathbf{\bar 3},\mathbf{1})$
\end{tabular}
\end{minipage}
\end{center}
\caption{Lepton and baryon number preserving fermionic operators in dim-6 SMEFT ~\cite{Grzadkowski:2010es}, together with the corresponding spurious transformations of the Wilson coefficients under $SU(3)_{\ell, e, \nu} \equiv SU(3)_\ell \times SU(3)_e \times SU(3)_\nu \subset G_{Lf}$, the non-abelian part of the MLFV group. Transformations under the $U(1)$ factors can be inferred from \cref{eqn:TransnuSMEFT}. Operators are organized into eight classes as in Ref.~\cite{Alonso:2013hga}, with classes 1-4 collecting bosonic operators not shown here. The subscripts $p, r, s, t$ denote flavor indices. Note that these SMEFT operators by definition do not involve right-handed neutrino fields $\nu$, and hence transform trivially under the $SU(3)_\nu$ factor.}
\label{tab:MLFVdim6SMEFTLNP}
\end{table}

\begin{table}[t]
\begin{center}
\fontsize{8.5pt}{11pt}\selectfont
\renewcommand{\arraystretch}{1.5}
\begin{tabular}[t]{c|c|c|c}
\multicolumn{2}{c}{$L$- and $B$-violating} & \multicolumn{1}{c}{$SU(3)_{\ell, e, \nu}$} & \# invariants under lepton parity $\mathcal{P}_\text{LN}$ \\
\hline
$Q_{duq}$ & $\epsilon_{\alpha\beta\gamma} \epsilon_{jk} \big[ (d_p^\alpha)^T C u_r^\beta \big] \big[ (q_s^{\gamma j})^T C \ell_t^k \big]$ & $\mathbf{(\bar{3}, 1, 1)}$ & 0 \\
$Q_{qqu}$ & $\epsilon_{\alpha\beta\gamma} \epsilon_{jk} \big[ (q_p^{\alpha j})^T C q_r^{\beta k} \big] \big[ (u_s^\gamma)^T C e_t \big]$ & $\mathbf{(1, \bar{3}, 1)}$ & 0 \\
$Q_{qqq}$ & $\epsilon_{\alpha\beta\gamma} \epsilon_{jn} \epsilon_{km} \big[ (q_p^{\alpha j})^T C q_r^{\beta k} \big] \big[ (q_s^{\gamma m})^T C \ell_t^n \big]$ & $\mathbf{(\bar{3}, 1, 1)}$ & 0 \\
$Q_{duu}$ & $\epsilon_{\alpha\beta\gamma} \big[ (d_p^\alpha)^T C u_r^\beta \big] \big[ (u_s^\gamma)^T C e_t \big]$ & $\mathbf{(1, \bar{3}, 1)}$ & 0 \\
\end{tabular}
\end{center}
\caption{Lepton and baryon number violating fermionic operators in dim-6 SMEFT ~\cite{Grzadkowski:2010es, Alonso:2014zka}, together with the corresponding spurious transformations of the Wilson coefficients under $SU(3)_{\ell, e, \nu} \equiv SU(3)_\ell \times SU(3)_e \times SU(3)_\nu \subset G_{Lf}$, the non-abelian part of the MLFV group. Transformations under the $U(1)$ factors can be inferred from \cref{eqn:TransnuSMEFT}. Here $C=i\gamma^2\gamma^0$ is the charge conjugation matrix. The subscripts $p, r, s, t$ denote flavor indices. Note that these SMEFT operators by definition do not involve right-handed neutrino fields $\nu$, and hence transform trivially under the $SU(3)_\nu$ factor. All these operators violate the lepton number $U(1)_\text{LN}$, as well as the lepton parity $\mathcal{P}_\text{LN} \equiv (-1)^\text{LN}$.}
\label{tab:MLFVdim6SMEFTLNV}
\end{table}

\begin{table}[t]
\begin{center}
\fontsize{8.5pt}{11pt}\selectfont
\hspace{-3cm}
\begin{minipage}[t]{2.7cm}
\renewcommand{\arraystretch}{1.5}
\begin{tabular}[t]{c|c|cc}
\multicolumn{2}{c}{$5: \psi^2H^3 + \hbox{h.c.}$}  & \text{Restriction} & \#\\
\hline
$Q_{eH}$ & $(H^\dag H)(\bar \ell_p e_r H)$ & $p=r$ & 3 \\
$Q_{uH}$ & $(H^\dag H)(\bar q_p u_r \widetilde H )$ & $-$ & $-$ \\
$Q_{dH}$ & $(H^\dag H)(\bar q_p d_r H)$ & $-$ & $-$\\
\end{tabular}
\end{minipage}

\vspace{0.4cm}
\hspace{-2.5cm}
\begin{minipage}[t]{5.2cm}
\renewcommand{\arraystretch}{1.5}
\begin{tabular}[t]{c|c|cc}
\multicolumn{2}{c}{$6:\psi^2 XH+\hbox{h.c.}$} &  \text{Restriction} & \#\\ 
\hline
$Q_{eW}$ & $(\bar \ell_p \sigma^{\mu\nu} e_r) \tau^I H W_{\mu\nu}^I$ & $p=r$ & 3 \\
$Q_{eB}$ & $(\bar \ell_p \sigma^{\mu\nu} e_r) H B_{\mu\nu}$ & $p=r$ & 3 \\
$Q_{uG}$ & $(\bar q_p \sigma^{\mu\nu} T^A u_r) \widetilde H \, G_{\mu\nu}^A$ & $-$ & $-$ \\
$Q_{uW}$ & $(\bar q_p \sigma^{\mu\nu} u_r) \tau^I \widetilde H \, W_{\mu\nu}^I$ & $-$ & $-$ \\
$Q_{uB}$ & $(\bar q_p \sigma^{\mu\nu} u_r) \widetilde H \, B_{\mu\nu}$ & $-$ & $-$ \\
$Q_{dG}$ & $(\bar q_p \sigma^{\mu\nu} T^A d_r) H\, G_{\mu\nu}^A$ & $-$ & $-$\\
$Q_{dW}$ & $(\bar q_p \sigma^{\mu\nu} d_r) \tau^I H\, W_{\mu\nu}^I$ & $-$ & $-$\\
$Q_{dB}$ & $(\bar q_p \sigma^{\mu\nu} d_r) H\, B_{\mu\nu}$& $-$ & $-$ \\
\end{tabular}
\end{minipage}
\hspace{2cm}
\begin{minipage}[t]{5.4cm}
\renewcommand{\arraystretch}{1.5}
\begin{tabular}[t]{c|c|cc}
\multicolumn{2}{c}{$7:\psi^2H^2 D$} &  \text{Restriction} & \#\\ 
\hline
$Q_{H \ell}^{(1)}$ & $(H^\dag i\overleftrightarrow{D}_\mu H)(\bar \ell_p \gamma^\mu \ell_r)$ & $p=r$ & 3 \\
$Q_{H \ell}^{(3)}$ & $(H^\dag i\overleftrightarrow{D}^I_\mu H)(\bar \ell_p \tau^I \gamma^\mu \ell_r)$ & $p=r$ & 3 \\
$Q_{H e}$ & $(H^\dag i\overleftrightarrow{D}_\mu H)(\bar e_p \gamma^\mu e_r)$ & $p=r$ & 3 \\
$Q_{H q}^{(1)}$ & $(H^\dag i\overleftrightarrow{D}_\mu H)(\bar q_p \gamma^\mu q_r)$& $-$ & $-$ \\
$Q_{H q}^{(3)}$ & $(H^\dag i\overleftrightarrow{D}^I_\mu H)(\bar q_p \tau^I \gamma^\mu q_r)$& $-$ & $-$\\
$Q_{H u}$ & $(H^\dag i\overleftrightarrow{D}_\mu H)(\bar u_p \gamma^\mu u_r)$&  $-$ & $-$\\
$Q_{H d}$ & $(H^\dag i\overleftrightarrow{D}_\mu H)(\bar d_p \gamma^\mu d_r)$& $-$ & $-$ \\
$Q_{H u d}$ + h.c. & $i(\widetilde H ^\dag D_\mu H)(\bar u_p \gamma^\mu d_r)$& $-$ & $-$\\
\end{tabular}
\end{minipage}

\vspace{0.4cm}
\hspace{-2.5cm}
\begin{minipage}[t]{4.75cm}
\renewcommand{\arraystretch}{1.5}
\begin{tabular}[t]{@{}c|c|cc}
\multicolumn{2}{c}{$8:(\bar LL)(\bar LL)$} &  \text{Restriction} & \#\\ 
\hline
\multirow{2}{*}{$Q_{\ell\ell}$} & \multirow{2}{*}{$(\bar \ell_p \gamma_\mu \ell_r)(\bar \ell_s \gamma^\mu \ell_t)$} 
  & $p=r\geq s=t$ & 6\\
 && $p=t>s=r$ & 3\\
$Q_{qq}^{(1)}$ & $(\bar q_p \gamma_\mu q_r)(\bar q_s \gamma^\mu q_t)$ & $-$ & $-$\\
$Q_{qq}^{(3)}$ & $(\bar q_p \gamma_\mu \tau^I q_r)(\bar q_s \gamma^\mu \tau^I q_t)$ & $-$ & $-$\\
$Q_{\ell q}^{(1)}$ & $(\bar \ell_p \gamma_\mu \ell_r)(\bar q_s \gamma^\mu q_t)$ & $p=r$ & 3\\
$Q_{\ell q}^{(3)}$ & $(\bar \ell_p \gamma_\mu \tau^I \ell_r)(\bar q_s \gamma^\mu \tau^I q_t)$ & $p=r$ & 3
\end{tabular}
\end{minipage}
\hspace{3cm}
\begin{minipage}[t]{4.75cm}
\renewcommand{\arraystretch}{1.5}
\begin{tabular}[t]{c|c|cc}
\multicolumn{2}{c}{$8:(\bar LL)(\bar RR)$} &  \text{Restriction} & \#\\ 
\hline
\multirow{2}{*}{$Q_{\ell e}$} & \multirow{2}{*}{$(\bar \ell_p \gamma_\mu \ell_r)(\bar e_s \gamma^\mu e_t)$} 
  & $p=r,\, s=t$ & 9\\
 && $p=t\neq s=r$ & 6\\
$Q_{\ell u}$ & $(\bar \ell_p \gamma_\mu \ell_r)(\bar u_s \gamma^\mu u_t)$ & $p=r$ & 3 \\
$Q_{\ell d}$ & $(\bar \ell_p \gamma_\mu \ell_r)(\bar d_s \gamma^\mu d_t)$ & $p=r$ & 3 \\
$Q_{qe}$ & $(\bar q_p \gamma_\mu q_r)(\bar e_s \gamma^\mu e_t)$ &  $p=r$ & 3\\
$Q_{qu}^{(1)}$ & $(\bar q_p \gamma_\mu q_r)(\bar u_s \gamma^\mu u_t)$ &  $-$ & $-$\\
$Q_{qu}^{(8)}$ & $(\bar q_p \gamma_\mu T^A q_r)(\bar u_s \gamma^\mu T^A u_t)$ &  $-$ & $-$\\
$Q_{qd}^{(1)}$ & $(\bar q_p \gamma_\mu q_r)(\bar d_s \gamma^\mu d_t)$ &  $-$ & $-$\\
$Q_{qd}^{(8)}$ & $(\bar q_p \gamma_\mu T^A q_r)(\bar d_s \gamma^\mu T^A d_t)$ & $-$ & $-$
\end{tabular}
\end{minipage}

\vspace{0.4cm}
\hspace{-1.5cm}
\begin{minipage}[t]{5.25cm}
\renewcommand{\arraystretch}{1.5}
\begin{tabular}[t]{c|c|cc}
\multicolumn{2}{c}{$8:(\bar RR)(\bar RR)$} &  \text{Restriction} & \#\\ 
\hline
$Q_{ee}$ & $(\bar e_p \gamma_\mu e_r)(\bar e_s \gamma^\mu e_t)$  
& $p=r\geq s=t$ & 6\\
$Q_{uu}$ & $(\bar u_p \gamma_\mu u_r)(\bar u_s \gamma^\mu u_t)$ & $-$ & $-$ \\
$Q_{dd}$ & $(\bar d_p \gamma_\mu d_r)(\bar d_s \gamma^\mu d_t)$ & $-$ & $-$ \\
$Q_{eu}$ & $(\bar e_p \gamma_\mu e_r)(\bar u_s \gamma^\mu u_t)$ & $p=r$ & 3 \\
$Q_{ed}$ & $(\bar e_p \gamma_\mu e_r)(\bar d_s\gamma^\mu d_t)$ & $p=r$ & 3 \\
$Q_{ud}^{(1)}$ & $(\bar u_p \gamma_\mu u_r)(\bar d_s \gamma^\mu d_t)$ & $-$ & $-$ \\
$Q_{ud}^{(8)}$ & $(\bar u_p \gamma_\mu T^A u_r)(\bar d_s \gamma^\mu T^A d_t)$ & $-$ & $-$ \\
\end{tabular}
\end{minipage}
\hspace{3cm}
\begin{minipage}[t]{5.5cm}
\renewcommand{\arraystretch}{1.5}
\begin{tabular}[t]{c|c|cc}
\multicolumn{2}{c}{$8:(\bar LR)(\bar L R)+\hbox{h.c.}$} &  \text{Restriction} & \#\\ 
\hline
$Q_{quqd}^{(1)}$ & $(\bar q_p^j u_r) \epsilon_{jk} (\bar q_s^k d_t)$  & $-$ & $-$\\
$Q_{quqd}^{(8)}$ & $(\bar q_p^j T^A u_r) \epsilon_{jk} (\bar q_s^k T^A d_t)$ &  $-$ & $-$\\
$Q_{\ell equ}^{(1)}$ & $(\bar \ell_p^j e_r) \epsilon_{jk} (\bar q_s^k u_t)$  & $p=r$ & 3\\
$Q_{\ell equ}^{(3)}$ & $(\bar \ell_p^j \sigma_{\mu\nu} e_r) \epsilon_{jk} (\bar q_s^k \sigma^{\mu\nu} u_t)$ & $p=r$ & 3\\
\end{tabular}

\vspace{0.5cm}
\begin{tabular}[t]{c|c|cc}
\multicolumn{2}{c}{$8:(\bar LR)(\bar RL)+\hbox{h.c.}$} &  \text{Restriction} & \#\\ 
\hline
$Q_{\ell edq}$ & $(\bar \ell_p^j e_r)(\bar d_s q_{tj})$ & $p=r$ & 3
\end{tabular}
\end{minipage}
\end{center}
\caption{Restrictions from $U(1)_e\times U(1)_\mu \times U(1)_\tau$ invariance on the lepton and baryon number preserving dim-6 SMEFT fermionic operators listed in \cref{tab:MLFVdim6SMEFTLNP}. These operators generate Lepton Universality Violating processes but no Charged Lepton Flavor Violations. Indicated in the table are the restricted structures of flavor indices $p, r, s, t$, together with the resulting total number of $U(1)_e\times U(1)_\mu \times U(1)_\tau$ invariants, where ``$-$'' means no restrictions needed and the resulting number will be the same as the dimension of the representation.}
\label{tab:MLFVdim6SMEFTLNPH}
\end{table}

\begin{table}[t]
\begin{center}
\fontsize{8.5pt}{11pt}\selectfont
\hspace{-3.5cm}
\begin{minipage}[t]{2.7cm}
\renewcommand{\arraystretch}{1.5}
\begin{tabular}[t]{c|c|c}
\multicolumn{2}{c}{$5: \psi^2H^3 + \hbox{h.c.}$}  & $SU(3)_{\ell, e, \nu}$ \\
\hline
$Q_{\ell\nu H}$ & $(\bar{\ell}_p \nu_r)\tilde{H}(H^\dag H)$ & $(\mathbf{3},\mathbf{1},\mathbf{\bar 3})$ \\
\end{tabular}
\end{minipage}

\vspace{0.4cm}
\hspace{-2.2cm}
\begin{minipage}[t]{5.2cm}
\renewcommand{\arraystretch}{1.5}
\begin{tabular}[t]{c|c|c}
\multicolumn{2}{c}{$6:\psi^2 XH+\hbox{h.c.}$} & $SU(3)_{\ell, e, \nu}$ \\
\hline
$Q_{\nu W}$ & $(\bar{\ell}_p \sigma^{\mu\nu} \nu_r) \tau^I \tilde{H} W_{\mu\nu}^I$ & $(\mathbf{3},\mathbf{1},\mathbf{\bar 3})$ \\
$Q_{\nu B}$ & $(\bar{\ell}_p \sigma^{\mu\nu} \nu_r) \tilde{H} B_{\mu\nu}$ & $(\mathbf{3},\mathbf{1},\mathbf{\bar 3})$ \\
\end{tabular}
\end{minipage}
\hspace{1.5cm}
\begin{minipage}[t]{5.4cm}
\renewcommand{\arraystretch}{1.5}
\begin{tabular}[t]{c|c|c}
\multicolumn{2}{c}{$7:\psi^2H^2 D$} & $SU(3)_{\ell, e, \nu}$ \\
\hline
$Q_{H \nu}$ & $(H^\dag i\overleftrightarrow{D}_\mu H)(\bar \nu_p \gamma^\mu \nu_r)$&  $(\mathbf{1},\mathbf{1},\mathbf{1}\oplus\mathbf{8})$\\
$Q_{H \nu e}$ + h.c. & $i(\widetilde H ^\dag D_\mu H)(\bar{\nu}_p \gamma^\mu e_r)$& $(\mathbf{1},\mathbf{\bar 3},\mathbf{3})$\\
\end{tabular}
\end{minipage}

\vspace{0.4cm}
\hspace{-2.0cm}
\begin{minipage}[t]{4.75cm}
\renewcommand{\arraystretch}{1.5}
\begin{tabular}[t]{c|c|c}
\multicolumn{2}{c}{$8:(\bar LL)(\bar RR)$} & $SU(3)_{\ell, e, \nu}$ \\
\hline
$Q_{\ell\nu}$ & $(\bar \ell_p \gamma_\mu \ell_r)(\bar \nu_s \gamma^\mu \nu_t)$ & $(\mathbf{1}\oplus\mathbf{8},\mathbf{1},\mathbf{1}\oplus\mathbf{8})$ \\
$Q_{q\nu}$ & $(\bar q_p \gamma_\mu q_r)(\bar \nu_s \gamma^\mu \nu_t)$ &  $(\mathbf{1},\mathbf{1},\mathbf{1}\oplus\mathbf{8})$\\
\end{tabular}
\end{minipage}

\vspace{0.4cm}
\hspace{-0.5cm}
\begin{minipage}[t]{5.25cm}
\renewcommand{\arraystretch}{1.5}
\begin{tabular}[t]{c|c|c}
\multicolumn{2}{c}{$8:(\bar RR)(\bar RR)$} & $SU(3)_{\ell, e, \nu}$ \\
\hline
$Q_{\nu\nu}$ & $(\bar \nu_p \gamma_\mu \nu_r)(\bar \nu_s \gamma^\mu \nu_t)$ & $(\mathbf{1},\mathbf{1},\mathbf{1}\oplus\mathbf{8}\oplus\mathbf{27})$ \\
$Q_{e\nu}$ & $(\bar e_p \gamma_\mu e_r)(\bar \nu_s \gamma^\mu \nu_t)$ & $(\mathbf{1},\mathbf{1}\oplus\mathbf{8},\mathbf{1}\oplus\mathbf{8})$ \\
$Q_{u\nu}$ & $(\bar u_p \gamma_\mu u_r)(\bar \nu_s \gamma^\mu \nu_t)$ & $(\mathbf{1},\mathbf{1},\mathbf{1}\oplus\mathbf{8})$ \\
$Q_{d\nu}$ & $(\bar d_p \gamma_\mu d_r)(\bar \nu_s \gamma^\mu \nu_t)$ & $(\mathbf{1},\mathbf{1},\mathbf{1}\oplus\mathbf{8})$ \\
$Q_{du\nu e}$ + h.c. & $(\bar d_p \gamma_\mu u_r)(\bar \nu_s \gamma^\mu e_t)$ & $(\mathbf{1},\mathbf{\bar 3},\mathbf{3})$ \\
\end{tabular}
\end{minipage}
\hspace{3.2cm}
\begin{minipage}[t]{5.5cm}
\renewcommand{\arraystretch}{1.5}
\begin{tabular}[t]{c|c|c}
\multicolumn{2}{c}{$8:(\bar LR)(\bar L R)+\hbox{h.c.}$} & $SU(3)_{\ell, e, \nu}$ \\
\hline
$Q_{\ell\nu\ell e}$ & $(\bar \ell_p^j \nu_r) \epsilon_{jk} (\bar \ell_s^k e_t)$  & $(\mathbf{\bar 3}\oplus\mathbf{6},\mathbf{\bar 3},\mathbf{\bar 3})$\\
$Q_{\ell\nu q d}$ & $(\bar \ell_p^j \nu_r) \epsilon_{jk} (\bar q_s^k d_t)$ &  $(\mathbf{3},\mathbf{1},\mathbf{\bar 3})$\\
$Q_{\ell d q\nu}$ & $(\bar \ell_p^j d_r) \epsilon_{jk} (\bar q_s^k \nu_t)$  & $(\mathbf{3},\mathbf{1},\mathbf{\bar 3})$\\
\end{tabular}

\vspace{0.3cm}
\begin{tabular}[t]{c|c|c}
\multicolumn{2}{c}{$8:(\bar LR)(\bar RL)+\hbox{h.c.}$} & $SU(3)_{\ell, e, \nu}$\\
\hline
$Q_{q u \nu \ell}$ & $(\bar{q}_p u_r)(\bar{\nu}_s \ell_t)$ & $(\mathbf{\bar 3},\mathbf{1},\mathbf{3})$\\
\end{tabular}
\end{minipage}
\end{center}
\caption{Lepton and baryon number preserving dim-6 $\nu$SMEFT operators involving right-handed neutrino fields $\nu$ \cite{Liao:2016qyd}, together with the corresponding spurious transformations of the Wilson coefficients under $SU(3)_{\ell, e, \nu} \equiv SU(3)_\ell \times SU(3)_e \times SU(3)_\nu \subset G_{Lf}$, the non-abelian part of the MLFV group. Transformations under the $U(1)$ factors can be inferred from \cref{eqn:TransnuSMEFT}. The subscripts $p, r, s, t$ denote flavor indices.}
\label{tab:MLFVdim6nuSMEFTLNP}
\end{table}

\begin{table}[t]
\begin{center}
\fontsize{8.5pt}{11pt}\selectfont
\renewcommand{\arraystretch}{1.5}
\begin{tabular}[t]{c|c|c|c}
\multicolumn{2}{c}{$L$-violating + h.c.} & \multicolumn{1}{c}{$SU(3)_{\ell, e, \nu}$} & \# invariants under lepton parity $\mathcal{P}_\text{LN}$ \\
\hline
$Q_{\nu\nu\nu\nu}$ & $(\nu_p C \nu_r)(\nu_s C \nu_t)$&  $(\mathbf{1},\mathbf{1},\mathbf{6})$ & $6$\\
\end{tabular}

\vspace{0.4cm}
\begin{tabular}[t]{c|c|c|c}
\multicolumn{2}{c}{$L$- and $B$-violating + h.c.} & \multicolumn{1}{c}{$SU(3)_{\ell, e, \nu}$} & \# invariants under lepton parity $\mathcal{P}_\text{LN}$ \\
\hline
$Q_{qqd\nu}$ & $\epsilon_{\alpha\beta\gamma}\epsilon_{jk}[(q_p^{\alpha j})^T C q_r^{\beta k}] [(d_s^\gamma)^T C \nu_t]$ & $(\mathbf{1},\mathbf{1},\mathbf{\bar 3})$ & 0 \\
$Q_{udd\nu}$ & $\epsilon_{\alpha\beta\gamma} [(u_p^{\alpha})^T C d_r^\beta] [(d_s^\gamma)^T C \nu_t]$ & $(\mathbf{1},\mathbf{1},\mathbf{\bar 3})$ & 0\\
\end{tabular}
\end{center}
\caption{Lepton number violating dim-6 $\nu$SMEFT operators involving right-handed neutrino fields $\nu$ \cite{Liao:2016qyd}, together with the corresponding spurious transformations of the Wilson coefficients under $SU(3)_{\ell, e, \nu} \equiv SU(3)_\ell \times SU(3)_e \times SU(3)_\nu \subset G_{Lf}$, the non-abelian part of the MLFV group. Transformations under the $U(1)$ factors can be inferred from \cref{eqn:TransnuSMEFT}. Here $C=i\gamma^2\gamma^0$ is the charge conjugation matrix. The subscripts $p, r, s, t$ denote flavor indices. All these operators violate the lepton number $U(1)_\text{LN}$; among these the operator $Q_{\nu\nu\nu\nu}$ preserves the lepton parity $\mathcal{P}_\text{LN} \equiv (-1)^\text{LN}$.}
\label{tab:MLFVdim6nuSMEFTLNV}
\end{table}

\bibliographystyle{JHEP}
\bibliography{ref}
\end{spacing}

\end{document}